\documentclass[journal=jacsat,manuscript=article]{achemso}

\usepackage[version=3]{mhchem} % Formula subscripts using \ce{}
\usepackage[utf8]{inputenc} % Add this line
\usepackage{graphicx}% Include figure files
\usepackage{dcolumn}% Align table columns on decimal point
\usepackage{bm}% bold math
\usepackage{amsmath}
\usepackage{enumitem}
\usepackage{hyperref} % add hypertext capabilities
\usepackage[user]{zref}
\usepackage{float}
\usepackage{placeins}
\usepackage{multirow}
\usepackage{array}
\usepackage{makecell}
\usepackage{tabularray}
\usepackage{float}
\usepackage{amsfonts}
\usepackage{booktabs} 
\newcommand{\te}[1]{\overline{\overline{\mathrm{#1}}}}
\newcommand{\ve}[1]{\mathbf{#1}}
\author{Hossein Allahverdizadeh}
\email{hossein.allahverdizadeh@epfl.ch}
\author{Karim Achouri}
\email{karim.achouri@epfl.ch}
\affiliation[EPFL]
{Institute of Electrical and Microengineering, École Polytechnique Fédérale de Lausanne (EPFL), Laboratory for Advanced Electromagnetics and Photonics, Lausanne, Switzerland}

\title[An \textsf{achemso} demo]
  {Multipolar Angular Scattering \\ of Substrated Metasurfaces}


\begin{document}


\begin{abstract}
Properly modeling and predicting the scattering response of a metasurface is a particularly challenging task. This has been shown to be especially difficult if the metasurface supports both local and nonlocal interactions, in the form of lattice coupling effects, multipolar contributions or bianisotropic responses. So far, existing methods to approach this problem have been restricted to normal incidence in a homogeneous background medium. We overcome these limitations by providing a rigorous and comprehensive formalism that accommodates both oblique incidence and the presence of different superstrate and substrate. This is achieved by extending our existing metasurface modeling framework to account for nonlocal and multipolar contributions up to the octupolar order and properly accounting for the scattering effects due to an inhomogeneous background medium. Additionally, our method is based on exact spherical multipole decomposition, which intrinsically accounts for toroidal contributions. We demonstrate the effectiveness of our approach by modeling the response of several dielectric and plasmonic metasurfaces that exhibit sharp spectral features including bound states in the continuum. Overall, our formalism yields excellent agreement with full-wave simulations.
\end{abstract}


\section{\label{sec:intro}Introduction  \protect }

Metasurfaces are planar arrays of subwavelength resonant unit cells that have evolved into a powerful platform for manipulating electromagnetic waves through careful design. A plethora of works have already shown that metasurfaces can independently control amplitude, phase, polarization, frequency, and momentum, enabling a vast array of applications~\cite{li2023universal,lin2014dielectric,achouri2021}. In addition, manipulating intrinsic resonances of single scatters alongside lattice coupling effects, allows us to engineer multi-resonant responses, supporting broadband and efficient wavefront manipulation~\cite{kim2024multi}. These developments highlight the versatility of metasurfaces in the fields of photonics and electromagnetics.

Besides their temporal spectral control, metasurfaces can also exhibit spatially nonlocal responses, meaning that their behavior depends not only on the local fields at each point but also on the angle of incidence or the transverse wavevector of the incoming wave~\cite{simovski2018}. As a consequence, nonlocal metasurfaces can distinguish between different momentum components of an incident wave~\cite{pfeiffer2013metamaterial,silveirinha2014nonlocal,zhou2020nonlocal,achouri2020a}, and support highly angle-selective resonances. This enables applications like angular filtering, vortex mode selection and generation~\cite{achouri2015general,achouri2021}, or the realization of bound states in the continuum (BIC)~\cite{zhen2014topological,doeleman2018experimental}. They can also be used to achieve generalized Kerker effects at oblique incidence, a concept which is often only considered for normally incident waves~\cite{PhysRevLett.122.193905}. On the other hand, nonlocal metasurfaces may also be used to realize angle-independent responses~\cite{yucel2024angle}, demonstrating that nonlocality does not only increase angular dispersion but may also be used to completely suppress it. This ability to engineer a metasurface angular scattering response becomes especially important when metasurfaces are designed to work under wide-angle illumination, where modal interference and guided-mode resonances strongly depend on the angle. By engineering the nonlocality of metasurfaces, we can thus realize complex angular responses. This enables a powerful approach for realizing applications such as beam steering, diffraction channel control~\cite{gong2023multipolar,liu2017beam}, or signal modulation across multiple channels~\cite{momeni2019generalized,abdolali2019parallel}, and multiplexing~\cite{jang2021} which are important in free-space communication, optical computing, and photonic devices. 

To understand and design metasurfaces, multipolar decomposition-based modal analysis has become a key method. It gives deep physical insight by identifying which electric or magnetic multipoles contribute and how they interfere to form the overall metasurface scattering response. This has been widely studied, from single particle responses to collective lattice effects~\cite{alaee2018electromagnetic,allayarov2024multiresonant,rahimzadegan2022comprehensive,babicheva2017resonant}.

One of the ongoing challenges in these methods is the correct identification and treatment of transverse contributions, especially when the illumination breaks in-plane symmetry (in the cases of oblique incidence) or when the metasurface is placed in an inhomogeneous environment (e.g., on a substrate). For these reasons, most existing techniques have been restricted to the case where the metasurface is illuminated at normal incidence and is placed inside a homogeneous medium~\cite{savinov2014toroidal,babicheva2017resonant,allayarov2024multiresonant} or to the case where the metasurface is only modeled with dipolar responses~\cite{albooyeh2011,achouri2020a,cuesta2025}. Additionally, when the presence of a substrate and/or oblique incidence is considered, the field inside the structure may include multipolar contributions with more complex angular behavior, including higher-order or mixed-parity modes, which are hard to isolate using primitive or irreducible Cartesian multipoles. In such situations, exact spherical multipoles are needed~\cite{nanz2016toroidal}. But interpreting effects in metasurface responses that account for transverse modes under oblique incidence emerging from exact spherical multipoles has been difficult~\cite{prokhorov2022resonant}.

To deal with these issues, we propose a general framework using exact spherical multipole expansions together with the Generalized Sheet Transition Conditions (GSTC)~\cite{achouri2015general,achouri2021,achouri2023spatial}. This approach brings two main advantages. First, the spherical basis naturally provides a complete and orthogonal set of multipolar moments. Second, by using the GSTC, we reduce the problem to surface current distributions, avoiding the need to solve for volumetric field profiles~\cite{achouri2021}. This becomes especially useful in practical metasurfaces with substrates or even multilayers provided that they are subwavelength. We also show that this method works even in cases with BIC, where the leakage is highly angle selective and the Q factors are extremely high~\cite{hsu2016bound,zhen2014topological}.

\begin{figure}[h!]
  \centering
  \includegraphics[width=0.8\textwidth]{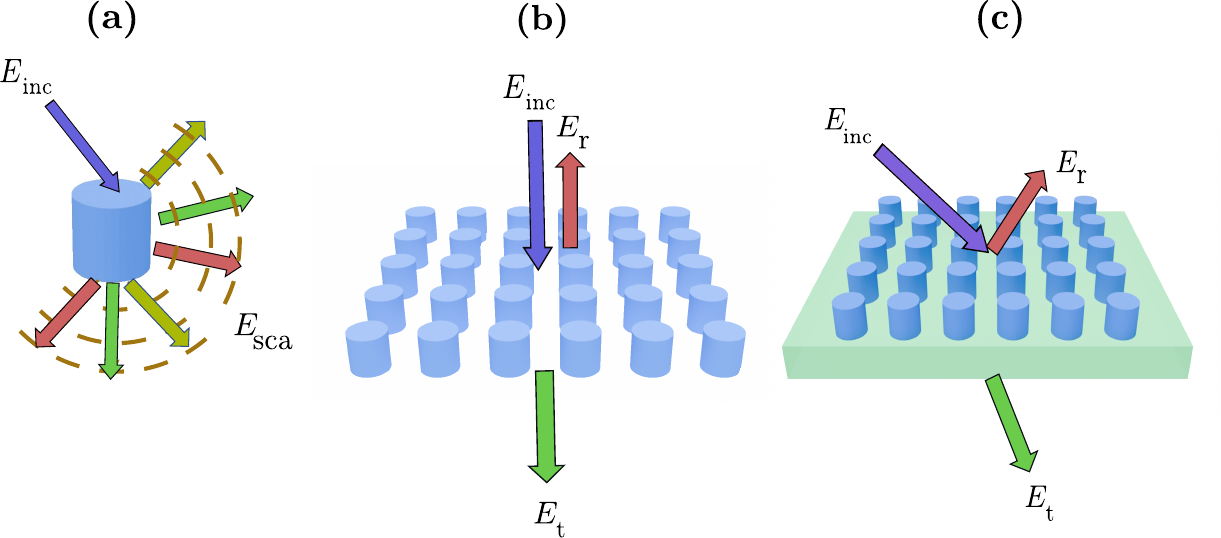} % Replace with your image file
  \caption{Increasing complexity of scattering scenarios. (a)~Scattering by a single particle. (b)~Scattering by a metasurface at normal incidence. (c)~Scattering by a metasurface at oblique incidence and in the presence of a substrate.}
  \label{fig_intro}
\end{figure}

In this paper, we start by reviewing the exact formulation of spherical multipoles, from a single particle to a periodic metasurface under normal incidence, as shown in Fig.~\ref{fig_intro}a,b. Then, we derive an equivalent surface current representation of the metasurface, including up to octupolar moments. Then, using the GSTC, we calculate the co-polarized oblique reflection and transmission coefficients for TE and TM waves, accounting for different substrate and superstrate materials, as suggested in Fig.~\ref{fig_intro}c.

To validate the robustness of our model, we consider several examples. We start by using our method to predict the TE and TM scattering coefficients for oblique incidence of metasurfaces made out of dispersive materials and embedded in free space. These predictions are then compared to full-wave simulations to asses their agreement. Then, we will examine configurations involving dielectric substrates and investigate the impact of the spatial origin of the equivalent surface current distribution since, in the presence of a substrate, which breaks normal mirror symmetry, it is not straightforward to know where the latter should be placed~\cite{kildishev2023art}. Next, we will apply our model to scenarios exhibiting bound states in the continuum and will show that we are able to successfully capture the emergence of narrow-band multipolar resonances. This behavior is analyzed through detailed multipolar decomposition. Finally, we will investigate a BIC configuration in which the structure is embedded asymmetrically within a dielectric medium occupying one half space. This setup leads to the emergence of a pair of distinct BIC, whose origin we identified and successfully validated with our modeling framework.

\section{Review of Spherical Multipoles Based Modeling}

The electric field scattered by an isolated particle, as the one shown in Fig.~\ref{fig_intro}a, may be expressed in terms of a multipolar decomposition. In what follows, we shall only consider an exact multipolar decomposition based on vector spherical harmonics as it intrinsically yields a complete and orthogonal set of irreducible multipolar moments valid beyond the long-wavelength approximation~\cite{alaee2018electromagnetic}. For particle centered at the origin of the system of coordinates, its scattered electric field, $\ve{E}_\textbf{s}(\ve{r})$, may be expressed as~\cite{rahimzadegan2022comprehensive}
\begin{equation}
\label{eq_EsExact}
\mathbf{E}_\text{s}(\mathbf{r})=\sum_{l=1}^{\infty} \sum_{m=-l}^{l} b_{l m}^{\mathrm{e}} \mathbf{N}_{l m}^{(3)}(k \mathbf{r})+b_{l m}^{\mathrm{m}} \mathbf{M}_{l m}^{(3)}(k \mathbf{r}),
\end{equation}
where $\ve{r}=x\hat{\mathbf{x}} + y\hat{\mathbf{y}} + z\hat{\mathbf{z}}$ is the observation point, $k$ is the wavenumber in the background medium, and $\{\mathbf{N}_{lm}, \mathbf{M}_{lm}\}$ are respectively the electric and magnetic vector spherical harmonics that are associated with the coefficients $\{b^\text{e}_{lm},b^\text{m}_{lm}\}$. All of these quantities are defined in Appendix B of the Supplementary Information. While~\eqref{eq_EsExact} provides an exact and rigorous expansion of the scattered field in spherical coordinates, it is often convenient to express it in Cartesian coordinates. For this purpose, we map the vector spherical harmonic functions $\{\mathbf{N}_{lm}, \mathbf{M}_{lm}\}$ into corresponding Cartesian multipoles using the relation Eq.~(B3) in Appendix~B of the Supplementary Information. Restricting ourselves to the first three multipolar orders, relation~\eqref{eq_EsExact} transforms into
\begin{equation}
\label{scat_direct}
\begin{split}
\mathbf{E}_{\mathrm{s}}(\mathbf{r}) &= \frac{k^2}{4 \pi} \frac{e^{ikr}}{r} \Bigg[
\hat{\mathbf{r}} \times \left( \frac{1}{\epsilon} \mathbf{p} \times \hat{\mathbf{r}} \right) 
+ \eta \left( \mathbf{m} \times \hat{\mathbf{r}} \right) - \frac{ik}{2\epsilon} \hat{\mathbf{r}} \times \left(\mathbf{Q}^{\text{(e)}} \times \hat{\mathbf{r}} \right) 
\\
&\qquad - \frac{{ik\eta}}{2}  \left( \mathbf{Q}^{\text{(m)}} \times \hat{\mathbf{r}} \right) - \frac{k^2}{6\epsilon} \hat{\mathbf{r}} \times \left(\mathbf{O}^{\text{(e)}} \times \hat{\mathbf{r}} \right) 
- \frac{k^2\eta}{6}  \left( \mathbf{O}^{\text{(m)}} \times \hat{\mathbf{r}} \right)
\Bigg],
\end{split}
\end{equation}
where $\ve{\hat{r}}=\ve{r}/|\ve{r}|$, and the electric and magnetic dipole $\{\ve{p},\ve{m}\}$, quadrupolar $\{\mathbf{Q}^{\text{(e)}},\mathbf{Q}^{\text{(m)}}\}$ and octupolar $\{\mathbf{O}^{\text{(e)}},\mathbf{O}^{\text{(m)}}\}$ moments are given in Table~\ref{tab:traceless_multipoles}.

We may now apply this formalism to the case of a 2D rectangular lattice lying in the $xy$-plane, as the one illustrated in Fig.~\ref{fig_intro}b. To do so, we sum the contributions from all individual scattering particles over the entire lattice. In spherical coordinates, this transforms~\eqref{eq_EsExact} into
\begin{equation}
\mathbf{E}_{\mathrm{s}}(\mathbf{r})=\sum_{\mathbf{R}} \sum_{l=1}^{\infty} \sum_{m=-l}^{l}\left\{b_{l m}^{\mathrm{e}} \mathbf{N}_{l m}^{(3)}(k(\mathbf{r}-\mathbf{R}))+b_{l m}^{\mathrm{m}} \mathbf{M}_{l m}^{(3)}(k(\mathbf{r}-\mathbf{R}))\right\} e^{i \mathbf{k}_{\|} \cdot \mathbf{R}},
\label{spheric,2}
\end{equation}
where $\mathbf{R}=n_1\hat{\mathbf{x}} + n_2\hat{\mathbf{y}}$ is a periodicity vector with $n_1$ and $n_2$ corresponding to the location of each particle in the lattice. We may now simplify this equation by considering that our array is subwavelength implying that only specular radiation ($0^\text{th}$-diffraction orders) reaches the far-field region. For this purpose, we start by expressing~\eqref{spheric,2} in reciprocal space by taking its Fourier transform leading to~\cite{rahimzadegan2022comprehensive}
\begin{equation}
\mathbf{E}_{s}(\mathbf{r})=\sum_{l=1}^{\infty} \sum_{m=-l}^{l} \frac{2 \pi i^{-l}}{A k} \sum_{\mathbf{G}}\left[b_{l m}^{\mathrm{e}} i \mathbf{Z}_{l m}\left(\mathbf{k}_{\mathbf{G}}^{ \pm}\right)+b_{l m}^{\mathrm{m}} \mathbf{X}_{l m}\left(\mathbf{k}_{\mathbf{G}}^{ \pm}\right)\right] \frac{e^{i \mathbf{k}_{\mathbf{G}}^{ \pm} \cdot \mathbf{r}}}{\left|k_{\mathbf{G}, z}^{ \pm}\right|},
\label{final_spherical}
\end{equation}
where $\mathbf{G}=\frac{2\pi n_1}{\Lambda_1}\hat{\mathbf{x}} +\frac{2\pi n_2}{\Lambda_2}\hat{\mathbf{y}}$ is the reciprocal lattice vector, $\Lambda_1$ and $\Lambda_2$ are the lattice periods along $x$ and $y$, respectively, and $A=\Lambda_1\Lambda_2$ is the unit-cell area. The functions $\mathbf{Z}_{l m}$ and $\mathbf{Z}_{l m}$ are defined in Appendix~B of the Supplementary Information. The diffraction wave vector, $\ve{k}_\ve{G}^\pm$ is defined as~\cite{rahimzadegan2022comprehensive}
\begin{equation}
    \mathbf{k}^{\pm}_{\mathbf{G}} = \mathbf{k}_{\parallel} + \mathbf{G} \pm \hat{\mathbf{z}} \sqrt{k^2 - (\mathbf{k}_{\parallel} + \mathbf{G})^2} 
= \mathbf{k}_{\parallel} + \mathbf{G} \pm k_{\mathbf{G},z} \, \hat{\mathbf{z}},
\end{equation}
where $\mathbf{k}_{\parallel}$ is defined as the transverse wave vector and $k_{\mathbf{G},z}$ as the vertical wave vector component.  Now, we set $\mathbf{G}=0$ in~\eqref{final_spherical} to satisfy the prescription of specular radiation and transform the resulting expression into Cartesian coordinates. The field scattered by the array, assuming a \emph{normally incident} plane wave, is finally obtained following the procedure described in~\cite{savinov2014toroidal} as
\begin{equation}
\begin{aligned}
\mathbf{E}_\text{s} &=
\mu c^2
\bigg\{
- ik \mathbf{p}_{\parallel} - \frac{ik}{c} \hat{\mathbf{z}} \times \mathbf{m}_{\parallel} + \frac{k^2}{2} \left( \mathbf{Q}^{\text{(e)}} \cdot \hat{\mathbf{z}} \right)_{\parallel}
- \frac{k^2}{2c} \hat{\mathbf{z}} \times \left( \mathbf{Q}^{\text{(m)}} \cdot \hat{\mathbf{z}} \right)_{\parallel}\\
&\qquad\qquad + \frac{ik^3}{6} \left[ \left( \mathbf{O}^{\text{(e)}} \cdot \hat{\mathbf{z}} \right) \cdot \hat{\mathbf{z}} \right]_{\parallel}
+ \frac{ik^3}{6c} \hat{\mathbf{z}} \times 
\left[ \left( \mathbf{O}^{\text{(m)}} \cdot \hat{\mathbf{z}} \right) \cdot \hat{\mathbf{z}} \right]_{\parallel}
\bigg\} e^{ikz}.
\end{aligned}
\end{equation}
It is clear that this analysis is only valid for the case of normal incidence and for a metasurface embedded in a homogeneous and uniform background medium. In this work, we aim to extend this formula to accommodate arbitrary oblique incidence and include the case of non-uniform  background media (different superstrate and substrate).

\begin{table}[H]
\centering
\caption{Exact electric and magnetic multipole moments~\cite{alaee2018electromagnetic}. The terms $B^{(\text{m},\text{e})}_{ijkl}$ and $\mathbf{V}^{(\text{m},\text{e})}$ are provided in Table~2 of Appendix~B in the Supplementary Information}.
\renewcommand{\arraystretch}{2.5}
\setlength{\tabcolsep}{12pt} % Adjust column spacing
\footnotesize
\begin{tabular}{c|p{13cm}}
\hline
\textbf{Order} & \textbf{Electric Multipoles Representation}  \\ \hline
1 &
\(
p_i = -\frac{1}{i\omega A} \int d^3 \mathbf{r} \, J_i j_0(kr) 
+ \frac{k^2}{2} \int d^3 \mathbf{r} \Big[ 3(\mathbf{r} \cdot \mathbf{J}) r_i - r^2 J_i \Big] j_2(kr)
\) \\ \hline
2 &
\(
Q_{ij}^\text{(e)}  = -\frac{1}{i \omega A} \bigg\{ 
 \int d^3 \mathbf{r} \bigg[
3 \Bigl( r_j\,J_i + r_i\,J_j \Bigr)
- 2 \Bigl( \mathbf{r}\cdot\mathbf{J} \Bigr) \delta_{ij}
\bigg] \frac{j_1(kr)}{kr} 
 + 2k^2 \int d^3 \mathbf{r} \bigg[
5\,r_i\,r_j\,\Bigl(\mathbf{r}\cdot\mathbf{J}\Bigr)
- \Bigl( r_i\,J_j + r_j\,J_i \Bigr)r^2-r^2\Bigl(\mathbf{r}\cdot\mathbf{J}\Bigr) \delta_{ij}
\bigg] \frac{j_3(kr)}{(kr)^3}
\bigg\}
\) \\ \hline
3 &
\(
O_{ijk}^\text{(e)}  = -\frac{15}{i\omega A}\int_V \frac{j_2(kr)}{(kr)^2}\Bigl[ \; 
 J_i\,r_j\,r_k + r_j\,J_i\,r_k + r_k\,r_i\,J_j -\; \Bigl( \delta_{ij}\,V_k^{\text{(e)}} + \delta_{ik}\,V_j^{\text{(e)}} + \delta_{jk}\,V_i^{\text{(e)}} \Bigr)
\Bigr]\, dV
\)\\ \hline

4 &
\(
X_{ijkl}^\text{(e)} = -\frac{105}{i\omega A}\int_V \frac{j_4(kr)}{(kr)^4}\Bigl[ \; 
 J_i\,r_j\,r_k\,r_l + r_i\,J_j\,r_k\,r_l 
+\; r_i\,r_j\,J_k\,r_l + r_i\,r_j\,r_k\,J_l
-\; B^{\text{(e)}}_{ijkl}
\Bigr]\, dV
\) \\ \hline

\textbf{Order} & \textbf{Magnetic Multipoles Representation}  \\ \hline
1 &
\(\displaystyle
m_i = \frac{3}{2A} \int d^3 \mathbf{r} \, \Big(\mathbf{r} \times \mathbf{J} \Big)_i \frac{j_1(kr)}{kr}
\) \\ \hline
2 &
\(\displaystyle
Q_{ij}^\text{(m)} = \frac{5}{A} \int d^3 \mathbf{r} \Big\{ r_i \Big(\mathbf{r} \times \mathbf{J} \Big)_j 
+ r_j \Big(\mathbf{r} \times \mathbf{J} \Big)_i \Big\} \frac{j_2(kr)}{(kr)^2}
\) \\ \hline
3 &
\(
O_{ijk}^\text{(m)}  = \frac{105}{4A} \int_V \frac{j_3(k r)}{(k r)^3} \Big[
\; \epsilon_{ilm}\, r_l\, J_m\, r_j\, r_k
+ r_i\, \epsilon_{jlm}\, r_l\, J_m\, r_k 
+ r_i\, r_j\, \epsilon_{klm}\, r_l\, J_m 
- \Bigl( \delta_{ij}\, V_k^{\text{(m)}} + \delta_{ik}\, V_j^{\text{(m)}} + \delta_{jk}\, V_i^{\text{(m)}} \Bigr)
\Big]\, dV
\) \\ \hline
4 &
\(\displaystyle
X_{ijkl}^\text{(m)} = \frac{189}{A} \int \frac{j_4(kr)}{(kr)^4} \Big[ \epsilon_{ipq}\, r_p\, J_q\, r_j\, r_k\, r_l 
 + r_i\, \epsilon_{jpq}\, r_p\, J_q\, r_k\, r_l 
 + r_i\, r_j\, \epsilon_{kpq}\, r_p\, J_q\, r_l 
 + r_i\, r_j\, r_k\, \epsilon_{lpq}\, r_p\, J_q 
 - B^{\text{(m)}}_{ijkl} \Big]\, dV
\) \\ \hline
\end{tabular}
\label{tab:traceless_multipoles}
\end{table}
It is important to note that the multipolar coefficients in Table~\ref{tab:traceless_multipoles} represent \textbf{irreducible Cartesian components}, which are obtained by directly mapping the corresponding spherical components to their Cartesian counterparts. For an irreducible representation, the tensor must be both \textbf{symmetric} and \textbf{traceless}. This implies that for a multipole of order \( n \) (corresponding to an \( n \times n \) tensor), only \( 2n+1 \) independent components are needed to fully define the irreducible tensor. Interestingly, these \( 2n+1 \) components correspond precisely to the modes characterized by \( l \)~(angular momentum) with \(-l \leq m \leq l\) (i.e. \( 2l+1 \) components). In a similar manner, the symmetric-traceless multipole representation can also be derived directly from a Cartesian expansion. This approach yields a symmetrized, traceless form that is particularly insightful for revealing the emergence of toroidal modes. Further details are provided in Appendix~B of the Supplementary Information.

In Table~\ref{tab:traceless_multipoles}, the definitions of the multipolar moments involve current terms denoted by \( \mathbf{J} \). In practice, to compute the multipolar components, one needs to define \( \mathbf{J} \). However, for dielectric particles, there are no real (conduction) currents present within the object. Instead, from numerical simulations, we can obtain the electric field inside the object and then calculate the equivalent current density distribution using
\begin{equation}
    \mathbf{J} = -i\omega\epsilon\left(\epsilon_\text{d} - \epsilon_\text{b}\right)\mathbf{E},
    \label{eq_vol_cur}
\end{equation}
where \(\epsilon_\text{d}\) is the object relative permittivity and \(\epsilon_\text{b}\) is the relative permittivity of the medium containing both the object and the incident field~\cite{paknys2016}.

\section{Far-Field Scattering Based on Boundary Conditions}

In order to overcome the limitations of previous modeling approaches and appropriately accommodate arbitrary oblique incidence and include the presence of non-uniform background media, we now consider a modeling technique for metasurfaces based on boundary conditions. Specifically, we shall use the Generalized Sheet Transition Conditions (GSTC), which have been shown to be excellent at modeling the angular response of a metasurface even in the presence of different substrate and superstrate~\cite{idemen2011discontinuities,achouri2022multipolar,tiukuvaara2023quadrupolar}. While successful, this model has so far only been used in conjunctions with effective material parameters (typically susceptibilities), which are numerically computed by solving a large system of equations based on the GSTC. This approach is straightforward to apply in the dipolar regime, where only few susceptibility components play a role. However, it becomes exponentially more complicated to apply when quadrupolar or even octupolar contributions are taken into account as it significantly increases the number of effective susceptibilities to solve for~\cite{achouri2022multipolar,tiukuvaara2023quadrupolar}. For this reason, we will not consider effective medium parameters in this work but rather directly use the expression of the multipole moments produced by each scattering particles composing the metasurface and that thus intrinsically include lattice coupling effects, as done in~\cite{savinov2014toroidal,babicheva2017resonant,allayarov2024multiresonant}.

Note that historically the GSTC have been limited to the dipolar regime~\cite{achouri2021}. Recently, they have been extended to include quadrupolar responses in~\cite{achouri2022multipolar}. However, to properly account for exotic scattering effects, such as bound states in the continuum (BIC), higher-order multipolar responses must be included~\cite{10.1117/12.3016150}. For this purpose, we shall next provide a derivation of the GSTC that includes multipolar responses up to the octupolar order. The method that we will use to derive the GSTC is based on a Green function approach developed in~\cite{achouri2022multipolar}, which requires expressing the metasurface response in terms of an equivalent surface current that is expanded based on spherical multipole moments. As a consequence, we start by deriving an expression of a metasurface equivalent current up to the octupolar order. 

\subsection{Equivalent Current for an Isolated Particle}

We now consider that an isolated particle may be modeled as an equivalent homogeneous current point source, \( \mathbf{J}_{\text{h}} \) expressed in [A$\cdot$m], for which the electric vector potential reads
\begin{equation}
\mathbf{A}(\mathbf{r}) = \frac{\mu}{4\pi} \, \mathbf{J}_{\text{h}} \, \frac{e^{ikr}}{r},
\label{poten_cordinate}
\end{equation}
where \( r = |\mathbf{r}| \). The current $\ve{J}_\text{h}$ is expanded in terms of multipolar moments using the procedure as outlined in~\cite{nanz2016toroidal,papas1988}, leading to
\begin{equation}
\mathbf{J}_{\text{h}} = \sum_{n=1}^{\infty} \frac{(-1)^{n-1}}{n!} \left[\nabla \times (\nabla^{n-1} \cdot\mathbf{M}_n) -  i\omega\nabla^{n-1} \cdot  \mathbf{P}_n \right],
\label{stef_potential}
\end{equation}
where $\{\mathbf{P}_n,\mathbf{M}_n\}$ are the $\text{n}^\text{th}$ electric and magnetic multipolar moments given in Eq.~(B34) in Appendix~B of the Supplementary Information. Restricting ourselves to the octupolar order ($n = 3$) and transforming the resulting equation into Cartesian coordinates, reduces~\eqref{stef_potential} to
\begin{equation}
\mathbf{J}_\mathrm{h} = -i\omega \mathbf{p} + \nabla \times \mathbf{m} + \frac{i\omega}{2} \nabla \cdot \mathbf{Q}^\text{(e)} - \frac{1}{2} \nabla \times (\nabla \cdot\mathbf{Q}^\text{(m)}) - \frac{i \omega}{6} {\nabla}\cdot 
\big( {\nabla} \cdot {{\mathbf{O}^\text{(e)}}}\big)+   \frac{1}{6} \nabla \times \left( \nabla \cdot\nabla \cdot \mathbf{O}^{\text{(m)}} \right)
.
\label{current_equivalent}
\end{equation}
\noindent
Now that we have established the equivalent current for an isolated particle, we can determine the corresponding equivalent surface current for a metasurface and subsequently derive the associated boundary conditions.

\subsection{Transverse Electric and Magnetic Field Boundary Conditions}

The equivalent electric current density of a spatially varying metasurface lying in the $xy$-plane at $z=0$ takes the general form~\cite{achouri2022multipolar}
\begin{equation}
\mathbf{J}(x, y,z)=\delta(z) \mathbf{J}_{\mathrm{s}}(x, y),
\label{current_basic}
\end{equation}
where $\ve{J}_\mathrm{s}$ is in [A/m]. This surface current may be connected to the isolated particle current given in~\eqref{current_equivalent} by dividing the latter by the unit-cell surface area, $A$, such that $\ve{J}_\mathrm{s} = \ve{J}_\mathrm{h}/A$. In the general case where the metasurface is spatially varying, it follows that the multipolar moments in~\eqref{current_equivalent} are different in each unit cell.

Now, to derive the octupolar extension of the GSTC, we first need to obtain the electric vector potential, $\mathbf{A}$, associated with $\mathbf{J}$, that must satisfy the Helmholtz equation
\begin{equation}
    \nabla^2 \mathbf{A} + k^2 \mathbf{A} = -\mu \mathbf{J}.
    \label{helmholmz}
\end{equation}
A solution to this equation may be found by transforming it into the Fourier domain~\cite{achouri2022multipolar}. To do so, we consider the inverse Fourier integral of the current given by
\begin{equation}
    \mathbf{J}_\mathrm{s} = \int_{-\infty}^{+\infty} \int_{-\infty}^{+\infty} 
\tilde{\mathbf{J}}_s e^{i(k_x x + k_y y)} \, dk_x \, dk_y.
\label{current}
\end{equation}
Substituting~\eqref{current} along with~\eqref{current_basic} into~\eqref{helmholmz} leads to~\cite{achouri2022multipolar}
\begin{equation}
\mathbf{\tilde A}=-\frac{\mu}{2 i k_z} \mathbf{\tilde J}_\mathrm{s} e^{i k_z|z|}.
\label{elec_potential}
\end{equation}
Note that, even though we have solved the Helmholtz equation assuming a homogeneous environment, this derivation remains valid even in the presence of different substrate and superstrate, as was demonstrated in~\cite{tiukuvaara2023quadrupolar,idemen2011discontinuities}. Now that we have the electric potential, we calculate the corresponding scattered electric and magnetic fields using~\cite{achouri2022multipolar}
\begin{equation}
\begin{aligned}
& \mathbf{E}= -\frac{1}{i\omega\mu\epsilon}\nabla(\nabla\cdot\mathbf{A})+i\omega\mathbf{A}= -\frac{1}{2 k_z \omega \epsilon}\left[\nabla \nabla+k^2 \te{I}\right] \cdot \mathbf{J}_{\mathrm{s}} e^{i k_z|z|}, \\
& \mathbf{H}= \frac{1}{\mu}\nabla\times\mathbf{A}=-\frac{1}{2i k_z} \nabla \times \mathbf{J}_{\mathrm{s}} e^{i k_z|z|},
\end{aligned}
\label{electric field and magnetic currents}
\end{equation}
where $\te{I}$ is identity matrix. By substituting~\eqref{current} into~\eqref{electric field and magnetic currents} and following the conventional textbook procedure for finding boundary conditions via pill-box integration, and substituting (\mbox{$\widetilde{\nabla}= i\mathbf{k}_\parallel+\bm{\partial_z}$}, where $\mathbf{k}_\parallel$ corresponds to the components of the wavevector parallel to the metasurface and $\bm{\partial_z}=\partial_z \hat{\mathbf{z}}$) we obtain
\begin{equation}
\begin{aligned}
\mathbf{E} = & 
- \frac{1}{2 k_z \omega\epsilon} 
\left[ 
\widetilde{\nabla}\widetilde{\nabla}
+ k^2 \te{I} 
\right] 
\cdot \tilde{\mathbf{J}}_\text{s} e^{i k_z |z|}, \\
\mathbf{H} = & 
- \frac{1}{2i k_z} \widetilde{\nabla}\times \tilde{\mathbf{J}}_\text{s} e^{i k_z |z|}.
\end{aligned}
\label{electric_open}
\end{equation}
By the same token, we transform~\eqref{current_equivalent} into
\begin{equation}
\tilde{\mathbf{J}}_\mathbf{s} = -i \omega \mathbf{p} 
+ \widetilde{\nabla} \times \mathbf{m} 
+ \frac{i \omega}{2} \widetilde{\nabla}\cdot {{\mathbf{Q}^\text{(e)}}}  - \frac{1}{2} \widetilde{\nabla} \times 
\big( \widetilde{\nabla} \cdot {{\mathbf{Q}^\text{(m)}}} \big)  - \frac{i \omega}{6} \widetilde{\nabla}\cdot 
\big( \widetilde{\nabla} \cdot {{\mathbf{O}^\text{(e)}}} \big)+   \frac{1}{6} \widetilde{\nabla} \times \left( \widetilde{\nabla} \cdot\widetilde{\nabla} \cdot \mathbf{O}^{\text{(m)}} \right).
\label{current_open}
\end{equation}
Now we have everything needed to compute the octupolar GSTC, which are obtained using
\begin{equation}
\begin{aligned}
\Delta \mathbf{E}_{\mathrm{\|}} & = \mathbf{E}_{\|}^{z=0^+} - \mathbf{E}_{\|}^{z=0^-}, \\
\Delta \mathbf{H}_{\mathrm{\|}} & = \mathbf{H}_{\|}^{z=0^+} - \mathbf{H}_{\|}^{z=0^-}. 
\end{aligned}
\label{field_current}
\end{equation}
By substituting~\eqref{current_open} into~\eqref{electric_open} and decomposing the resulting expressions in terms of \(\bm{\partial_z}^n\), defined as \(\frac{\partial^n}{\partial z^n} \hat{\mathbf{z}}\), we retain only the odd-order derivatives. The even-order derivatives of the term \(e^{j k_z |z|}\) cancel out at the \(z^+\) and \(z^-\) boundaries since they have the same sign at \(z=0\). Alternatively, due to the opposite signs of the odd-order derivatives at the top and bottom sides, we can consider only one of them and multiply our final expressions by a factor of 2. This leads to the GSTC expressions

\begin{equation}
\begin{aligned}
\bm{\Delta E_{\|}} = & \frac{-1}{k_z \omega \epsilon} 
\Bigg\{
    \Big( 
        - \bm{k}_\parallel \bm{k}_\parallel 
        + \bm{\partial_z} \bm{\partial_z} 
        + k^2  \mathbf{I}
    \Big) 
    \cdot 
    \Big[ 
        \bm{\partial_z} \times \bm{m} 
        + \frac{i \omega}{2} \bm{\partial_z} \cdot \bm{Q}^{(\mathrm{e})} 
        - \frac{1}{2} i \bm{k}_\parallel \times (\bm{\partial_z} \cdot \bm{Q}^{(\mathrm{m})}) \\
        & - \frac{1}{2} \bm{\partial_z} \times (i \bm{k}_\parallel \cdot \bm{Q}^{(\mathrm{m})}) 
        + \frac{i \omega}{6} i \bm{k}_\parallel \cdot (\bm{\partial_z} \cdot \bm{O}^{(\mathrm{e})}) 
        + \frac{i \omega}{6} \bm{\partial_z} \cdot (i \bm{k}_\parallel \cdot \bm{O}^{(\mathrm{e})}) \\
        & + \frac{1}{2} i \bm{k}_\parallel \times \left( i \bm{k}_\parallel \cdot \bm{\partial_z} \cdot \bm{O}^{(\mathrm{m})} \right)
        + \frac{1}{2} i \bm{k}_\parallel \times \left( \bm{\partial_z} \cdot i \bm{k}_\parallel \cdot \bm{O}^{(\mathrm{m})} \right) \\
        & + \frac{1}{2} \bm{\partial_z} \times \left( i \bm{k}_\parallel \cdot i \bm{k}_\parallel \cdot \bm{O}^{(\mathrm{m})} \right)
        + \frac{1}{2} \bm{\partial_z} \times \left( \bm{\partial_z} \cdot \bm{\partial_z} \cdot \bm{O}^{(\mathrm{m})} \right)
    \Big] \\
    & + 
    \Big(
         i \bm{k}_\parallel \bm{\partial_z} 
        + \bm{\partial_z} i \bm{k}_\parallel
    \Big)
    \cdot 
    \Big[ 
        -i \omega \bm{p} 
        + i \bm{k}_\parallel \times \bm{m} 
        + \frac{i \omega}{2} i \bm{k}_\parallel \cdot \bm{Q}^{(\mathrm{e})} 
        - \frac{1}{2} i \bm{k}_\parallel \times (i \bm{k}_\parallel \cdot \bm{Q}^{(\mathrm{m})}) \\
        & - \frac{1}{2} \bm{\partial_z} \times (\bm{\partial_z} \cdot \bm{Q}^{(\mathrm{m})}) 
        - \frac{i \omega}{6} i \bm{k}_\parallel \cdot (i \bm{k}_\parallel \cdot \bm{O}^{(\mathrm{e})}) 
        - \frac{i \omega}{6} \bm{\partial_z} \cdot (\bm{\partial_z} \cdot \bm{O}^{(\mathrm{e})}) \\
        & + \frac{1}{2} i \bm{k}_\parallel \times (i \bm{k}_\parallel \cdot i \bm{k}_\parallel \cdot \bm{O}^{(\mathrm{m})}) 
        + \frac{1}{2} i \bm{k}_\parallel \times (\bm{\partial_z} \cdot \bm{\partial_z} \cdot \bm{O}^{(\mathrm{m})}) \\
        & + \frac{1}{2} \bm{\partial_z} \times (i \bm{k}_\parallel \cdot \bm{\partial_z} \cdot \bm{O}^{(\mathrm{m})}) 
        + \frac{1}{2} \bm{\partial_z} \times (\bm{\partial_z} \cdot i \bm{k}_\parallel \cdot \bm{O}^{(\mathrm{m})})
    \Big]
\Bigg\}.
\end{aligned}
\label{huge_e}
\end{equation}
\begin{equation}
\begin{aligned}
\mathbf{\Delta H_{\|}} = \frac{-1}{i k_z} \Bigg\{ & 
    i \mathbf{k}_\parallel \times \Bigg[
        \bm{\partial_z} \times \mathbf{m} 
        + \frac{i \omega}{2} \bm{\partial_z} \cdot \mathbf{Q}^{(\mathrm{e})} 
        - \frac{1}{2} i \mathbf{k}_\parallel \times (\bm{\partial_z} \cdot \mathbf{Q}^{(\mathrm{m})}) \\
        & - \frac{1}{2} \bm{\partial_z} \times (i \mathbf{k}_\parallel \cdot \mathbf{Q}^{(\mathrm{m})}) 
        - \frac{i \omega}{6} i \mathbf{k}_\parallel \cdot (\bm{\partial_z} \cdot \mathbf{O}^{(\mathrm{e})}) 
        - \frac{i \omega}{6} \bm{\partial_z} \cdot (i \mathbf{k}_\parallel \cdot \mathbf{O}^{(\mathrm{e})}) \\
        & + \frac{1}{2} i \mathbf{k}_\parallel \times \left( i \mathbf{k}_\parallel \cdot \bm{\partial_z} \cdot \mathbf{O}^{(\mathrm{m})} \right)
        + \frac{1}{2} i \mathbf{k}_\parallel \times \left( \bm{\partial_z} \cdot i \mathbf{k}_\parallel \cdot \mathbf{O}^{(\mathrm{m})} \right) \\
        & + \frac{1}{2} \bm{\partial_z} \times \left( i \mathbf{k}_\parallel \cdot i \mathbf{k}_\parallel \cdot \mathbf{O}^{(\mathrm{m})} \right)
        + \frac{1}{2} \bm{\partial_z} \times \left( \bm{\partial_z} \cdot \bm{\partial_z} \cdot \mathbf{O}^{(\mathrm{m})} \right)
    \Bigg] \\
    & + \bm{\partial_z} \times \Bigg[
        -i \omega \mathbf{p} 
        + i \mathbf{k}_\parallel \times \mathbf{m} 
        + \frac{i \omega}{2} i \mathbf{k}_\parallel \cdot \mathbf{Q}^{(\mathrm{e})} 
        + \frac{1}{2} \mathbf{k}_\parallel \times (\mathbf{k}_\parallel \cdot \mathbf{Q}^{(\mathrm{m})}) \\
        & - \frac{1}{2} \bm{\partial_z} \times (\bm{\partial_z} \cdot \mathbf{Q}^{(\mathrm{m})}) 
        + \frac{i \omega}{6} \mathbf{k}_\parallel \cdot (\mathbf{k}_\parallel \cdot \mathbf{O}^{(\mathrm{e})}) 
        - \frac{i \omega}{6} \bm{\partial_z} \cdot (\bm{\partial_z} \cdot \mathbf{O}^{(\mathrm{e})}) \\
        & + \frac{1}{2} i \mathbf{k}_\parallel \times \left( i \mathbf{k}_\parallel \cdot i \mathbf{k}_\parallel \cdot \mathbf{O}^{(\mathrm{m})} \right)
        + \frac{1}{2} \bm{\partial_z} \times \left( i \mathbf{k}_\parallel \cdot \bm{\partial_z} \cdot \mathbf{O}^{(\mathrm{m})} \right) \\
        & + \frac{1}{2} \bm{\partial_z} \times \left( \bm{\partial_z} \cdot i \mathbf{k}_\parallel \cdot \mathbf{O}^{(\mathrm{m})} \right)
    \Bigg]
\Bigg\}.
\label{huge_h}
\end{aligned}
\end{equation}

\subsubsection{Oblique Scattering of Co-Polarized TE and TM Waves}

We now compute the fields scattered by a metasurface surrounded by different media on both sides. To do so, we use the GSTC expressions given in~\eqref{huge_e} and~\eqref{huge_h} and, for convenience, replace their right-hand sides by $\boldsymbol\psi_{E}$ and $\boldsymbol\psi_{H}$, respectively, while expressing the field differences $\Delta \mathbf{E}_{\|}$ and $\Delta \mathbf{H}_{\|}$ in terms of incident $(i)$, reflected $(r)$ and transmitted $(t)$ fields. This leads to
\begin{equation}
\begin{array}{rll}
\Delta \mathbf{E}_{\|} = \boldsymbol\psi_{E} & \longrightarrow & \mathbf{E}_{\|,\mathrm{t}} - \mathbf{E}_{\|,i} - \mathbf{E}_{\|,\mathrm{r}} = \boldsymbol\psi_{E}, \\[1mm]
\Delta \mathbf{H}_{\|} = \boldsymbol\psi_{H} & \longrightarrow & \mathbf{H}_{\|,\mathrm{t}} - \mathbf{H}_{\|,i} - \mathbf{H}_{\|,\mathrm{r}} = \boldsymbol\psi_{H}.
\end{array}
\label{boundary_field}
\end{equation}
To obtain reflection and transmission coefficients, we solve these equations by expressing the tangential magnetic fields in terms of tangential electric fields using~\cite{albooyeh2016electromagnetic,shalin2010broadband,shalin2009optical}
\begin{equation}
\mathbf{E}_{\|} = \mp \mathbf{Z}_{l} \cdot \Big( \mathbf{n} \times \mathbf{H}_{\|} \Big)
\quad \text{and} \quad
\mathbf{n} \times \mathbf{H}_{\|} = \mp \mathbf{Y}_{l} \cdot \mathbf{E}_{\|},
\label{impedance}
\end{equation}
where the admittance, $\mathbf{Y}_{l}$, and impedance, $\mathbf{Z}_{l}$, tensors are provided in Eq.~(C44) in Appendix~C of the Supplementary Information. Here, the subscript $l=\{1,2\}$ denotes the top ($+z$) and bottom ($-z$) media, respectively. Substituting~\eqref{impedance} into~\eqref{boundary_field} and solving for the reflected and transmitted fields yields
\begin{subequations}
\label{eq_Ert}
\begin{align}
\mathbf{E}_{\|,\mathrm{t}} &= 2 \mathbf{Z}_{2}\cdot\Big(\mathbf{Z}_{1}+\mathbf{Z}_{2}\Big)^{-1} \cdot \mathbf{E}_{\|,i} 
+ \mathbf{Z}_{2}\cdot\Big(\mathbf{Z}_{1}+\mathbf{Z}_{2}\Big)^{-1} \cdot \Big[\boldsymbol\psi_{E} - \mathbf{Z}_{1}\cdot\Big(\hat{\mathbf{z}} \times \boldsymbol\psi_{H}\Big)\Big],\label{E_t}\\
\mathbf{E}_{\|,\mathrm{r}} &= \Big(\mathbf{Z}_{2}-\mathbf{Z}_{1}\Big)\cdot\Big(\mathbf{Z}_{1}+\mathbf{Z}_{2}\Big)^{-1} \cdot \mathbf{E}_{\|,i}
- \mathbf{Z}_{1}\cdot\Big(\mathbf{Z}_{1}+\mathbf{Z}_{2}\Big)^{-1} \cdot \Big[\boldsymbol\psi_{E} + \mathbf{Z}_{2}\cdot\Big(\hat{\mathbf{z}} \times \boldsymbol\psi_{H}\Big)\Big].
\end{align}
\label{E_r}
\end{subequations}
The derivation of these equations is provided in Appendix~C of the Supplementary Information. Note that in the particular case where the two media on both sides of the metasurface are identical, then (\(\mathbf{Z}_1=\mathbf{Z}_2=\mathbf{Z}\)), reducing~\eqref{eq_Ert} to
\begin{equation}
\begin{aligned}
\mathbf{E}_{\|,\mathrm{t}} &= \mathbf{E}_{\|,i} + \frac{1}{2}\Big[\boldsymbol\psi_{E} - \mathbf{Z}\cdot\Big(\hat{\mathbf{z}} \times \boldsymbol\psi_{H}\Big)\Big], \\
\mathbf{E}_{\|,\mathrm{r}} &= -\frac{1}{2}\Big[\boldsymbol\psi_{E} + \mathbf{Z}\cdot\Big(\hat{\mathbf{z}} \times \boldsymbol\psi_{H}\Big)\Big].
\end{aligned}
\label{no_material_impedance}
\end{equation}

Now, as an illustration of this method, we provide the co-polarized reflection and transmission coefficients assuming wave propagation in the \(xz\)-plane for which the wave vector is $\mathbf{k} = k_x\,\hat{\mathbf{x}} + k_z\,\hat{\mathbf{z}}$. In what follows, we only treat the case where the media on both sides of the metasurface are identical and provide the general relations for an asymmetric environment in Appendix~E of the Supplementary Information.

In the case of TE-polarized waves, the only non-zero field components are \(E_y\), \(H_x\) and \(H_z\). This implies that the only contributing components in~\eqref{huge_e} and~\eqref{huge_h} are the \(y\)-component of \(\Delta \mathbf{E}\) and the \(x\)-component of \(\Delta \mathbf{H}\). After some algebraic manipulations, we find that
\begin{subequations}
\label{psi_TE_EH}
\begin{align}
\psi_{E}^{\text{TE}} &= \bm{\psi}_{E}^{\text{TE}}\cdot \mathbf{\hat y}=\tilde{M}_x 
+ \tilde{Q}_{yz}^\text{(e)}
+ \tilde{Q}_{xx}^\text{(m)} + \tilde{Q}_{zz}^\text{(m)}
+ \tilde{O}_{yzx}^\text{(e)},\label{psi_TE_E}\\
\psi_{H}^{\text{TE}} &=\bm{\psi}_{H}^{\text{TE}} \cdot \mathbf{\hat x}= \tilde{P}_y 
+ \tilde{M}_z 
+ \tilde{Q}_{xz}^\text{(e)}
+ \tilde{Q}_{yx}^\text{(e)}
+ \tilde{O}_{yxx}^\text{(e)}
+ \tilde{O}_{yzz}^\text{(e)}.\label{psi_TE_H}
\end{align}
\end{subequations}
In these equations, we have, for convenience, redefined the multipolar elements, which now appear as tilded variables. The mapping relations between these tilded variables, which include physical constants, and the corresponding untilded elements are provided in Table~7 of Appendix~D in the Supplementary Information.

For simplicity, we consider a normalized incident electric field, i.e., \(E_y^\text{i} = 1\), which directly leads to the reflection and the transmission coefficients
\begin{subequations}
\label{RT_TE} 
\begin{align}
r^{\text{TE}} &= \frac{E_y^\text{r}}{E_y^\text{i}}= \frac{1}{2} \Biggl( \psi_{E}^{\text{TE}} - Z_{yy}\psi_{H}^{\text{TE}} \Biggr), \label{R_TE} \\
t^{\text{TE}} & = \frac{E_y^\text{t}}{E_y^\text{i}}= 1 - \frac{1}{2} \Biggl( \psi_{E}^{\text{TE}} + Z_{yy}\psi_{H}^{\text{TE}} \Biggr). \label{T_TE_ex}
\end{align}
\end{subequations}

The same procedure is now applied to the case of TM-polarized waves. By considering the \(x\)-component of \(\Delta \mathbf{E}\) and the \(y\)-component of \(\Delta \mathbf{H}\), we obtain
\begin{subequations}
\label{psi_TM_EH}
\begin{align}
\psi_E^{\text{TM}} &= \bm{\psi}_E^{\text{TM}}\cdot \mathbf{\hat x}=
\tilde{M}_y 
+ \tilde{P}_z 
+ \tilde{Q}_{xz}^\text{(e)} 
+ \tilde{Q}_{xy}^\text{(m)}
+ \tilde{O}_{xxz}^\text{(e)} 
+ \tilde{O}_{zzz}^\text{(e)},
\label{psi_e_tm}\\
\psi_{H}^{\text{TM}} &=\bm{\psi}_H^{\text{TM}}\cdot \mathbf{\hat y}= \tilde{P}_x 
+ \tilde{Q}_{yz}^\text{(m)} + \tilde{Q}_{zy}^\text{(m)}
+ \tilde{Q}_{xx}^\text{(e)} + \tilde{Q}_{zz}^\text{(e)}
+ \tilde{O}_{yzx}^\text{(e)}
+ \tilde{O}_{xxx}^\text{(e)}
+ \tilde{O}_{xzz}^\text{(e)}.
\label{psi_h_tm}
\end{align}
\end{subequations}
Following the same procedure as for TE-polarized waves, the TM-polarized reflection and transmission coefficients are
\begin{subequations}
\begin{align}
&r^{\text{TM}} = \frac{H_y^r}{H_y^\text{i}}= \frac{\eta}{2} \Bigg( 
\psi_{H}^{\text{TM}} - Y_{yy} \psi_{E}^{\text{TM}}
\Bigg),
\label{r_tm_1}\\
& t^{\text{TM}} =\frac{H_y^t}{H_y^\text{i}}= 
1 - \frac{\eta}{2} \Bigg( 
\psi_{H}^{\text{TM}} - Y_{yy}  \psi_{E}^{\text{TM}}  \Bigg).
\label{t_tm_1}
\end{align}
\end{subequations}
where $\eta=\eta_0/n_\text{b}$ with $n_\text{b}$ being the refractive index of the background medium. Note that $Z_{yy}$ and $Y_{yy}$ are defined in Appendix~E of the Supplementary Information.
\begin{equation}
    {Z_{yy}} = \frac{1}{{{Y_{yy}}}} = \frac{{\eta k}}{{{k_z}}}.
\end{equation}

\section{Results and Verification Examples}

To demonstrate the validity of our GSTC based modeling framework, we now consider various relevant scenarios. This includes cases where different multipole moments arise under varying incident angles, polarization states (TE and TM) and in the presence of a substrate. Finally, we will examine more complex situations, such as bound states in the continuum (BIC), and demonstrate how they emerge from the collective contributions of higher-order multipoles.

\subsection{Cases With Uniform Background}

\begin{figure}[h!]
  \centering
  \includegraphics[width=0.5\textwidth]{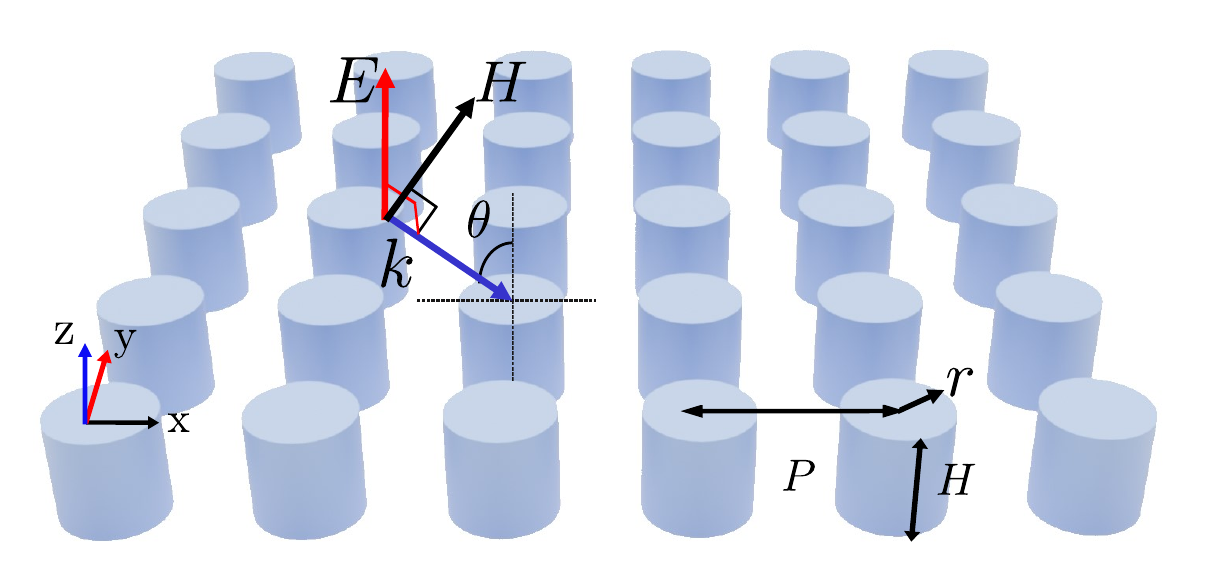} % Replace with your image file
  \caption{Schematic of nano-cylinders with $P=225$ nm and $H$ as the height of the cylinders under TE illumination.}
  \label{fig:te_schematic}
\end{figure}

We start by verifying the validity of our method for the case of the scattering of an obliquely incident TE-polarized wave by a metasurface made of cylindrical scattering particles. Here, we simulate and retrieve the reflection and transmission coefficients assuming that the nano-cylinders are either composed of dispersive gold or amorphous silicon. Based on Fig.~\ref{fig:te_schematic}, the assumed periodicity is \(P = 225\)~nm and the cylinder radius is \(r = 100\)~nm. The metasurface is illuminated at different oblique incidence angles (\(\theta\)) and is embedded in air.

\begin{figure}[h!]
    \centering
    \includegraphics[width=1\textwidth]{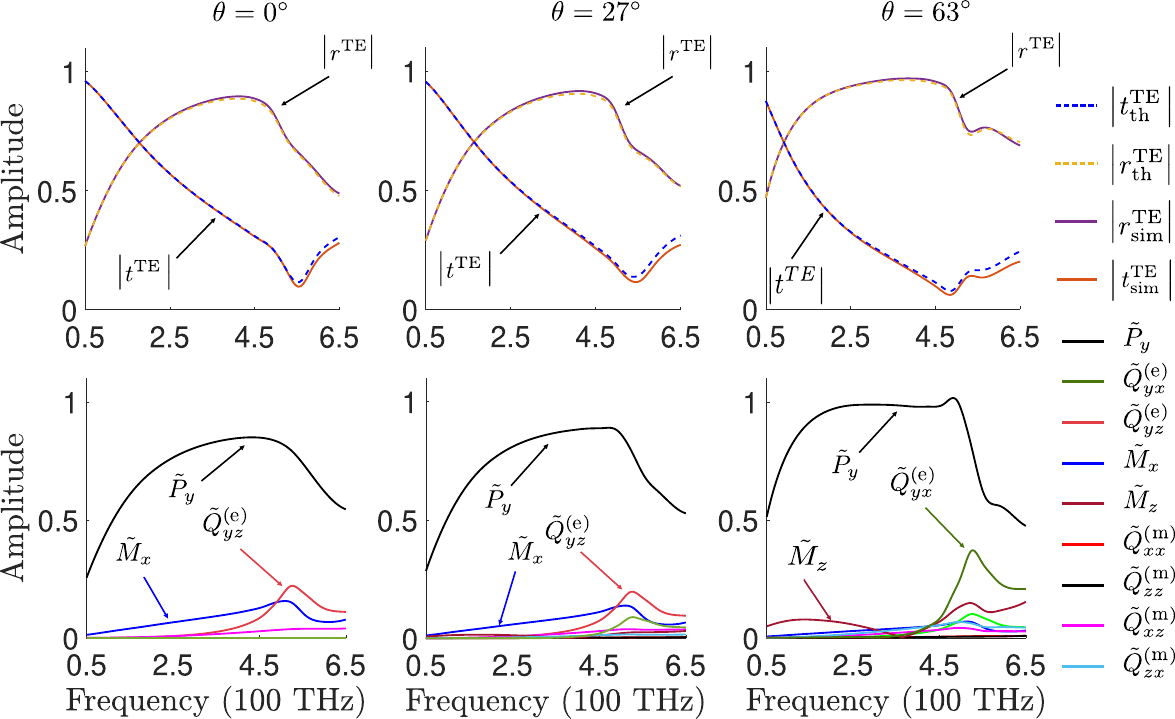} % Replace 'example-image' with your file name
    \caption{Simulation and theoretical results for gold nano-cylinders for $P=225$~nm and $H=95$~nm and $r=100$~nm under TE illumination for $\theta=0^\circ,27^\circ,63^\circ$.}
    \label{TE_95_GOLD1} % Label for referencing the figure
\end{figure}

\begin{figure}[h!]
    \centering
\includegraphics[width=1\textwidth]{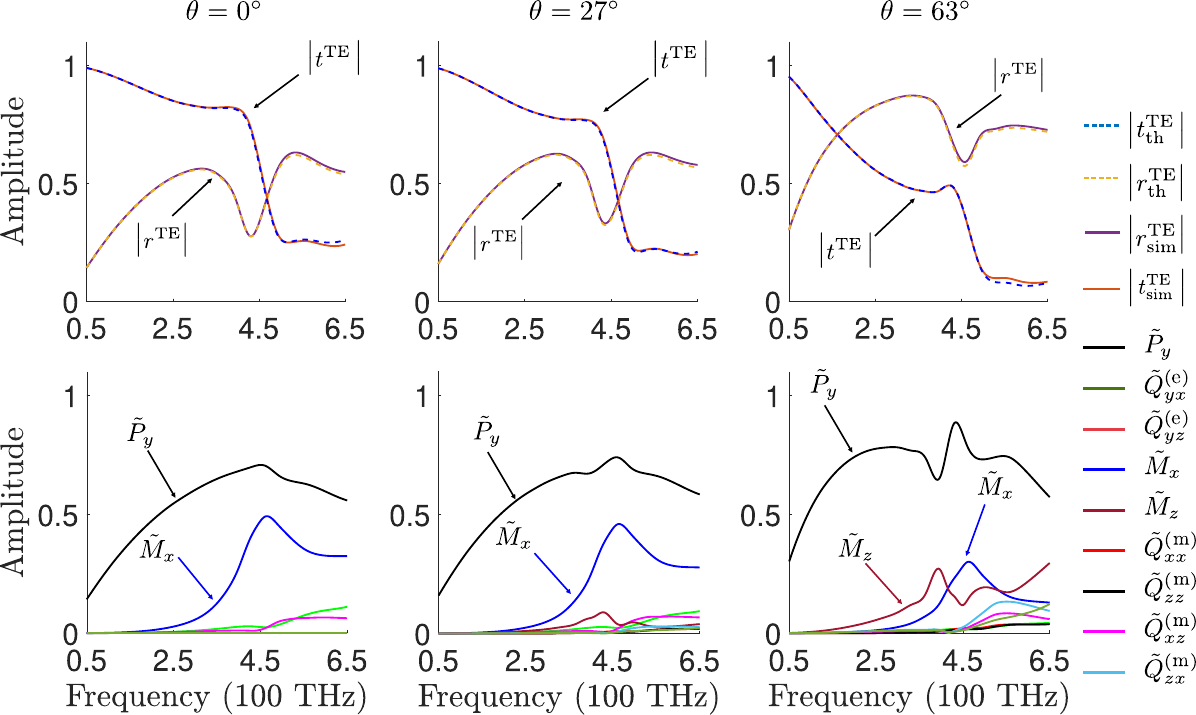} % Replace 'example-image' with your file name
    \caption{Simulation and theoretical results for amorphous silicon nano-cylinders for $P=225$~nm and $H=95$~nm and $r=100$~nm under TE illumination for $\theta=0^\circ,27^\circ,63^\circ$.}
    \label{TE_siam_95} % Label for referencing the figure
\end{figure}

In Figs.~\ref{TE_95_GOLD1} and \ref{TE_siam_95}, we present both COMSOL simulations and theoretical results for the reflection and transmission for the two metasurfaces composed of amorphous silicon and gold nano-cylinders, respectively. Theoretical results have been obtained from the combination of~\eqref{RT_TE} with~\eqref{psi_TE_EH}. In these expressions, the tilded multipole moments have been computed by performing the integrations in Table~\ref{tab:traceless_multipoles} using the equivalent volume current density~\eqref{eq_vol_cur} defined from COMSOL-simulated fields inside the cylinders for the incident angles $\theta =\{0^\circ, 27^\circ, 63^\circ\}$, and subsequently applying the mapping relations in Table~6 of Appendix~D in the Supplementary Information. In these figures, the subscripts ``sim'' and ``th'' correspond to the COMSOL simulated results and the theoretical predictions from the GSTC, respectively. For each case, we also provide the tilded multipole moment amplitudes to show which ones contribute the most to the overall metasurface scattering response. In Appendix~F of the Supplementary Information, we provide the results for different heights, incidence angles as well as for the case of TM polarization.

For both cases, we see that our method agrees very well with the COMSOL simulations over the entire frequency range and even for large angles of incidence. From the multipolar decompositions, we also see that while the scattering response is generally dominated by the electric dipole moment $\tilde{P}_y$, the contributions from other multipoles are significant and strongly vary with respect to the incidence angle. For instance, we see that the magnetic moment $\tilde{M}_z$ starts to contribute only at large angles of incidence, which highlights the importance of including normal polarization components in the model -- something that was often ignored in other modeling techniques~\cite{achouri2020a,achouri2021}. Note that in both examples, the contributions from octupolar responses are negligible.

\subsection{Cases that Include a Substrate}

\begin{figure}[h!]
\centering
\includegraphics[width=0.8\textwidth]{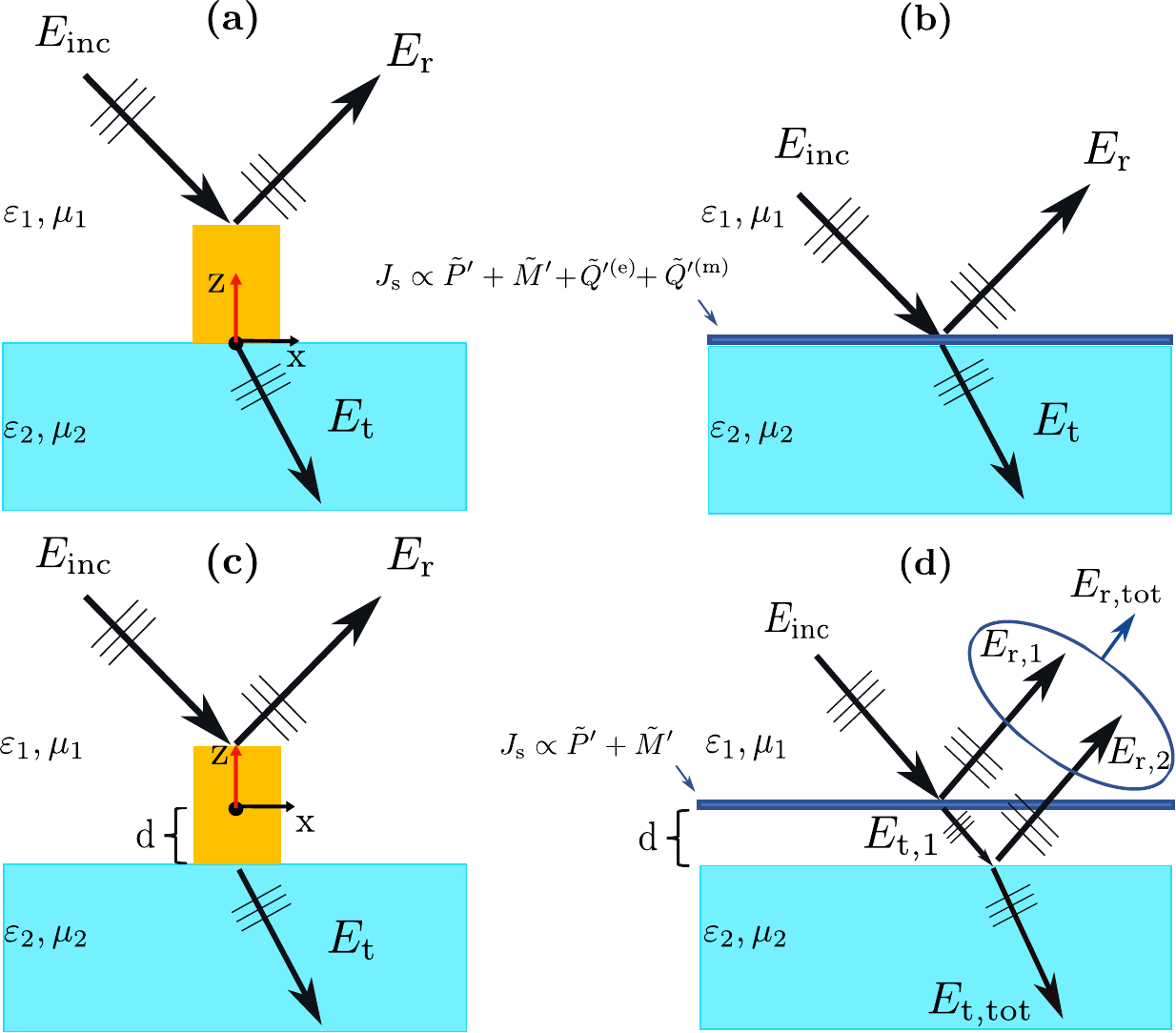} % Replace 'example-image' with your file name
\caption{\textbf{Different choices for the origin of the boundary conditions.} In (a) and (b), the boundary origin is at the interface between the different media, resulting in the presence of higher-order multipole moments. In (c) and (d), the boundary origin is at the center of mass of the object and only a few multipole moments are required to model the metasurface response. In (d), the reflection and transmission coefficients must be evaluated by taking into account the distance between the boundary and the actual interface and the additional scattering at the latter. Here, $E_{\text{r},1}$ and $E_{\text{t},1}$ are the reflected and transmitted electric fields from the metasurface, which is fully embedded in the top material, calculated directly from~\eqref{RT_TE}. The transmitted field $E_{\text{t},1}$ then reflects and refracts at the interface producing the reflected field $E_{\text{r},2}$ and the transmitted field $E_{\text{t,tot}}$. The total reflected field $E_{\text{r,tot}}$ is obtained by combining $E_{\text{r},1}$ and $E_{\text{r},2}$.}
\label{origin}
\end{figure}

We now demonstrate that our model remains valid in the presence of a substrate. The relevant formulas are provided in Appendix~E of the Supplementary Information. The problem that occurs when considering the presence of a substrate is where the GSTC boundary should be placed and where should be the origin of the system of coordinates used to perform the multipolar decomposition. We consider two different cases: 1)~the GSTC boundary is placed at the interface between the substrate and the superstrate, as illustrated in Fig.~\ref{origin}a,b; 2)~it is positioned at the center of mass of the object, as illustrated in Fig.~\ref{origin}c,d. More details on the optimal boundary position for substrate effects can be found in~\cite{evlyukhin2015substrate}, along with advanced studies on more complex geometries in~\cite{ustimenko2025optimal,kildishev2023art}.

Both boundary positions present their own challenges. In the first scenario, where the boundary is placed at the material interface, the origin of integration is strongly off-centered, which results in the appearance of higher-order multipole moments that are necessary to properly model the scattering response of the object~\cite{ustimenko2025optimal,kildishev2023art}. In the other case, placing the boundary at the object center of mass leads to a phase mismatch, due to the distance $d$ between the boundary and the material interface, and to additional scattering at this interface, as shown in Fig.~\ref{origin}d. Overall, it is generally better to place the boundary at the center of mass of the object, since fewer multipole orders are required to model the metasurface response, even if evaluating the total scattered fields is not as straightforward as in the other scenario, as we shall now describe.

To compute the fields in Fig.~\ref{origin}b, we simply use the scattered field equations Eq.~(E54) in Appendix~E of the Supplementary Information along with the normalized multipole moments in Table~7 of Appendix~E in the Supplementary Information and Table~8 in Appendix~E of the Supplementary Information . On the other hand, to evaluate the total fields $E_{\text{r,tot}}$ and $E_{\text{t,tot}}$ in Fig.~\ref{origin}d, we start by computing the fields $E_{\text{r},1}$ and $E_{\text{t},1}$ that are scattered by the metasurface. To do so, we take advantage of the fact that the metasurface is surrounded by identical media on both sides, which allows us to directly use~\eqref{RT_TE} with the material parameters corresponding to that of medium ``1''. We thus have
\begin{subequations}
\begin{align}
E_{\text{r},1} &= \frac{1}{2} \left( \psi_{E}^{\text{TE}} - Z_{yy}\psi_{H}^{\text{TE}} \right), \label{R1_TE} \\
E_{\text{t},1} & = 1 - \frac{1}{2} \left( \psi_{E}^{\text{TE}} + Z_{yy}\psi_{H}^{\text{TE}} \right), \label{T_TE_ex1}
\end{align}
\end{subequations}
where the terms $\psi_{E}^{\text{TE}}$ and $\psi_{H}^{\text{TE}}$ are computed using the normalized multipole moments in Table~7 of Appendix~D in the Supplementary Information.
Now, after propagating a distance $d$, the field $E_{\text{t},1}$ transmitted from the metasurface gets reflected and refracted by the substrate according to the Fresnel reflection and transmission coefficients~\cite{novotny2012principles}
\begin{subequations}
\label{eq_fresnel}
\begin{align}
r_\text{s} &= \frac{k_{z,2} - k_{z,1}}{k_{z,2} + k_{z,1}}, \\
t_\text{s} &= \frac{2k_{z,1}}{k_{z,2} + k_{z,1}}.
\end{align}
\end{subequations}
By applying the proper phase correction to compensate for the optical path difference between the GSTC and the substrate, we find that the total reflected and transmitted fields are, using~\eqref{eq_fresnel}, given by
\begin{subequations}
\label{eq_E_rttot}
\begin{align}
E_{\text{r},\text{tot}} &= E_\text{r,1} + {r_\text{s}}E_{\text{t},1}{e^{ - 2i{k_{z,1}}{d}}},\\
E_{\text{t},\text{tot}} &= \sqrt {\frac{{k_{z,2}}}{{{k_{z,1}}}}} t_\text{s}{e^{-i{k_{z,1}}d}}E_\text{t,1}.\label{E_t_t}
\end{align}
\end{subequations}
Note that in~\eqref{E_t_t}, the term $\sqrt {\frac{{k_{z,2}}}{{{k_{z,1}}}}}$ corresponds to power normalization, as detailed in Eq.~(E59) of Appendix~E in the Supplementary Information.
\\

\begin{figure}[h!]
    \centering
    \includegraphics[width=0.5\textwidth]{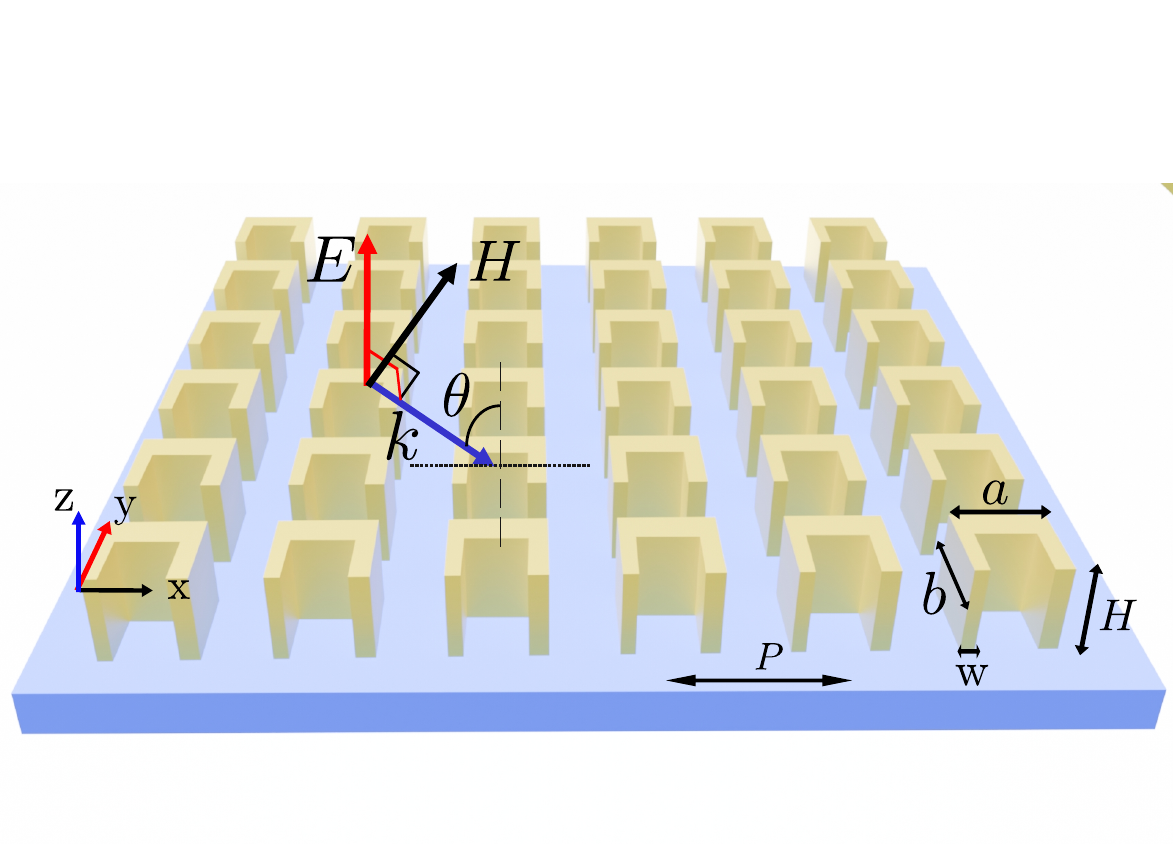}
    \caption{TE-illuminated metasurface made of gold C-shaped particles on a substrate of refractive index $n=1.49$. The top medium is air and $P=225$~nm, $a=b=100$~nm, $w=a/4$ and $H=115$~nm.}
    \label{strip1}
\end{figure}

We now demonstrate the validity of our model by comparing its scattering predictions with COMSOL simulations. For this purpose, we consider the metasurface illustrated in Fig.~\ref{strip1} consisting of C-shaped gold particles lying on a substrate with refractive index \(n = 1.49\).

\begin{figure}[h!]
\centering
\includegraphics[width=1\textwidth]{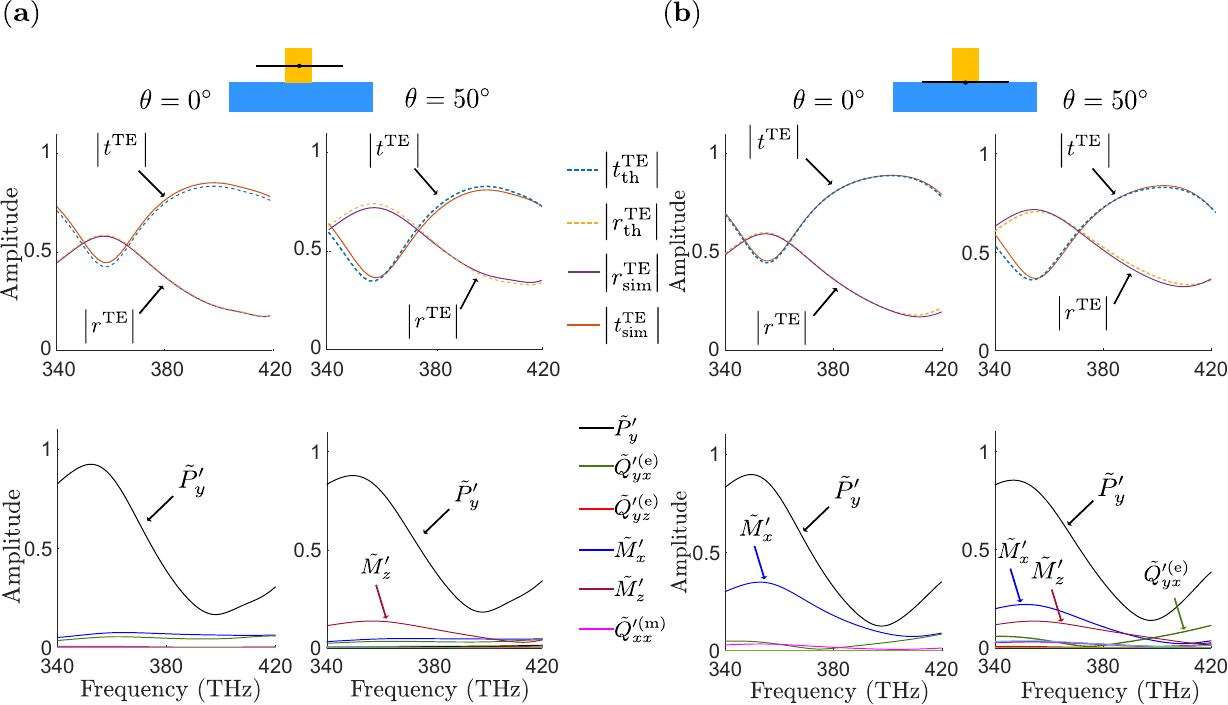}
\caption{Scattering by a metasurface in an asymmetric environment. (Top) Reflection and transmission coefficients for a nano-gold C-shaped metasurface on a glass substrate under TE illumination. (Bottom) Corresponding multipole moments. (a)~Case where the boundary is at the center of mass of the object and the fields are obtained using~\eqref{eq_E_rttot}. (b)~Case where the boundary is at the material interface.}
\label{bottom}
\end{figure}

The resulting comparison between the GSTC method and COMSOL simulations is provided in Fig.~\ref{bottom}, where we have considered two different angles of incidence \mbox{$\theta=\{0^\circ,50^\circ\}$}. In Fig.~\ref{bottom}a, the GSTC boundary (as well as the system of coordinates used to compute the multipole moments) is placed at the center of mass of the object, whereas it is placed at the material interface in Fig.~\ref{bottom}b. From these results, we see that we obtain excellent agreement between the predictions and simulations regardless of where the boundary is placed and of the angle of incidence. As expected, placing the boundary at the material interface instead of the center of the mass of the object leads to the presence of higher-order multipole moments, which we are nonetheless able to accommodate thanks to our octupolar GSTC model.

\subsection{GSTC Modeling of Bound States in the Continuum}

So far, we have applied our model to metasurfaces with and without substrate and under normal and oblique incidence. In this section, we extend this analysis to the case of metasurfaces supporting bound states in the continuum (BIC), which is more complex to model due to the presence high-Q factor resonances stemming from the interference of a large number of multipole moments.

\begin{figure}[h!]
    \centering
    \includegraphics[width=0.8\textwidth]{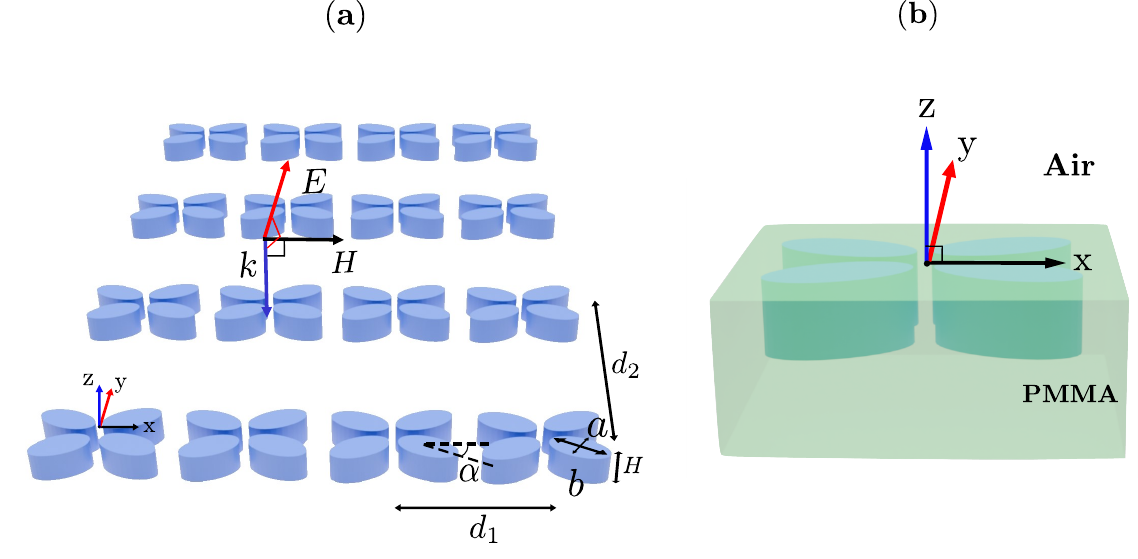}
    \caption{Dielectric metasurface made out of nano-silicon elliptical scatterers with $d_1=900$~nm, $d_2=1300$~nm, $a=530$~nm, $b=170$~nm and $H=540$~nm under normal incidence ($\theta=0$). In (a), the metasurface is fully embedded in air. In (b), it is in an asymmetric environment with PMMA in the lower half space and air in the upper half space.}
    \label{bic_without}
\end{figure}

To validate our approach, we replicate the key results from~\cite{liu2018extreme} to which we apply our modeling framework. The analysis in~\cite{liu2018extreme} investigates the fine-tuning of symmetry-protected BIC via a geometrical perturbation of the metasurface scattering particles. Since symmetry-protected BIC only occur at normal incidence, we shall also restrict our analysis to this prescription. We start by examining a dielectric metasurface made out of four elliptical scatterers, where a geometric perturbation is introduced by varying the angle~\(\alpha\), as depicted in Fig.~\ref{bic_without}a. This angular parameter allows breaking the symmetry of the mode profile within the scatterers leading to a modulation of the resonance Q factor~\cite{liu2018extreme}. If $\alpha=0^\circ$, then BIC are achieved (infinite Q factor), whereas for small values of $\alpha$ quasi-BIC are obtained. Theoretical predictions, numerical results, and associated multipolar decomposition, showing excellent agreement for this case, are provided in Fig.~S17  of Appendix~G in the Supplementary Information.

\begin{figure}[h!]
\centering
\includegraphics[width=0.8\textwidth]{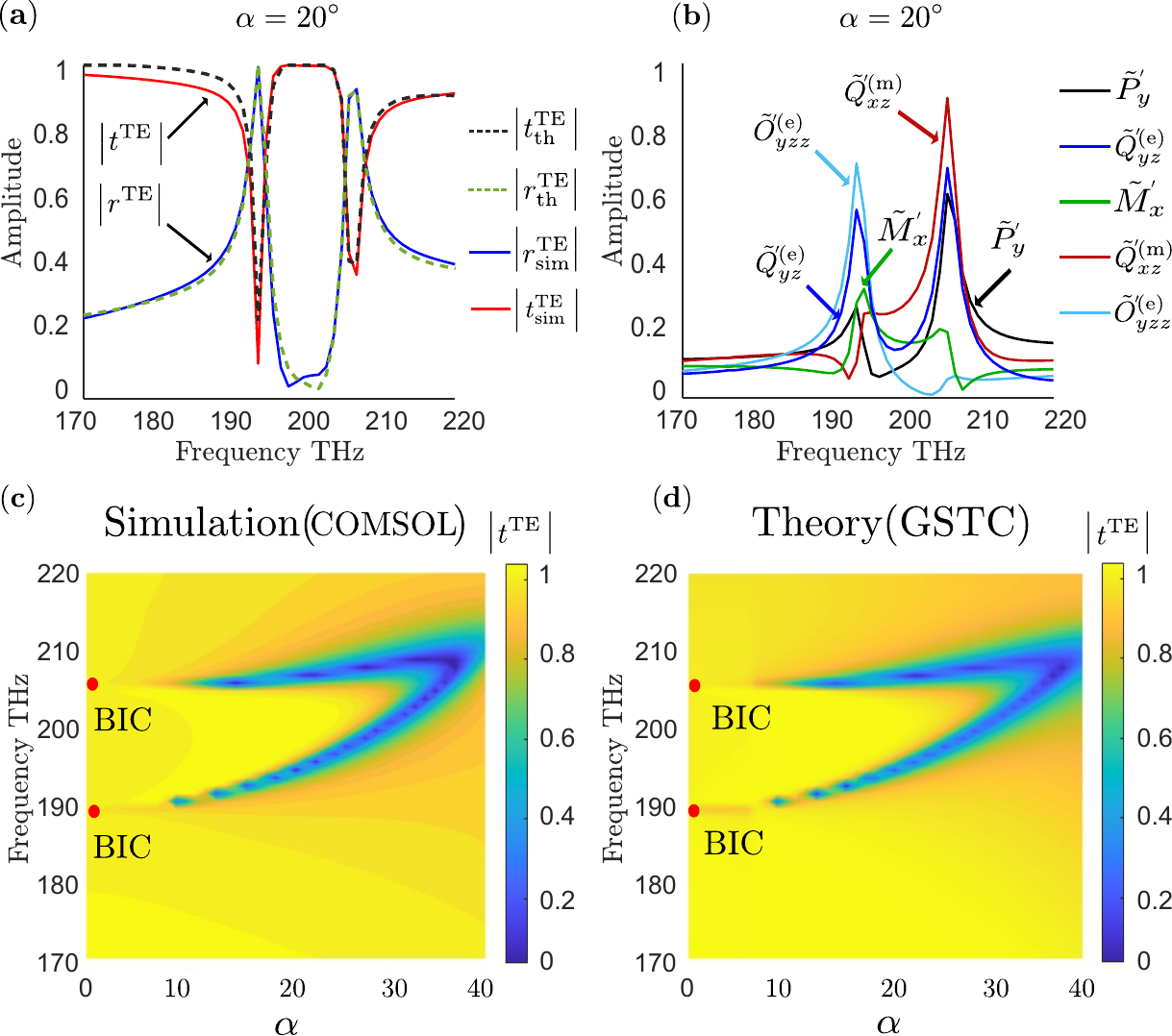}
\caption{Scattering analysis for the metasurface in Fig.~\ref{bic_without}b. (a)~Reflection and transmission coefficients for $\alpha=20^\circ$ obtained via COMSOL and our GSTC method. (b)~Corresponding multipolar decomposition. (c)-(d)~Transmission amplitudes for different geometrical perturbations ($\alpha$) obtained via COMSOL and our GSTC method, respectively.}
\label{bic_embeded}
\end{figure}

In what follows, we consider the case depicted in Fig.~\ref{bic_without}b, where the metasurface is embedded in an asymmetric dielectric environment, PMMA, in the lower half space and air in the upper half space. The presence of PMMA makes the problem more complex to model, which strengthen the claim that our approach is valid for modeling such scenarios. Figure~\ref{bic_embeded}a compares reflection and transmission spectra obtained via COMSOL and our GSTC model for $\alpha=20^\circ$. Figure~\ref{bic_embeded}b provides the corresponding multipolar decomposition. The transmission amplitude in terms of frequency and varying $\alpha$ parameters are plotted in Fig.~\ref{bic_embeded}c, for COMSOL results, and in Fig.~\ref{bic_embeded}d using our GSTC method. These results demonstrate that, despite the existence of a large number of multipolar moments, the predictions from our method are again in very good agreement with COMSOL simulations. As can be seen, as the geometric perturbation parameter \(\alpha\) increases, radiation leakage becomes stronger, the quality factor drops, and the resonance becomes broader. In Fig.~\ref{bic_embeded}c, the presence of the PMMA substrate leads to inversion symmetry breaking and thus to the emergence of higher-order multipoles including those with different parity leading to the apparition of two different types of (quasi-)BIC~\cite{liu2018extreme,poleva2023multipolar,achouri2023spatial}.

It is worth noting that in structures with \(C_{2v}\) symmetry, dipole-based BIC cannot form at normal incidence~\cite{gladyshev2023inverse} because there is no dipole moment that could, for instance, cancel out $\tilde{M}_x$ in~\eqref{psi_TE_E} under TE illumination. Additionally,  the dipole moment \(\tilde{M}_z\) in~\eqref{psi_TE_H} is zero at normal incidence since it depends on \(k_x\), as shown in Table~6 of Appendix~D in the Supplementary Information. This implies that there is also no dipole moment that could cancel out $\tilde{P}_y$ in~\eqref{psi_TE_H}. It follows that with such symmetry, only accidental BIC are allowed to exist~\cite{gladyshev2023inverse}.

In the structure of Fig.~\ref{bic_without}b, the broken \(C_4\) and inversion symmetries lead to the excitation of quadrupolar and even octupolar terms. This is clearly shown in Fig.~\ref{bic_embeded}b, which presents significantly more multipolar components as the one shown in Fig.~S17.b of Appendix~G in the Supplementary Information, where inversion symmetry is not broken. Importantly, these multipolar moments do not require oblique incidence and thus arise naturally from symmetry breaking. A deeper discussion of the role of symmetry in the emergence of multipoles is provided in~\cite{achouri2023spatial,poleva2023multipolar}.

It is worth noting that even at small values of~\(\alpha\), approaching the pure BIC condition, multipolar decomposition remains insightful. For instance, let us consider the case where $\alpha=1^\circ$ and whose results are presented in Fig.~S17 of Appendix~G in the Supplementary Information. We see that the scattering parameters in Fig.~S17.a of Appendix~G in the Supplementary Information, show very little indication of the presence of a BIC resonance due to extremely limited radiation leakage. Nevertheless, the extracted multipolar moments remain well defined and detectable revealing a sharp interplay among higher-order multipoles. This observation underscores the importance of near-field multipolar analysis for characterizing weakly radiative or symmetry-protected modes that evade detection through conventional scattering metrics.

\section{Conclusion}

In this work, we developed a rigorous analytical and numerical framework to describe the electromagnetic response of periodic metasurfaces based on exact multipolar expansions. Starting from the level of a single scatterer, we extended the formulation to metasurfaces using equivalent surface currents and demonstrated that this approach remains accurate even when higher-order multipoles such as quadrupoles and octupoles are included. In contrast to other methods presented so far, our approach naturally accommodates oblique incidence and the presence of different substrate and superstrate thanks to the GSTC.

We validated the GSTC approach by considering several examples, which all demonstrated to be in excellent agreement with full-wave simulations. One of these cases is that of a well known BIC configurations that exhibits controllable quasi-BIC conditions based on a geometrical perturbation parameter. We showed that, in this case, the emergence of higher-order multipole moments are due to structural asymmetry of the scatterers. Additionally, we presented that, even when far-field scattering parameters do not show the existence of BIC resonances, the corresponding multipolar moments, which were computed based on near-field integrations, reveal significant narrow bandwidth resonances. This provides a new perspective for understanding the underlying physics of bound states in the continuum emerging in multipolar metasurfaces.

This work shows that the proposed GSTC method is very well suited for modeling the angular response of metasurfaces. This opens up new ways for future developments in angular multiplexing and topological photonics using spatially uniform but angular dependent metasurfaces whose scattering response relies on non-local interactions arising from multipolar contributions of the metasurface scatterers and corresponding lattice coupling effects.

\newpage

\renewcommand{\theequation}{A\arabic{equation}}% Define equation numbering format for Appendix A
\renewcommand{\thefigure}{S\arabic{figure}}% Define figure numbering format for supplementary material
\renewcommand{\thesubsection}{\thesection\arabic{subsection}}

\renewcommand{\theequation}{A\arabic{equation}}
\section{\label{irreducable}Appendix A: Irreducible representation 
for higher-order multipoles from primitive ones} The general approach for performing multipolar decomposition in Cartesian coordinates has been to use a Taylor expansion of the electric vector potential. This approach leads to the so-called primitive multipoles~\cite{gurvitz2019high}. However, primitive multipoles are typically not expressed in terms of symmetric and traceless tensors. This is a problem because the physical interpretation of such tensors under rotations or translations is not well defined~\cite{nanz2016toroidal}. For this reason, it is common to apply symmetrization and detracing procedures to obtain physically meaningful quantities that take the form of irreducible Cartesian multipoles~\cite{gurvitz2019high}. Interestingly, these symmetrization and detracing techniques brings about toroidal modes. The primitive multipoles without toroidal modes up to the octupolar order are, from reference~\cite{gurvitz2019high}, shown in Table~\ref{Tab_2}.

\begin{table}[H]
\centering
\small
\caption{Primitive Cartesian Multipoles}
\renewcommand{\arraystretch}{1.8} % Adjust row spacing
\setlength{\arraycolsep}{2pt} % Adjust column spacing
\begin{tabular}{|c|p{10cm}|}
\hline
\textbf{Order} & \textbf{Electric Type Multipoles} \\ \hline
1 &
\( p_i = \frac{i}{\omega} \int j_i \, dV \) \\ \hline
2 &
\( {Q}_{jk}^\text{(e)} = \frac{i}{\omega} 
\int \left( r_j J_k + r_k J_j - \frac{2}{3} \delta_{jk} (\mathbf{r} \cdot \mathbf{J}) \right) dv\) \\ \hline
3 &
\(
O_{jkl}^\text{(e)} = \frac{i}{\omega} 
\int \Big( r_k r_j J_i + r_k r_i J_j + r_i r_j J_k \Big) dv 
- \frac{1}{5} \Big( 
\delta_{jk} {Q}_{vvl} 
+ \delta_{jl} {Q}_{vvk} 
+ \delta_{lk} {Q}_{vvj} 
\Big).
\) \\ \hline
%4 &
%\( {{X}}_{jkpl}^\text{(e)} =\frac{i}{\omega} 
%\int \Big( 
%r_l r_k r_j J_i + r_l r_k r_i J_j + r_l r_i r_j %J_k + r_i r_k r_j J_l 
%\Big) dv- \frac{1}{7} \Big( \delta_{pj} %{S}_{vvkl} + \delta_{pk} {S}_{vvjl} + \delta_{jk} %{S}_{vvpl} 
%+ \delta_{jl} {S}_{vvpk} + \delta_{kl} {S}_{vvpj} + \delta_{pl} {S}_{vvjk} \Big) -\frac{1}{15} \Big( 
%\delta_{pj} \delta_{kl} {S}_{uuvv} + \delta_{pk} \delta_{jl} {S}_{uuvv} + \delta_{pl} \delta_{jk} {S}_{uuvv} \Big) \) \\ \hline
\textbf{Order} & \textbf{Magnetic Type Multipoles} \\ \hline
1 &
\( m_i = \frac{1}{2} \int (\mathbf{r} \times \mathbf{J})_i \, dV \) \\ \hline
2 &
\(S_{ij} = \frac{1}{3} 
\int \Big( 
(\mathbf{r} \times \mathbf{J})_i r_j + (\mathbf{r} \times \mathbf{J})_j r_i 
\Big) dv.
 \) \\ \hline
3 &
\( O_{jpl}^\text{(m)}=\frac{1}{3} \Big( O_{jpl}^\text{(m)} + O_{pjl}^\text{(m)} + O_{lpj}^\text{(m)} \Big)
- \frac{1}{5} \Big( 
\delta_{jp} {O}_{vvl}^\text{(m)} + \delta_{jl} {O}_{vvp}^\text{(m)} + \delta_{pl} {O}_{vvj}^\text{(m)} 
\Big) + \frac{1}{3} \epsilon_{ajp} \Big( {R}_{al} + \frac{1}{2} \epsilon_{\beta al} G_\beta \Big) 
- \frac{1}{3} \epsilon_{ajl} \Big( {R}_{ap} + \frac{1}{2} \epsilon_{\beta ap} G_\beta \Big)\quad \text{where} \quad{O}_{vvj}^\text{(m)} = \int r^2 (\mathbf{r} \times \mathbf{J})_j dv,
{R}_{al} = \frac{3}{8} \int r^2 \Big( J_l r_\alpha + J_\alpha r_l \Big) - 2 (\mathbf{J} \cdot \mathbf{r}) r_\alpha r_l \, dv,
G_\beta = -\frac{3}{4} \int r^2 (\mathbf{r} \times \mathbf{J})_\beta \, dv \) \\ \hline

\end{tabular}
\label{Tab_2}
\end{table}

The corresponding toroidal expression formulas are presented in Table~\ref{table3}.

\begin{table}[H]
\centering
\small
\caption{The Set of the Toroidal Moments}
\renewcommand{\arraystretch}{1.8} % Adjust row spacing
\setlength{\arraycolsep}{2pt} % Adjust column spacing
\begin{tabular}{|c|p{12cm}|}
\hline
\textbf{Order} & \textbf{Toroidal Electric Multipoles} \\ \hline
1 &
\(
T_j^\text{(e)} = \frac{1}{10} \int \Big( (\mathbf{J} \cdot \mathbf{r}) r_j - 2r^2 J_j \Big) \, dV
\) \\ \hline
1 &
\(
T_j^{(2e)} = \frac{1}{280} \int \Big( 3r^4 J_j - 2r^2 (\mathbf{r} \cdot \mathbf{J}) r_j \Big) \, dV
\) \\ \hline
2 &
\(
T_{jk}^{(Q^\text{(e)})} = \frac{1}{42} \int \Big( 4(\mathbf{r} \cdot \mathbf{J}) r_j r_k 
+ 2 (\mathbf{J} \cdot \mathbf{r}) r_k \delta_{jk} 
- 5r^2 (r_j J_k + r_k J_j) \Big) \, dV
\) \\ \hline
3 &
\(
T_{jkl}^{(O^\text{(e)})} = \frac{1}{300} \int \Big( 35(\mathbf{r} \cdot \mathbf{J}) r_j r_k r_l 
- 20r^2 (r_j J_k r_l + r_k J_j r_l + r_l J_k r_j) 
+ (\delta_{jk} \delta_{lp} + \delta_{jl} \delta_{kp} + \delta_{kl} \delta_{jp}) 
\Big[ (\mathbf{r} \cdot \mathbf{J}) r_p r_j + 4r_p r^4 J_l \Big] \Big) \, dV
\) \\ \hline
\textbf{Order} & \textbf{Toroidal Magnetic Multipoles} \\ \hline
1 &
\(
T_j^\text{(m)} = \frac{i \omega}{20} \int r^2 (\mathbf{r} \times \mathbf{J}) \, dV
\) \\ \hline
2 &
\(
T_{jp}^{(Q^\text{(m)})} = \frac{i \omega}{42} \int r^2 \Big[ r_j (\mathbf{r} \times \mathbf{J})_p 
+ (\mathbf{r} \times \mathbf{J})_j r_p \Big] \, dV
\) \\ \hline
\end{tabular}
\label{table3}
\end{table}

We know from~\cite{gurvitz2019high} that the electric field scattered by a single particle has the form
\begin{equation} % This ensures numbering
\begin{aligned} % Align allows multi-line equations
\mathbf{E} &= \frac{k^2}{4 \pi \varepsilon} \frac{e^{ikR}}{R} 
\Bigg\{
\mathbf{\hat{R}} \times
\Bigg[
\left( \mathbf{p} + \frac{ik \varepsilon}{c} \mathbf{T}^\text{(e)} + \frac{ik^3 \varepsilon^2}{c} \mathbf{T}^{(2e)} \right) \times \mathbf{\hat{R}}
\Bigg] \\
&\quad + \frac{ik}{2} \mathbf{\hat{R}} \times
\Bigg[
\mathbf{\hat{R}} \times
\Bigg(
\Big(
\mathbf{Q}^\text{(e)} + \frac{ik \varepsilon}{c} \mathbf{T}^{(Q^\text{(e)})}
\Big) \cdot \mathbf{\hat{R}}
\Bigg)
\Bigg] \\
&\quad + \frac{k^2}{6} \mathbf{\hat{R}} \times
\Bigg[
\mathbf{\hat{R}} \times
\Bigg(
\Big( \mathbf{O}^\text{(e)} + \frac{ik \varepsilon}{c} \mathbf{T}^{(O^\text{(e)})} \Big) \cdot \mathbf{\hat{R}} \cdot \mathbf{\hat{R}}
\Bigg)
\Bigg] \\
%&\quad + \frac{ik^3}{24} \mathbf{\hat{z}} \times
%\Bigg[
%\Big( \mathbf{L}^{e} \cdot \mathbf{\hat{z}} \cdot \mathbf{\hat{z}} \cdot \mathbf{\hat{z}} \Big) \times \mathbf{\hat{z}}
%\Bigg] \\
%&\quad + \frac{k^4}{120} \mathbf{\hat{z}} \times
%\Bigg[
%\Big( \mathbf{D} \cdot \mathbf{\hat{z}} \cdot \mathbf{\hat{z}} \cdot \mathbf{\hat{z}} \cdot \mathbf{\hat{z}} \Big) \times \mathbf{\hat{z}}
%\Bigg] \\
&\quad + \frac{1}{c}
\Bigg[
\Big( \mathbf{m} + \frac{ik \varepsilon}{c} \mathbf{T}^\text{(m)} \Big) \times \mathbf{\hat{R}}
\Bigg] \\
&\quad + \frac{ik}{2c} \mathbf{\hat{R}} \times
\Bigg[
\Big( \mathbf{Q}^\text{(m)} + \frac{ik \varepsilon}{c} \mathbf{T}^{(Q^\text{(m)})} \Big) \cdot \mathbf{\hat{R}}
\Bigg].
%&\quad + \frac{k^2}{6c}
%\Bigg[
%\mathbf{\hat{z}} \times \Big( \mathbf{X} \cdot \mathbf{\hat{z}} \cdot \mathbf{\hat{z}} \Big)
%+ \frac{ik^3}{24c} \Big( \mathbf{Z} \cdot \mathbf{\hat{z}} \cdot \mathbf{\hat{z}} \cdot \mathbf{\hat{z}} \Big) \times \mathbf{\hat{z}}
%\Bigg]
%\Bigg\}
\end{aligned}
\label{scat_primi_cleaned} % Keep label for referencing
\end{equation}

This can be generally expressed as
\begin{equation}
{\bf{E}} = \frac{{{k^2}}}{{4\pi {\varepsilon}}}\frac{{{e^{ikR}}}}{R}\sum\limits_{j = 1}^\infty  {\frac{{{{( - ik)}^{j - 1}}}}{{j!}}}\left\{  {\mathbf{\hat{R}}} \times \left[\left(\mathbf{P_j} \cdot \bigotimes_{j-1}{{\mathbf{\hat{R}}}}\right) \times {\mathbf{\hat{R}}}\right] + \frac{1}{c} \left(\mathbf{M_j} \cdot \bigotimes_{j-1}{{\mathbf{\hat{R}}}}\right) \times {\mathbf{\hat{R}}} \right\},
\end{equation}
where \( \mathbf{P}_{j} \) and \( \mathbf{M}_{j} \) are the multipole coefficients defined in~\eqref{PM_general}, \( j \) is the multipolar order, and \( k = \omega \sqrt{\varepsilon \mu} \) is the wave number. In this expression, $\bigotimes_{j-1}{{\mathbf{\hat{R}}}}$ represents tensor products such that, for instance, $\bigotimes_{j=2}{{\mathbf{\hat{R}}}} = \mathbf{\hat{R}}\otimes\mathbf{\hat{R}}=\mathbf{\hat{R}}\mathbf{\hat{R}}$. Note that this expression is similar to~Eq.~(2) of the main text and is only valid under normal incidence.

\newpage
\renewcommand{\theequation}{B\arabic{equation}}
\section{\label{mapping_transformation}Appendix B: Mapping between multipolar representations in Cartesian and spherical coordinates}
The Cartesian components \( T_{pqr} \) may be connected to the spherical components \( T_{l}^m \) as ~\cite{normand1982relations}

\begin{equation}
(T_l)_{pqr} = \frac{(-i)^q}{2^{p+q}} \sum_{\substack{m=0 \\ p+q+m \text{ even}}}^{p+q}
\frac{C(p, q, m)}{1 + \delta_{m0}}
\sqrt{\frac{2^l (l-m)!(l+m)!}{(2l)!}}
\Big[ T_{lm} + (-1)^p T_{l,-m} \Big],
\label{transformer}
\end{equation}
where
\begin{equation}
C(p, q, m) = \sum_{n=0}^\infty \frac{n! (p - n)!}{\Big[\frac{1}{2}(p + q + m) - n\Big]! \Big[\frac{1}{2}(q - p - m) + n\Big]!} \cdot \frac{(-1)^n p! q!}{\Big[\frac{1}{2}(p + q + m)\Big]!}.
\end{equation}
and \( \delta_{m0} \) is the Kronecker delta, \( l = p+q+r \), where \( p, q, r \) are the exponents of Cartesian coordinates \( x, y, z \), and $(T_l)_{pqr} = (T_l)_{x^p y^q z^r}$.
For dipoles (e.g., the electric dipole \( p_z \)), we have
\begin{equation}
T_x = C_1 \left( \frac{T_{1,-1}^\omega - T_{11}^\omega}{\sqrt{2}} \right),
\end{equation}
\begin{equation}
T_y = C_1 \left( \frac{T_{1,-1}^\omega + T_{11}^\omega}{i\sqrt{2}} \right),
\end{equation}
\begin{equation}
T_z = C_1 T_{10}^\omega.
\end{equation}

For quadrupoles, we have
\begin{equation}
T_{xx} = C_2 \left( \frac{T_{22}^\omega + T_{2,-2}^\omega}{2} - \frac{1}{\sqrt{6}} T_{20}^\omega \right),
\end{equation}
\begin{equation}
T_{xy} = T_{yx} = C_2 \left( \frac{T_{2,-2}^\omega - T_{22}^\omega}{2i} \right),
\end{equation}
\begin{equation}
T_{xz} = T_{zx} = C_2 \left( \frac{T_{2,-1}^\omega - T_{21}^\omega}{2} \right),
\end{equation}
\begin{equation}
T_{yz} = T_{zy} = C_2 \left( \frac{T_{2,-1}^\omega + T_{21}^\omega}{2i} \right),
\end{equation}
\begin{equation}
T_{yy} = C_2 \left[ - \left( \frac{T_{22}^\omega + T_{2,-2}^\omega}{2} \right) - \frac{1}{\sqrt{6}} T_{20}^\omega \right],
\end{equation}
\begin{equation}
T_{zz} = C_2 \left( \frac{2}{\sqrt{6}} T_{20}^\omega \right) = -T_{xx} - T_{yy}.
\end{equation}
and for octupoles as
\begin{align}
    T_{xxx} &= -T_{yyx} - T_{zzx} = C_3 \frac{1}{10\sqrt{2}} 
    \left[ \sqrt{15} \left( T_{31}^{\omega} - T_{3,-1}^{\omega} \right) + 5 \left( T_{3,-3}^{\omega} - T_{33}^{\omega} \right) \right], \\[10pt]
    T_{yyy} &= -T_{txy} - T_{zzy} = C_3 \frac{i}{10\sqrt{2}} 
    \left[ \sqrt{15} \left( T_{31}^{\omega} + T_{3,-1}^{\omega} \right) + 5 \left( T_{3,-3}^{\omega} + T_{33}^{\omega} \right) \right], \\[10pt]
    T_{zzz} &= -T_{xxz} - T_{yyz} = C_3 \sqrt{\frac{2}{5}} T_{30}^{\omega}, \\[10pt]
    T_{xxy} &= T_{xyx} = T_{yxx} = C_3 \frac{i}{30\sqrt{2}} 
    \left[ \sqrt{15} \left( T_{31}^{\omega} - T_{3,-1}^{\omega} \right) - 15 \left( T_{3,-3}^{\omega} + T_{33}^{\omega} \right) \right], \\[10pt]
    T_{xxz} &= T_{xzx} = T_{zxx} = C_3 \left[ \frac{1}{2\sqrt{3}} \left( T_{32}^{\omega} + T_{3,-2}^{\omega} \right) - \frac{1}{\sqrt{10}} T_{30}^{\omega} \right], \\[10pt]
    T_{xyy} &= T_{yyx} = T_{yxy} = C_3 \frac{1}{30\sqrt{2}} 
    \left[ \sqrt{15} \left( T_{31}^{\omega} - T_{3,-1}^{\omega} \right) - 15 \left( T_{3,-3}^{\omega} - T_{33}^{\omega} \right) \right], \\[10pt]
    T_{xzz} &= T_{zxz} = T_{zzx} = C_3 \sqrt{\frac{2}{15}} \left( T_{3,-1}^{\omega} - T_{31}^{\omega} \right), \\[10pt]
    T_{yyz} &= T_{yzy} = T_{zyy} = -C_3 \left[ \frac{1}{2\sqrt{3}} \left( T_{32}^{\omega} + T_{3,-2}^{\omega} \right) + \frac{1}{\sqrt{10}} T_{30}^{\omega} \right], \\[10pt]
    T_{yzz} &= T_{zyz} = T_{zzy} = -C_3 i \sqrt{\frac{2}{15}} \left( T_{3,-1}^{\omega} + T_{31}^{\omega} \right), \\[10pt]
    T_{xyz} &= T_{xzy} = T_{yxz} = T_{yzx} = T_{zxy} = T_{zyx} = C_3 \frac{i}{2\sqrt{3}} \left( T_{32}^{\omega} - T_{3,-2}^{\omega} \right).
\end{align}

In all the above expressions, the \( C_n \) terms are constant coefficients that can be derived by comparing the long-wavelength approximations of the spherical and primitive multipole terms. The \( T_l^m \) terms correspond to the spherical multipole moments and originate from~Eq.(4) of the main text. These can be studied separately for electric and magnetic multipoles using the subscripts \( \{\text{e}, \text{m}\} \), where \( l \) indicates the multipole order. Specifically, we have \( T_{l,m}^e = b_{l,m}^e \) and \( T_{l,m}^m = b_{l,m}^m \). The general forms of \( b_{l,m}^e \) and \( b_{l,m}^m \) from~\cite{alaee2018electromagnetic} are

\begin{align}
\sqrt{\left(\frac{2\pi}{4\pi}\right)^3} b^e_{lm} 
&= \sum_{{\ell} {m}} (-i)^{{\ell}} 
\int d\hat{\mathbf{p}} \, \mathbf{Z}^\dagger_{lm} (\hat{\mathbf{p}}) Y_{{\ell} {m}} (\hat{\mathbf{p}}) 
\int d^3 \mathbf{r} \, \mathbf{J} (\mathbf{r}) Y_{{\ell} {m}}^* (\hat{\mathbf{r}}) j_{{\ell}} (kr), \\
\sqrt{\left(\frac{2\pi}{4\pi}\right)^3} b^m_{lm} 
&= \sum_{{\ell} {m}} (-i)^{{\ell}} 
\int d\hat{\mathbf{p}} \, \mathbf{X}^\dagger_{lm} (\hat{\mathbf{p}}) Y_{{\ell} {m}} (\hat{\mathbf{p}}) 
\int d^3 \mathbf{r} \, \mathbf{J} (\mathbf{r}) Y_{{\ell} {m}}^* (\hat{\mathbf{r}}) j_{{\ell}} (kr).
\label{coefff}
\end{align}
For example, let us consider the electric quadrupole. By following the above procedure for calculating higher-order multipoles using the \( T_{l}^m \) coefficients, we obtain the quadrupole expression as
\begin{align}
Q_{ij}^\text{(e)} &= C_2^e\frac{ik}{2\pi \sqrt {10}} \bigg\{
\int d^3 \mathbf{r} \bigg[ 
3 \big(r_j J_i^\omega + r_i J_j^\omega\big) 
- 2 \big(r \cdot \mathbf{J}_\omega\big) \delta_{ij} 
\bigg] 
\frac{j_1(kr)}{kr} \nonumber \\
&\quad + 2k^2 \int d^3 \mathbf{r} \bigg[ 
5 r_i r_j \big(r \cdot \mathbf{J}_\omega\big) 
- \big(r_i J_j + r_j J_i\big) r^2 
- r^2 \big(r \cdot \mathbf{J}_\omega\big) \delta_{ij} 
\bigg] 
\frac{j_3(kr)}{(kr)^3} 
\bigg\}.\label{quad}
\end{align}
Now, to obtain the \( C_n \) coefficients and also demonstrate the origin of the toroidal moments, we apply the long-wavelength approximation and use the small-argument approximation for the spherical Bessel functions, as follows
\begin{align}
\frac{j_1(kr)}{kr} &\approx \frac{1}{3} \left( 1 - \frac{(kr)^2}{10} \right), \\
\frac{j_3(kr)}{(kr)^3} &\approx \frac{1}{105},
\end{align}
which reduce the long-wavelength approximation for quadrupole moment to~\cite{alaee2018electromagnetic}
\begin{align}
&Q_{ij}^\text{(e)}\approx C_2^e\frac{ik}{6\pi \sqrt {10}} \bigg\{
\int d^3 \mathbf{r} 
\bigg[
3 \big(r_j J_i^\omega + r_i J_j^\omega\big) 
- 2 \big(r \cdot \mathbf{J}_\omega\big) \delta_{ij} 
\bigg] \nonumber \\
&\quad + k^2 \left(\frac{1}{14}\right) 
\int d^3 \mathbf{r} 
\bigg[
4 r_i r_j \big(r \cdot \mathbf{J}_\omega\big) 
- 5 r^2 \big(r_i J_j + r_j J_i\big) 
+ 2 r^2 \big(r \cdot \mathbf{J}_\omega\big) \delta_{ij} 
\bigg] 
\bigg\}.
\label{quad_ap}
\end{align}
By comparing the first part of~\eqref{quad_ap} with the primitive expression for the electric quadrupole moment given in Table~\ref{Tab_2}, we find that
\begin{equation}
C_2^e = \frac{{2\pi \sqrt {10} }}{{c{k^2}}}. 
\end{equation}
By comparing the first and second terms of~\eqref{quad_ap}, we observe that the first term corresponds to the primitive quadrupole moment, as also listed in Table~\ref{Tab_2}, given by
\begin{equation}
 {Q}_{jk}^\text{(e)} = \frac{i}{\omega} 
\int \left( r_j J_k + r_k J_j - \frac{2}{3} \delta_{jk} (\mathbf{r} \cdot \mathbf{J}) \right) dV,
\end{equation}
and the corresponding quadrupole toroidal moment in Table~\ref{table3} as
\begin{equation}
    \mathbf{T}_{jk,s}^{Q^\text{(e)}} = \frac{1}{42} \int \Big( 4(\mathbf{r} \cdot \mathbf{J}) r_j r_k 
+ 2 (\mathbf{J} \cdot \mathbf{r}) r_k \delta_{jk} 
- 5r^2 (r_j J_k + r_k J_j) \Big) \, dV
\end{equation}
which allows us to finally conclude that
\begin{equation}
    \mathbf{Q}_{jk}^\text{(e)} \approx \mathbf{Q}_{jk}^\text{(e)} + \frac{ik \epsilon_d}{c} \mathbf{T}_{jk,s}^{Q^\text{(e)}}. 
\end{equation}
Note that by comparing~Eq.(2) of the main text and~\eqref{scat_primi_cleaned}, we ensure consistency with the multipoles provided in Table 1 of the main text. This consistency is achieved through minor adjustments to the equations presented in Ref.~\cite{alaee2018electromagnetic}.

By using the generalized Cartesian-to-spherical transformation provided in~\eqref{transformer}, we can obtain the generalized definitions for the higher-order electric (\( \mathbf{P}_n \)) and magnetic (\( \mathbf{M}_n \)) multipoles as
\begin{equation}
    \begin{cases}
        \mathbf{P}_n = C_n^e\,\mathbf{T}_n^e, \\
        \mathbf{M}_n = C_n^m\,\mathbf{T}_n^m.
    \end{cases}
    \label{PM_general}
\end{equation}
Here, \( \mathbf{P}_n \) denotes the \( n^\text{th} \) order electric multipoles, and \( \mathbf{M}_n \) denotes the \( n^\text{th} \) order magnetic multipoles. The coefficients \( C_n^e \) and \( C_n^m \) are consistency factors obtained from comparing the spherical and primitive representations discussed earlier.

It is also worth noting that, to simplify calculations, we use the fact that multipole tensors are symmetric. For symmetric tensors of rank \( n \), the following relation holds
\begin{equation}
    T_{i_1 i_2 \cdots i_n} = T_{\pi(i_1) \pi(i_2) \cdots \pi(i_n)}
\end{equation}
where \( \pi \) represents any permutation of the indices \( i_1, i_2, \ldots, i_n \). This is exactly why we can express the indices in the form \( \{ x^p y^q z^r \} \) in~\eqref{transformer}, which can be shown as follows
\begin{equation}
\{ {i_1},{i_2},{i_3},....{i_n}\}  = \{ \pi ({i_1}),\pi ({i_2}), \cdots \pi ({i_n})\}  = \{ {x^p}{y^q}{z^r}\}.
\end{equation}
The traceless property of this tensors also implies
\begin{equation}
T_{\left\{ {q = r = 0} \right\}}^n + T_{\left\{ {p = r = 0} \right\}}^n + T_{\left\{ {q = p = 0} \right\}}^n = 0.
\end{equation}
Finally, some useful relationships for connecting the spherical basis, which have been used in the main text, are given below
\begin{equation}
\mathbf{M}_{jm}^{(3)} = h_j(kr)\,\mathbf{X}_{jm}\bigl(kr,\theta,\phi\bigr),
\quad
\mathbf{N}_{jm}^{(3)} = \frac{1}{k}\,\nabla \times \mathbf{M}_{jm}^{(3)}.
\label{vsh}
\end{equation}
\begin{equation}
\hat{\mathbf{r}} \times \mathbf{X}_{jm}\bigl(kr,\theta,\phi\bigr)
= \gamma_{jm}\Bigl[r_{jm}(\cos\theta)\,\hat{\boldsymbol{\theta}}
+ i\,n_{jm}(\cos\theta)\,\hat{\boldsymbol{\phi}}\Bigr]
\,e^{j\,m\,\phi}
= \mathbf{Z}_{jm}\bigl(kr,\theta,\phi\bigr).
\label{vsh1}
\end{equation}
\subsubsection{Definitions of terms in Table~1}
\label{some terms definition}

\begin{table}[H]
\centering
\footnotesize
\begin{tabular}{c|p{13cm}}
\hline
ET for electric &
\(\displaystyle
B_{ijkl}^{(\text{e})} = \frac{1}{7} \left(\delta_{ij} \hat{C}^{(\text{e})}_{kl} + \delta_{ik} \hat{C}^{(\text{e})}_{jl} 
+\delta_{il} \hat{C}^{(\text{e})}_{jk} +\delta_{jk} \hat{C}^{(\text{e})}_{il} +\delta_{jl} \hat{C}^{(\text{e})}_{ik} +\delta_{kl} \hat{C}^{(\text{e})}_{ij} \right),
\)

\(\displaystyle
\hat{\mathbf{C}}^{(\text{e})} = 2 (\mathbf{J} \cdot \mathbf{r}) \, \mathbf{rJrJ} + r^2 (\mathbf{Jr} + \mathbf{rJ}) - \frac{4}{3} r^2 (\mathbf{r} \cdot \mathbf{J}) \, \bar{\bar{I}}, \quad
\mathbf{V}^{(\text{e})} = \frac{1}{5} \left[ 2 (\mathbf{r} \cdot \mathbf{J}) \mathbf{r} + r^2 \mathbf{J} \right]
\) \\ \hline

ET for magnetic &
\(\displaystyle
B_{ijkl}^{(\text{m})} = \frac{1}{7A} \left(\delta_{ij} \hat{C}^{(\text{m})}_{kl} + \delta_{ik} \hat{C}^{(\text{m})}_{jl} 
+ \delta_{il} \hat{C}^{(\text{m})}_{jk} + \delta_{jk} \hat{C}^{(\text{m})}_{il} + \delta_{jl} \hat{C}^{(\text{m})}_{ik} + \delta_{kl} \hat{C}^{(\text{m})}_{ij} \right),
\)

\(\displaystyle
\hat{\mathbf{C}}^{(\text{m})} = r^2 \left[ (\mathbf{r} \times \mathbf{J}) \mathbf{r} + \mathbf{r} (\mathbf{r} \times \mathbf{J}) \right], \quad
\mathbf{V}^{(\text{m})} = \frac{1}{5} r^2 (\mathbf{r} \times \mathbf{J})
\) \\ \hline
\end{tabular}
\caption{Electric and magnetic extra tensor terms used in the expansion.}
\label{extra_terms}
\end{table}

\subsubsection{Comparison between exact spherical multipoles and irreducible representation and origin of toroidal moments}
\label{Comparison Between Exact Spherical Multipoles and Irreducible Representation and present origin of Toroidal Modes}
By comparing~Eq.(2) of the main text and~\eqref{scat_primi_cleaned}, we see, using the long-wave approximation of multipoles in Table~\ref{Tab_2}, that we have~\cite{alaee2018electromagnetic}
\begin{equation}
\begin{aligned}
    \mathbf{p}_j^s &\approx \mathbf{p}_j + \frac{ik \varepsilon_d}{c} \mathbf{T}_{j,s}^\text{(e)} + \frac{ik^3 \varepsilon_d^2}{c} \mathbf{T}_{j,s}^{(2\text{e})}, \\
    \mathbf{m}_j^s &\approx \mathbf{m}_j + \frac{ik \varepsilon_d}{c} \mathbf{T}_{j,s}^\text{(m)}, \\
\mathbf{Q}_{jk}^\text{(e)} &\approx \mathbf{Q}_{jk}^\text{(e)} + \frac{ik \varepsilon_d}{c} \mathbf{T}_{jk,s}^{(Q^\text{(e)})}, \\
    \mathbf{S}_{jp}^\text{(m)} &\approx \mathbf{S}_{jp}^\text{(m)} + \frac{ik \varepsilon_d}{c} \mathbf{T}_{jp,s}^{(Q^\text{(m)})},
    \label{approxim_spherical_quad}
\end{aligned}
\end{equation}
where the superscript ``s'' is related to the exact spherical multipolar moments given in ~Table 1 of the main text and the other terms correspond to the primitive and toroidal multipolar moments given in Table~\ref{Tab_2} and Table~\ref{table3}, respectively.

In the general decomposition of primitive multipoles into symmetric (s), antisymmetric (a) and diagonal parts, a rank-2 tensor \( T_{ij} \) can be decomposed as
\begin{equation}
    T_{ij} = T_{ij}^\text{s} + T_{ij}^\text{a} + \operatorname{Tr}[T_{ij}],
\end{equation}
where the symmetric and antisymmetric components are given by
\begin{equation}
    T_{ij}^\text{s} = \frac{1}{2} (T_{ij} + T_{ji}), \quad 
    T_{ij}^\text{a} = \frac{1}{2} (T_{ij} - T_{ji}).
\end{equation}

The symmetric part (\( T_{ij}^\text{s} \)) corresponds to electric multipoles, while the antisymmetric part (\( T_{ij}^\text{a} \)) corresponds to magnetic multipoles and the diagonal part (\( \text{Tr}[T_{ij}] \)) corresponds to toroidal and charge mean values~\cite{nanz2016toroidal}.

In a general case, we have
\begin{equation}
T_{k-2} \propto \text{Tr}(T_{ij}) \epsilon_{ijk}, \quad 
M_{k-1} \propto \epsilon_{ijk}, T_{ij}^\text{a} \quad  \text{and} \quad  P_k \propto T_{ij}^\text{s}.
\end{equation}

It is also interesting to note that toroidal terms can be obtained directly by comparing with the spherical multipole expressions. However, when using the method of symmetrization and detracing, the toroidal or magnetic modes often appear from higher-order tensors. For example, to extract the toroidal dipole mode, one must symmetrize and detrace the octupolar terms, as shown below~\cite{riccardi2022}.

\begin{table}[H]
    \centering
    \caption{Overview of the multipole moments emerging in different orders of $\vec{A}$}
    \label{tab:multipole_moments}
    \renewcommand{\arraystretch}{1.5}
    \begin{tabular}{l c c c}
        \toprule
        \multicolumn{1}{c}{} & \multicolumn{3}{c}{$n$} \\ 
        \cmidrule(lr){2-4}
        type & 0 & 1 & 2 \\
        \midrule
        el.  & $\mathbf{p}$ & $\mathbf{Q}^\text{(e)}$ & $\mathbf{O}^\text{(e)}$ \\
        mag. & - & $\mathbf{m}$ & $\mathbf{Q}^\text{(m)}$ \\
        tor. & - & - & $\mathbf{T}^\text{(e)}$ \\
        \bottomrule
    \end{tabular}
\end{table}

\newpage
\renewcommand{\theequation}{C\arabic{equation}}
\section{\label{fields_inhomo}Appendix C: Evaluation of reflected and transmitted electric fields for different substrate and superstrate media}
To find the corresponding electric and magnetic fields in the case where the structure is embedded between different substrate and superstrate materials, we need to apply the boundary conditions at the material interface. These conditions are then solved using the equivalent impedances for TE and TM polarizations at the top and bottom regions.

The impedance components defined in~Eq.(22) of the main text for TE and TM polarizations can be written as~\cite{9100112}
\begin{equation}
\mathbf{Z}_{l} = Z_{l}^{\text{TM}}\,\frac{\mathbf{k}_{\|}\mathbf{k}_{\|}}{k_{\|}^2} 
+ Z_{l}^{\text{TE}}\,\frac{\mathbf{n}\times\mathbf{k}_{\|}\times\mathbf{k}_{\|}}{k_{\|}^2},
\quad \text{and} \quad
\mathbf{Y}_{l} = \mathbf{Z}_{l}^{-1}.
\label{impedance1}
\end{equation}
where $\ve{n}$ is unit vector normal to the interface and \(l = 1,2\) corresponds to the top and bottom layer environments and
\begin{equation}
Z_{l}^{\text{TM}} = \eta_{l} \sqrt{1 - \frac{k_\parallel^2}{k_{l}^2}}, \quad
Z_{l}^{\text{TE}} = \frac{\eta_{l}}{\sqrt{1 - \frac{k_\parallel^2}{k_{l}^2}}},
\label{impedance2}
\end{equation}
where \(\eta_{l}\) is the impedance and \(k_{l}\) is the wavevector in the corresponding environment defined as
\begin{equation}
k_{l} = n_l k, \quad \text{and} \quad \eta_l = \sqrt{\frac{\mu_l}{\varepsilon_l}}.
\label{impedance3}
\end{equation}

By applying the transverse impedance formula~Eq.(22) of the main text and using~Eq.(21) of the main text, we have
\begin{equation}
\begin{aligned}
\mathbf{E}_{\|, \mathrm{t}} - \mathbf{E}_{\|, i} - \mathbf{E}_{\|, \mathrm{r}} &= \boldsymbol\psi_{E}, \\
\hat{\mathbf{z}} \times \mathbf{Y}_{2} \cdot \mathbf{E}_{\|, \mathrm{t}} - \hat{\mathbf{z}} \times \mathbf{Y}_{1} \cdot \mathbf{E}_{\|, i} + \hat{\mathbf{z}} \times \mathbf{Y}_{1} \cdot \mathbf{E}_{\|, \mathrm{r}} &= \boldsymbol\psi_{H}.
\end{aligned}
\end{equation}
Here, the indices 1 and 2 correspond to the \(-z\) and \(+z\) regions, respectively. Pre-multiplying the second equation by \(\hat{\mathbf{z}} \times\) and rewriting the admittances as impedances, we obtain
\begin{equation}
\begin{aligned}
\mathbf{E}_{\|, \mathrm{t}} - \mathbf{E}_{\|, i} - \mathbf{E}_{\|, \mathrm{r}} &= \boldsymbol\psi_{E}, \\
\mathbf{Z}_{1} \cdot \mathbf{E}_{\|, \mathrm{t}} - \mathbf{Z}_{2} \cdot \mathbf{E}_{\|, i} + \mathbf{Z}_{2} \cdot \mathbf{E}_{\|, \mathrm{r}} &= - \mathbf{Z}_{1} \cdot \mathbf{Z}_{2} \cdot \Bigl( \hat{\mathbf{z}} \times \boldsymbol\psi_{H} \Bigr).
\end{aligned}
\end{equation}
Solving these equations for the reflected and transmitted electric fields yields
\begin{equation}
\begin{aligned}
\mathbf{E}_{\|, \mathrm{t}} &= 2\,\mathbf{Z}_{2} \cdot \Bigl(\mathbf{Z}_{1}+\mathbf{Z}_{2}\Bigr)^{-1} \cdot \mathbf{E}_{\|, i} 
+ \mathbf{Z}_{2} \cdot \Bigl(\mathbf{Z}_{1}+\mathbf{Z}_{2}\Bigr)^{-1} \cdot \Bigl[\boldsymbol\psi_{E} - \mathbf{Z}_{1} \cdot \Bigl(\hat{\mathbf{z}} \times \boldsymbol\psi_{H}\Bigr)\Bigr],
\end{aligned}
\label{E_t1}
\end{equation}
\begin{equation}
\begin{aligned}
\mathbf{E}_{\|, \mathrm{r}} &= \Bigl(\mathbf{Z}_{2} - \mathbf{Z}_{1}\Bigr) \cdot \Bigl(\mathbf{Z}_{1}+\mathbf{Z}_{2}\Bigr)^{-1} \cdot \mathbf{E}_{\|, i} 
- \mathbf{Z}_{1} \cdot \Bigl(\mathbf{Z}_{1}+\mathbf{Z}_{2}\Bigr)^{-1} \cdot \Bigl[\boldsymbol\psi_{E} + \mathbf{Z}_{2} \cdot \Bigl(\hat{\mathbf{z}} \times \boldsymbol\psi_{H}\Bigr)\Bigr].
\end{aligned}
\label{E_r1}
\end{equation}
By using~Eq.(22) of the main text and by following a similar procedure, the transverse magnetic fields are
\begin{equation}
\mathbf{H}_{\|,\mathrm{t}} = 2 \Big(\mathbf{Z}_1 + \mathbf{Z}_2\Big)^{-1}\cdot\mathbf{Z}_1\cdot\mathbf{H}_{\|,i} 
+ \Big(\mathbf{Z}_1 + \mathbf{Z}_2\Big)^{-1}\cdot\Big[\mathbf{Z}_1\cdot\boldsymbol\psi_H - \boldsymbol\psi_E \times \hat{\mathbf{z}}\Big],
\label{mag_with_impe}
\end{equation}
\begin{equation}
\mathbf{H}_{\|,\mathrm{r}} = \Big(\mathbf{Z}_1 + \mathbf{Z}_2\Big)^{-1}\cdot\Big(\mathbf{Z}_2-\mathbf{Z}_1\Big)\cdot\mathbf{H}_{\|,i}
+ \Big(\mathbf{Z}_1 + \mathbf{Z}_2\Big)^{-1}\cdot\Big[\boldsymbol\psi_E \times \hat{\mathbf{z}} + \mathbf{Z}_2\cdot\boldsymbol\psi_H\Big].
\label{mag_with_impe1}
\end{equation}
For identical media, these formulas reduce to
\begin{equation}
\begin{aligned}
\mathbf{H}_{\|,\mathrm{t}} &= \mathbf{H}_{i} + \frac{1}{2}\Big[\boldsymbol\psi_H - \mathbf{Y}\cdot\Big(\boldsymbol\psi_E \times \hat{\mathbf{z}}\Big)\Big], \\
\mathbf{H}_{\|,\mathrm{r}} &= \frac{1}{2}\Big[\boldsymbol\psi_H + \mathbf{Y}\cdot\Big(\boldsymbol\psi_E \times \hat{\mathbf{z}}\Big)\Big].
\end{aligned}
\label{H_free}
\end{equation}

\newpage
\renewcommand{\theequation}{D\arabic{equation}}
\section{\label{map_table}Appendix D: Mapping relations for multipoles in a homogeneous medium}
In order to simplify the equations~Eq.(19) of the main text and~Eq.(20) of the main text in terms of multipoles, we denote them by $\boldsymbol{\psi}_E$ and $\boldsymbol{\psi}_H$, respectively. They are presented in~Eq.(25) of the main text and~Eq.(27) of the main text, where the multipolar terms are defined in Table~\ref{table_te_tm_cores}.
\begin{table}[H]
\centering
\renewcommand{\arraystretch}{1.5} % Increase vertical spacing between rows
\footnotesize
\begin{tabular}{|c|c|c|c|}
\hline
\multicolumn{2}{|c|}{\textbf{TE Multipoles}} & \multicolumn{2}{c|}{\textbf{TM Multipoles}} \\
\hline
\textbf{Original Multipole} & \textbf{Tilde Form} & \textbf{Original Multipole} & \textbf{Tilde Form} \\
\hline
$ i\omega \mu M_x$ & $\tilde{M}_x$ & $ i \omega\mu M_y$ & $\tilde{M}_y$ \\
\hline
$\frac{k^2}{2\varepsilon} Q^\text{(e)}_{yz}$ & $\tilde{Q}^\text{(e)}_{yz}$ & $\frac{k^2-k_x^2}{2 \varepsilon} Q^\text{(e)}_{xz}$ & $\tilde{Q}^\text{(e)}_{xz}$ \\
\hline
$ -\frac{\omega \mu}{2}k_x Q^\text{(m)}_{xx} $ & $\tilde{Q}^\text{(m)}_{xx}$ & $\frac{-i}{\varepsilon} k_x P_z$ & $\tilde{P}_z$ \\
\hline
$ -\frac{\omega \mu}{2}k_x Q^\text{(m)}_{zz} $ & $\tilde{Q}^\text{(m)}_{zz}$ & $i\omega{P}_x$ & $\tilde{P}_x$ \\
\hline
$\frac{i k^2 k_x}{3\varepsilon} O^\text{(e)}_{yzx}$ & $\tilde{O}^\text{(e)}_{yzx}$ & $\frac{{\omega \mu {k_x}}}{2}{Q^\text{(m)}_{xy}}$ & $\tilde{Q}^\text{(m)}_{xy}$ \\
\hline
$-i\omega P_y $ & $\tilde{P}_y$ & $k_x\frac{{2k^2-k_x^2}}{{6\varepsilon }}{O^\text{(e)}_{zxx}}$ & $\tilde{O}^\text{(e)}_{zxx}$ \\
\hline
$ \frac{k_x^2-k_{z}^2}{2} Q^\text{(m)}_{xz}$ & $\tilde{Q}^\text{(m)}_{xz}$ & $ \frac{-i k_{z}^2 k_x}{6 \varepsilon} O^\text{(e)}_{zzz}$ & $\tilde{O}^\text{(e)}_{zzz}$ \\
\hline
$\frac{-\omega k_x}{2} Q^\text{(e)}_{yx} $ & $\tilde{Q}^\text{(e)}_{yx}$ & $\frac{k_{x}^2+k_{x}k_{z}}{2} Q^\text{(m)}_{yz}$ & $\tilde{Q}^\text{(m)}_{yz}$ \\
\hline
$-i k_x M_z$ & $\tilde{M}_z$ & $\frac{\omega}{2}k_x Q^\text{(e)}_{xx} $ & $\tilde{Q}^\text{(e)}_{xx}$ \\
\hline
$- \frac{i \omega k_x^2}{6} O^\text{(e)}_{yxx} $ & $\tilde{O}^\text{(e)}_{yxx}$  & $ -\frac{\omega}{2} k_x Q^\text{(e)}_{zz} $ & $\tilde{Q}^\text{(e)}_{zz}$ \\
\hline
$\frac{i \omega k_{z}^2}{6} O^\text{(e)}_{yzz}$ & $\tilde{O}^\text{(e)}_{yzz}$ &  $\frac{i \omega k_x^2}{3} O^\text{(e)}_{yzx} $ & $\tilde{O}^\text{(e)}_{yzx}$ \\
\hline
 &  & $- \frac{i \omega k_x^2}{6} O^\text{(e)}_{xxx}$ & $\tilde{O}^\text{(e)}_{xxx}$ \\
\hline
 &  & $\frac{i \omega k_{z,2}}{6} O^\text{(e)}_{xzz}$ & $\tilde{O}^\text{(e)}_{xzz}$ \\
\hline
\end{tabular}
\caption{Mapping relations from multipolar decomposition moments to corresponding TE and TM polarizations.}
\label{table_te_tm_cores}
\end{table}

\newpage
\renewcommand{\theequation}{E\arabic{equation}}
\section{\label{sec_structure_dim}Appendix E: Reflected and transmitted electric fields embedded in different media}
By using the impedance formalism described in Appendix C for an interface between different media and considering~Eq.(23) of the main text, we derive the TE electric fields
\begin{subequations}
\label{eq_mat_E_rt}
\begin{align}
E_y^\text{r} &= \left( \frac{\frac{\eta_2 k_2}{k_{z,2}} - \frac{\eta_1 k_1}{k_{z,1}}}{\frac{\eta_2 k_2}{k_{z,2}} + \frac{\eta_1 k_1}{k_{z,1}}} \right) E_y^\text{i} 
- \left( \frac{\frac{\eta_1 k_1}{k_{z,1}}}{\frac{\eta_2 k_2}{k_{z,2}} + \frac{\eta_1 k_1}{k_{z,1}}} \right) \psi_E^\text{TE}
- \left( \frac{\frac{\eta_2 \eta_1 k_1 k_2}{k_{z,2} k_{z,1}}}{\frac{\eta_2 k_2}{k_{z,2}} + \frac{\eta_1 k_1}{k_{z,1}}} \right) \psi_H^\text{TE}\label{mat_E_r},\\
E_y^\text{t} &= 2 \left( \frac{\frac{\eta_2 k_2}{k_{z,2}}}{\frac{\eta_2 k_2}{k_{z,2}} + \frac{\eta_1 k_1}{k_{z,1}}} \right) E_y^\text{i} 
+ \left( \frac{\frac{\eta_2 k_2}{k_{z,2}}}{\frac{\eta_2 k_2}{k_{z,2}} + \frac{\eta_1 k_1}{k_{z,1}}} \right) \psi_E^\text{TE}
- \left( \frac{\frac{\eta_2 \eta_1 k_1 k_2}{k_{z,2} k_{z,1}}}{\frac{\eta_2 k_2}{k_{z,2}} + \frac{\eta_1 k_1}{k_{z,1}}} \right) \psi_H^\text{TE},
\label{mat_E_t}
\end{align}
\end{subequations}
and, by a similar procedure, the magnetic fields are
\begin{subequations}
\begin{align}
H_y^\text{r} &= \left( \frac{\frac{\eta_2 k_2}{k_{z,2}} - \frac{\eta_1 k_1}{k_{z,1}}}{\frac{\eta_2 k_2}{k_{z,2}} + \frac{\eta_1 k_1}{k_{z,1}}} \right) H_y^\text{i} 
+ \left( \frac{\frac{\eta_1 k_1}{k_{z,1}}}{\frac{\eta_2 k_2}{k_{z,2}} + \frac{\eta_1 k_1}{k_{z,1}}} \right) \psi_H^\text{TM}
- \left( \frac{1}{\frac{\eta_2 k_2}{k_{z,2}} + \frac{\eta_1 k_1}{k_{z,1}}} \right) \psi_E^\text{TM},\label{mat_h_r}\\
H_y^\text{t} &= 2 \left( \frac{\frac{\eta_1 k_1}{k_{z,1}}}{\frac{\eta_2 k_2}{k_{z,2}} + \frac{\eta_1 k_1}{k_{z,1}}} \right) H_y^\text{i} 
+ \left( \frac{\frac{\eta_1 k_1}{k_{z,1}}}{\frac{\eta_2 k_2}{k_{z,2}} + \frac{\eta_1 k_1}{k_{z,1}}} \right) \psi_H^\text{TM}
- \left( \frac{1}{\frac{\eta_2 k_2}{k_{z,2}} + \frac{\eta_1 k_1}{k_{z,1}}} \right) \psi_E^\text{TM}.\label{mat_h_t}
\end{align}
\end{subequations}
The corresponding reflection and transmission coefficients are defined in the same way as in free space as
\begin{subequations}
\begin{align}
r^{\text{TE}} &= \frac{E_y^\text{r}}{E_y^\text{i}}, \quad \tilde t^{\text{TE}} = \frac{E_y^\text{t}}{E_y^\text{i}}, \label{T_TE1}\\[1mm]
r^{\text{TM}} &= \frac{H_y^\text{r}}{H_y^\text{i}}, \quad \tilde t^{\text{TM}} = \frac{H_y^\text{t}}{H_y^\text{i}}. \label{r_t_tm}
\end{align}
\end{subequations}
We used the symbol \( \sim \) in the above transmission formulas to emphasize that the corresponding transmission, in the presence of a substrate, should be examined with respect to power conservation. This is because, according to Snell's law, the transmitted wave propagates at a different angle when passing into a medium with a different refractive index. As a result, additional corrections may be required to ensure that power is conserved.

To compute the scattering power coefficients (i.e., reflectance and transmittance), we use the time-averaged Poynting vector, given by
\begin{equation}
\langle S_z \rangle 
= \frac{1}{2}\,\mathrm{Re}\bigl\{\mathbf{E}_0 \times \mathbf{H}^*\bigr\}
= \frac{1}{2}\,\mathrm{Re}\Biggl\{
\bigl(\mathbf{E}_0\,e^{i\mathbf{k}\cdot\mathbf{r}}\bigr)
\times
\Bigl(\frac{\mathbf{k}\times \mathbf{E}_0}{\omega\,\mu}\,e^{i\mathbf{k}\cdot\mathbf{r}}\Bigr)^*
\Biggr\}
= \frac{|\mathbf{E}_0|^2}{2\,k\,\eta_0}\,e^{-2\mathrm{Im}(\mathbf{k})\,\cdot\mathbf{r}}\,\mathrm{Re}\left(\frac{k_z}{\mu_r}\right).
\end{equation}
The reflectance and transmittance may now be defined from the normal component of the Poynting vectors
at $z = 0$ as
\begin{subequations}
\begin{align}
R &= \frac{\langle S_{z,r}(z=0)\rangle}{\langle S_{z,i}(z=0)\rangle}
= \frac{|E_r|^2}{|E_i|^2}\,\frac{\mathrm{Re}\Bigl(\frac{k_{z,1}}{\mu_1}\Bigr)}
 {\mathrm{Re}\Big(\frac{k_{z,1}}{\mu_1}\Big)} = \left|r^\text{TE/TM}\right|^2, \\
T &= \frac{\langle S_{z,t}(z=0)\rangle}{\langle S_{z,i}(z=0)\rangle}
= \frac{|E_t|^2}{|E_i|^2}\,\frac{\mathrm{Re}\Big(\frac{k_{z,2}}{\mu_2}\Big)}
{\mathrm{Re}\Bigl(\frac{k_{z,1}}{\mu_1}\Bigr)}=|\tilde{t}^\text{TE/TM}|^2\text{Re}\left(\frac{{k}_{z,2}}{\mu_2}\right)\text{Re}\left(\frac{\mu_1}{{k}_{z,1}}\right).
\end{align}
\end{subequations}
As can be seen, the reflectance remains unchanged but the transmission coefficient, in the absence of magnetic response and loss, should be corrected as
\begin{equation}
    \left| t^\text{TE/TM} \right| = \left| {{{\tilde t}^\text{TE/TM}}} \right|\sqrt {\frac{{k_{z,2}}}{{k_{z,1}}}}. 
    \label{corrected_t}
\end{equation}
As shown in~Eq.(23) of the main text, the electric field associated with each multipole order differs when the structure is embedded in a medium other than air. To clearly indicate this difference, we use a tilde-primed notation for the corresponding multipolar quantities.

The corresponding electric fields normalizations, in the case where different substrate or superstrate are present, are summarized in Table~\ref{table_te_cores} for TE multipoles and in Table~\ref{table_tm_cores_modified} for TM multipoles.
\begin{table}[H]
\centering
\renewcommand{\arraystretch}{2.5} % Increase vertical spacing between rows
\footnotesize
\setlength{\tabcolsep}{10pt} % Adjust column spacing
\begin{tabular}{|c|c|}
\hline
\textbf{Electric field normalization} & \textbf{Tilde Form} \\
\hline
%--- Rows 1 to 6: Multiply by M_1 ---
\(
-\left(\frac{\frac{\eta_1 k_1}{k_{z,1}}}{\frac{\eta_2 k_2}{k_{z,2}}+\frac{\eta_1 k_1}{k_{z,1}}}\right)
\Bigl[-i\omega \mu M_x\Bigr]
\)
 & 
\(\widetilde{{M}}'_x\) \\
\hline
\(
-\left(\frac{\frac{\eta_1 k_1}{k_{z,1}}}{\frac{\eta_2 k_2}{k_{z,2}}+\frac{\eta_1 k_1}{k_{z,1}}}\right)
\Bigl[\frac{k_1^2}{2\varepsilon}\,Q^\text{(e)}_{yz}\Bigr]
\)
 & 
\(\widetilde{{Q}}^{'\text{(e)}}_{yz}\) \\
\hline
\(
-\left(\frac{\frac{\eta_1 k_1}{k_{z,1}}}{\frac{\eta_2 k_2}{k_{z,2}}+\frac{\eta_1 k_1}{k_{z,1}}}\right)
\Bigl[-\frac{\omega \mu}{2}k_x\,Q^\text{(m)}_{xx}\Bigr]
\)
 & 
\(\widetilde{{Q}}^{'\text{(m)}}_{xx}\) \\
\hline
\(
-\left(\frac{\frac{\eta_1 k_1}{k_{z,1}}}{\frac{\eta_2 k_2}{k_{z,2}}+\frac{\eta_1 k_1}{k_{z,1}}}\right)
\Bigl[-\frac{\omega \mu}{2}k_x\,Q^\text{(m)}_{zz}\Bigr]
\)
 & 
\(\widetilde{{Q}}^{'\text{(m)}}_{zz}\) \\
\hline
\(
-\left(\frac{\frac{\eta_1 k_1}{k_{z,1}}}{\frac{\eta_2 k_2}{k_{z,2}}+\frac{\eta_1 k_1}{k_{z,1}}}\right)
\Bigl[\frac{i\,k_1^2\,k_x}{3\varepsilon}\,O^\text{(e)}_{yzx}\Bigr]
\)

 & 
\(\widetilde{{O}}^{'\text{(e)}}_{yxz}\) \\
\hline
%--- Rows 7 to 13: Multiply by M_2 ---
\(
-\left(\frac{\frac{\eta_2\,\eta_1 k1,k_2}{k_{z,2}\,k_{z,1}}}{\frac{\eta_2 k_2}{k_{z,2}}+\frac{\eta_1 k_1}{k_{z,1}}}\right)
\Bigl[-i\omega P_y\Bigr]
\)
 & 
\(\widetilde{{P}}'_y\) \\
\hline
\(
-\left(\frac{\frac{\eta_2\,\eta_1 k1,k_2}{k_{z,2}\,k_{z,1}}}{\frac{\eta_2 k_2}{k_{z,2}}+\frac{\eta_1 k_1}{k_{z,1}}}\right)
\Bigl[\frac{k_x^2-k_{z,1}^2}{2}\,Q^\text{(m)}_{xz}\Bigr]
\)
 & 
\(\widetilde{{Q}}^{'\text{(m)}}_{zx}\) \\
\hline
\(
-\left(\frac{\frac{\eta_2\,\eta_1 k1,k_2}{k_{z,2}\,k_{z,1}}}{\frac{\eta_2 k_2}{k_{z,2}}+\frac{\eta_1 k_1}{k_{z,1}}}\right)
\Bigl[-\frac{\omega k_x}{2}\,Q^\text{(e)}_{yx}\Bigr]
\)
 & 
\(\widetilde{{Q}}^{'\text{(e)}}_{yx}\) \\
\hline
\(
-\left(\frac{\frac{\eta_2\,\eta_1 k1,k_2}{k_{z,2}\,k_{z,1}}}{\frac{\eta_2 k_2}{k_{z,2}}+\frac{\eta_1 k_1}{k_{z,1}}}\right)
\Bigl[-i k_x\,M_z\Bigr]
\)
 & 
\(\widetilde{{M}}'_z\) \\
\hline
\(
-\left(\frac{\frac{\eta_2\,\eta_1 k1,k_2}{k_{z,2}\,k_{z,1}}}{\frac{\eta_2 k_2}{k_{z,2}}+\frac{\eta_1 k_1}{k_{z,1}}}\right)
\Bigl[-\frac{i\omega k_x^2}{6}\,O^\text{(e)}_{yxx}\Bigr]
\)
 & 
\(\widetilde{{O}}^{'\text{(e)}}_{yxx}\) \\
\hline
\(
-\left(\frac{\frac{\eta_2\,\eta_1 k1,k_2}{k_{z,2}\,k_{z,1}}}{\frac{\eta_2 k_2}{k_{z,2}}+\frac{\eta_1 k_1}{k_{z,1}}}\right)
\Bigl[\frac{i\omega k_{z,1}^2}{6}\,O^\text{(e)}_{yzz}\Bigr]
\)
 & 
\(\widetilde{{O}}^{'\text{(e)}}_{yzz}\) \\
\hline
\end{tabular}
\caption{Multipoles and their corresponding mappings for TE polarization.}
\label{table_te_cores}
\end{table}

\begin{table}[H]
\centering
\renewcommand{\arraystretch}{2.5} % Increase vertical spacing between rows
\footnotesize
\setlength{\tabcolsep}{10pt} % Adjust column spacing
\begin{tabular}{|c|c|}
\hline
\textbf{Magnetic field normalization} & \textbf{Tilde Form } \\
\hline
$-\left(\frac{1}{\frac{\eta_2 k_2}{k_{z,2}}+\frac{\eta_1 k_1}{k_{z,1}}}\right)\Bigl[i\omega \mu M_y\Bigr]$ & $\tilde{M}_y'$ \\
\hline
$-\left(\frac{1}{\frac{\eta_2 k_2}{k_{z,2}}+\frac{\eta_1 k_1}{k_{z,1}}}\right)\Bigl[-\frac{k_1^2}{2\varepsilon}Q_{xz}\Bigr]$ & $\tilde{Q}_{xz}'$ \\
\hline
$-\left(\frac{1}{\frac{\eta_2 k_2}{k_{z,2}}+\frac{\eta_1 k_1}{k_{z,1}}}\right)\Bigl[\frac{i}{\varepsilon}k_x P_z\Bigr]$ & $\tilde{P}_z'$ \\
\hline
$-\left(\frac{1}{\frac{\eta_2 k_2}{k_{z,2}}+\frac{\eta_1 k_1}{k_{z,1}}}\right)\Bigl[-\frac{i}{\varepsilon}k_x^2\,i\,Q_{zx}\Bigr]$ & $\tilde{Q}_{zx}'$ \\
\hline
$-\left(\frac{1}{\frac{\eta_2 k_2}{k_{z,2}}+\frac{\eta_1 k_1}{k_{z,1}}}\right)\Bigl[\frac{\omega \mu\,k_x}{2}S_{xy}\Bigr]$ & $\tilde{S}_{xy}'$ \\
\hline
$-\left(\frac{1}{\frac{\eta_2 k_2}{k_{z,2}}+\frac{\eta_1 k_1}{k_{z,1}}}\right)\Bigl[\frac{i\,k_x\bigl(k_1^2-k_x^2\bigr)}{3\varepsilon}\,O_{xxz}\Bigr]$ & $\tilde{O}_{xxz}'$ \\
\hline
$-\left(\frac{1}{\frac{\eta_2 k_2}{k_{z,2}}+\frac{\eta_1 k_1}{k_{z,1}}}\right)\Bigl[-\frac{k_x^2\,k_z}{6\varepsilon}\,O_{zxx}\Bigr]$ & $\tilde{O}_{zxx}'$ \\
\hline
$-\left(\frac{1}{\frac{\eta_2 k_2}{k_{z,2}}+\frac{\eta_1 k_1}{k_{z,1}}}\right)\Bigl[-\frac{i\,k_{z,1}\,k_x}{6\varepsilon}\,O_{zzz}\Bigr]$ & $\tilde{O}_{zzz}'$ \\
\hline
$\frac{\frac{\eta_1 k_1}{k_{z,1}}}{\frac{\eta_2 k_2}{k_{z,2}}+\frac{\eta_1 k_1}{k_{z,1}}}\Bigl[i\omega P_x\Bigr]$ & $\tilde{P}_x'$ \\
\hline
$\frac{\frac{\eta_1 k_1}{k_{z,1}}}{\frac{\eta_2 k_2}{k_{z,2}}+\frac{\eta_1 k_1}{k_{z,1}}}\Bigl[-\frac{k^2}{2}S_{yz}\Bigr]$ & $\tilde{S}_{yz}'$ \\
\hline
$\frac{\frac{\eta_1 k_1}{k_{z,1}}}{\frac{\eta_2 k_2}{k_{z,2}}+\frac{\eta_1 k_1}{k_{z,1}}}\cdot\Bigl[\frac{k^2}{2}k_x^2S_{yz}\Bigr]$ & $\tilde{S}_{yz}'$ \\
\hline
$\frac{\frac{\eta_1 k_1}{k_{z,1}}}{\frac{\eta_2 k_2}{k_{z,2}}+\frac{\eta_1 k_1}{k_{z,1}}}\Bigl[\frac{\omega}{2}k_xQ_{xx}\Bigr]$ & $\tilde{Q}_{xx}'$ \\
\hline
$\frac{\frac{\eta_1 k_1}{k_{z,1}}}{\frac{\eta_2 k_2}{k_{z,2}}+\frac{\eta_1 k_1}{k_{z,1}}}\Bigl[-\frac{\omega}{2}k_xQ_{zz}\Bigr]$ & $\tilde{Q}_{zz}'$ \\
\hline
$\frac{\frac{\eta_1 k_1}{k_{z,1}}}{\frac{\eta_2 k_2}{k_{z,2}}+\frac{\eta_1 k_1}{k_{z,1}}}\Bigl[\frac{i\omega k_x^2}{3}O_{yzx}\Bigr[$ & $\tilde{O}_{yzx}'$ \\
\hline
$\frac{\frac{\eta_1 k_1}{k_{z,1}}}{\frac{\eta_2 k_2}{k_{z,2}}+\frac{\eta_1 k_1}{k_{z,1}}}\Bigl[-\frac{i\omega k_x^2}{6}O_{xxx}\Bigr]$ & $\tilde{O}_{xxx}'$ \\
\hline
$\frac{\frac{\eta_1 k_1}{k_{z,1}}}{\frac{\eta_2 k_2}{k_{z,2}}+\frac{\eta_1 k_1}{k_{z,1}}}\Bigl[\frac{i\omega k_{z,1}}{6}O_{xzz}\Bigr]$ & $\tilde{O}_{xzz}'$ \\
\hline
\end{tabular}
\caption{Multipoles and their corresponding mappings for TM polarization.}
\label{table_tm_cores_modified}
\end{table}

\newpage
\renewcommand{\theequation}{F\arabic{equation}}
\section{\label{additinal}Appendix F: Additional results for oblique TE and TM polarized incident wave}
\subsubsection{Gold nano-cylinders of different heights under TE polarization}
\begin{figure}[H]
    \centering
    \includegraphics[width=1\textwidth]{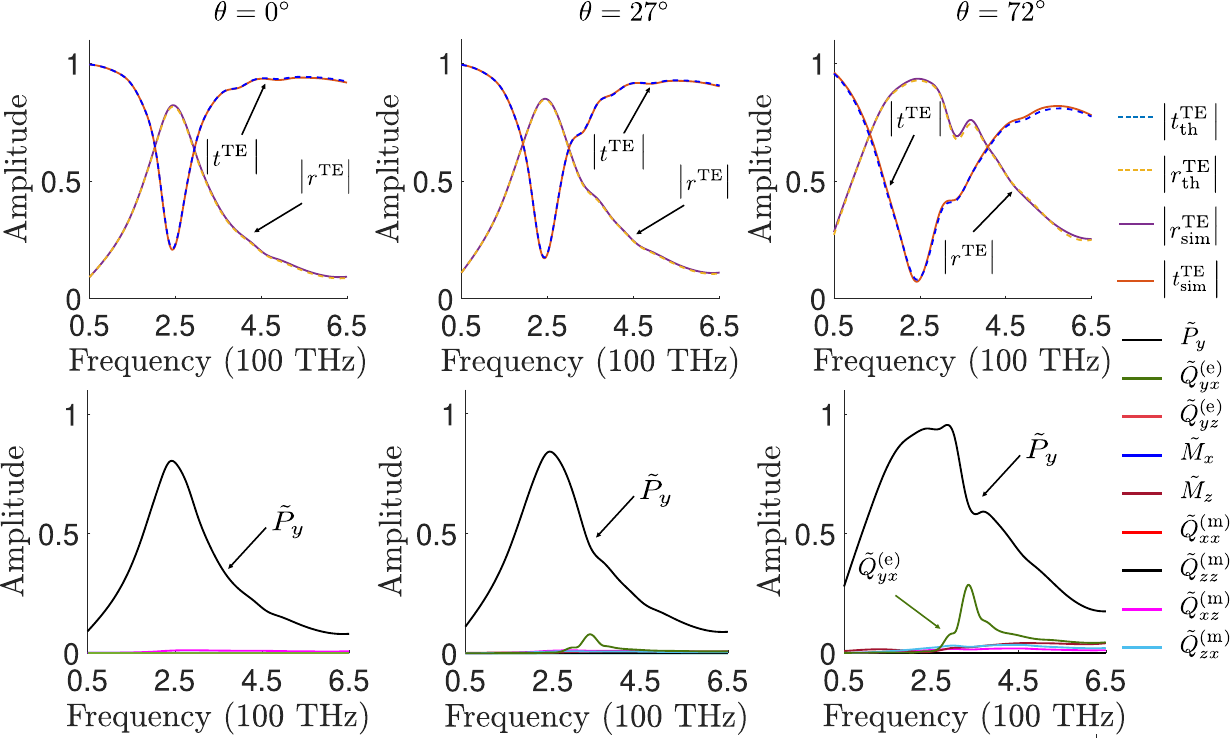} % Replace 'example-image' with your file name
    \caption{Simulation and theoretical results for gold nano-cylinders for $P=225$~nm and $H=5$~nm and $r=100$~nm under TE illumination for $\theta=0^\circ,27^\circ,72^\circ$.}
    \label{gold_te_5} % Label for referencing the figure
\end{figure}
\FloatBarrier
\begin{figure}[H]
    \centering
    \includegraphics[width=1\textwidth]{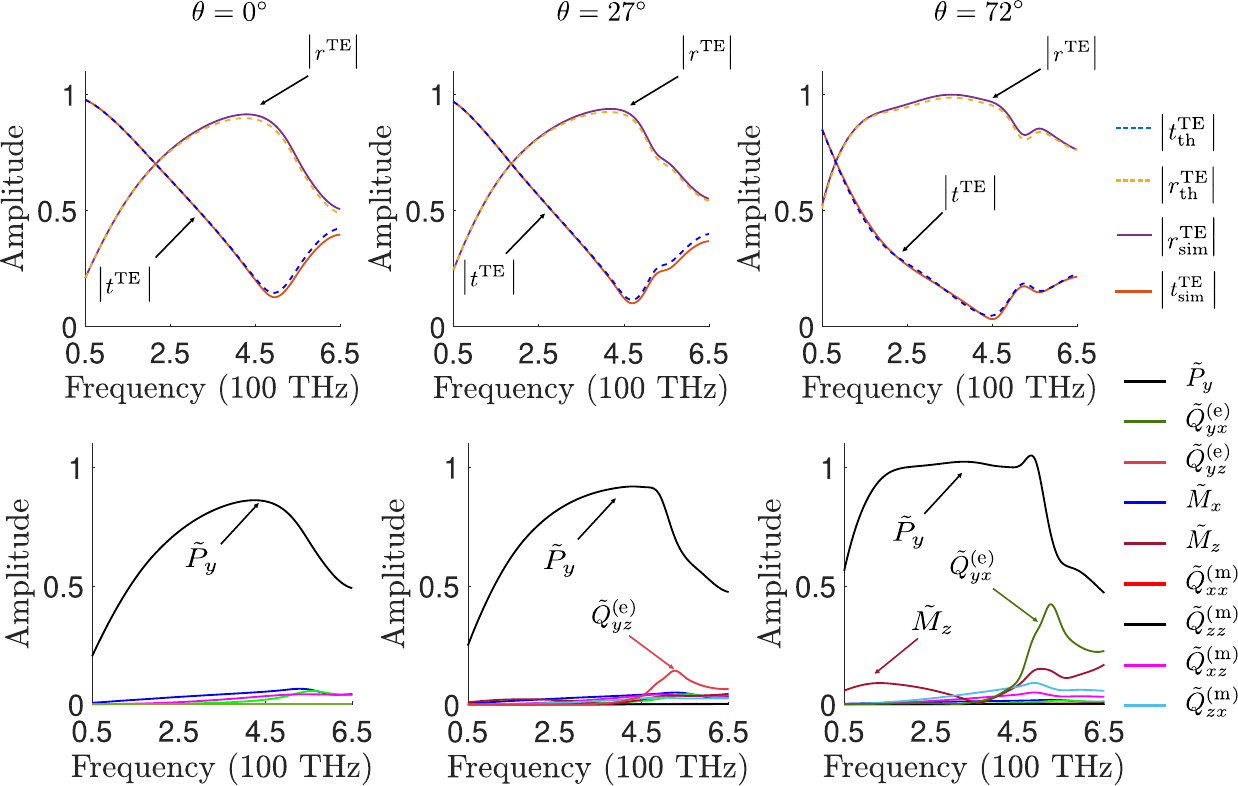} % Replace 'example-image' with your file name
    \caption{Simulation and theoretical results for gold nano-cylinders for $P=225$~nm and $H=65$~nm and $r=100$~nm under TE illumination for $\theta=0^\circ,27^\circ,72^\circ$.}
    \label{gold_te_65} % Label for referencing the figure
\end{figure}

\subsubsection{Amorphous silicon  nano-cylinders of different heights under TE polarization}
\FloatBarrier
\begin{figure}[ht]
    \centering
    \includegraphics[width=1\textwidth]{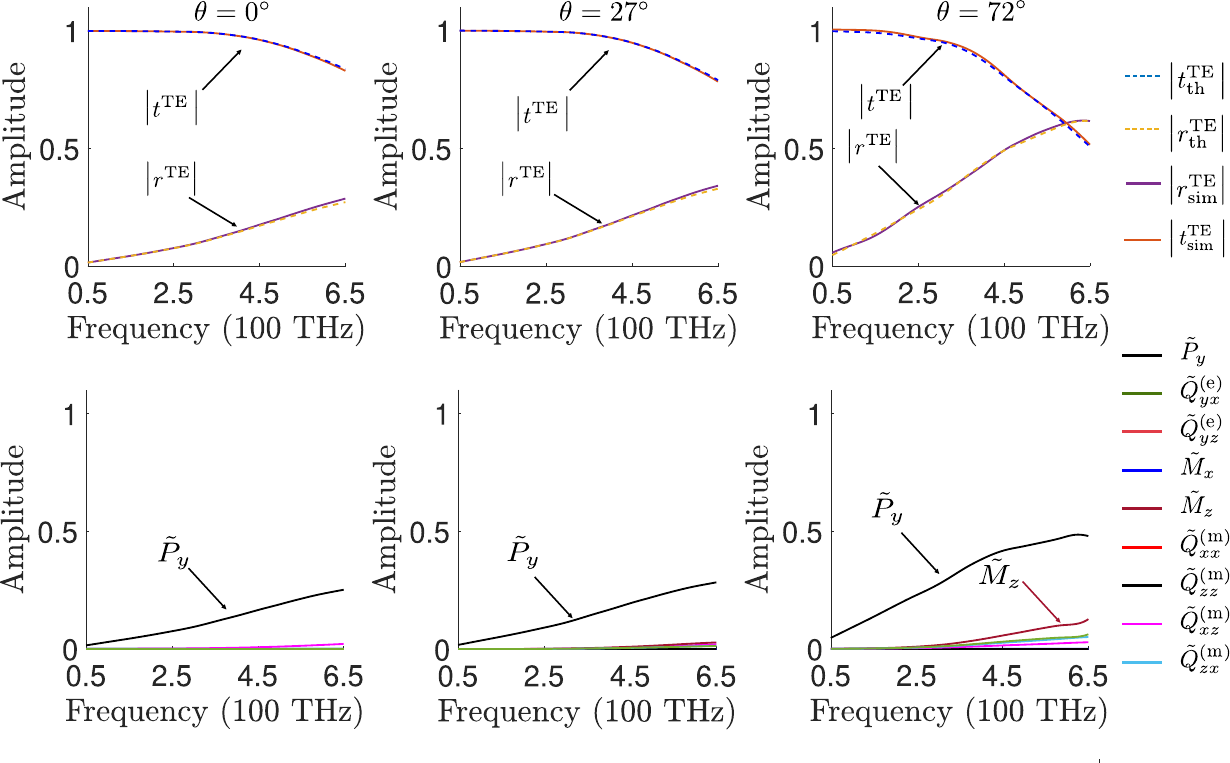} % Replace 'example-image' with your file name
    \caption{Simulation and theoretical results for amorphous silicon nano-cylinders for $P=225$~nm and $H=5$~nm and $r=100$~nm under TE illumination for $\theta=0^\circ,27^\circ,72^\circ$.}
    \label{siam_te_5} % Label for referencing the figure
\end{figure}
\FloatBarrier
\begin{figure}[H]
    \centering
    \includegraphics[width=1\textwidth]{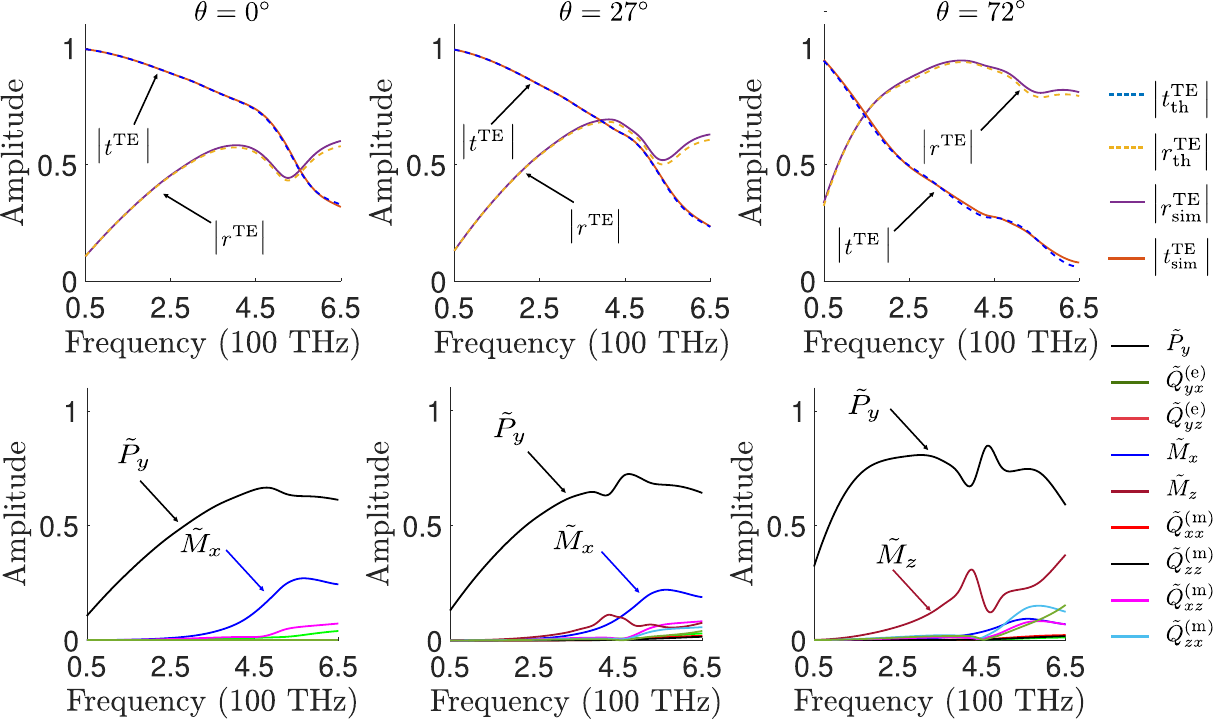} % Replace 'example-image' with your file name
    \caption{Simulation and theoretical results for amorphous silicon nano-cylinders for $P=225$~nm and $H=65$~nm and $r=100$~nm under TE illumination for $\theta=0^\circ,27^\circ,72^\circ$.}
    \label{siam_te_65} % Label for referencing the figure
\end{figure}

\subsubsection{Gold and amorphous silicon nano-cylinders under TM polarization}

\begin{figure}[H]
\centering
\includegraphics[width=0.5\textwidth]{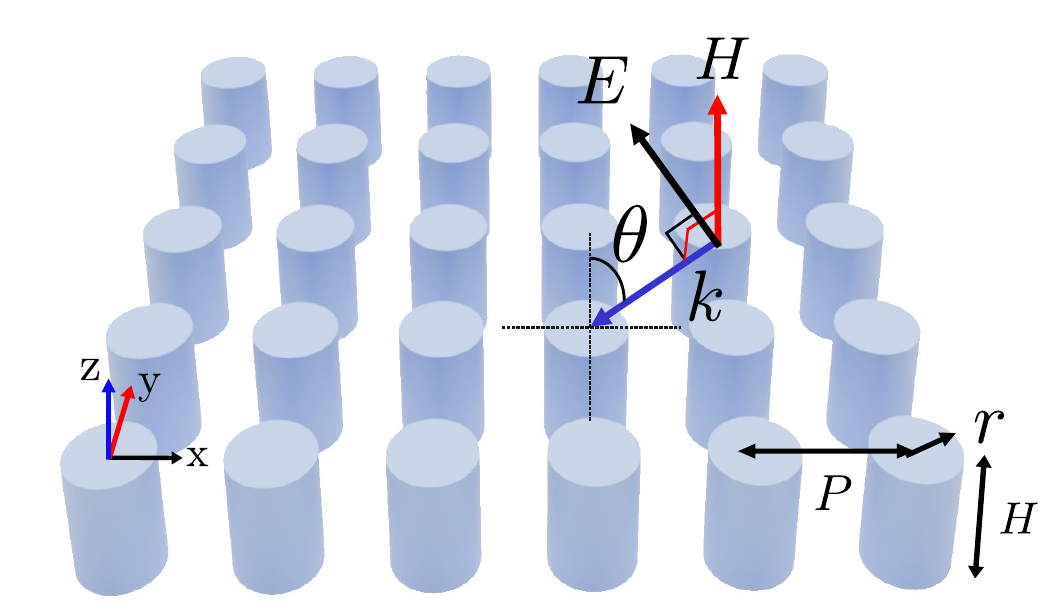} % Replace with your image file
\caption{Metasurface made out of nano-cylinders under TM illumination.}
\label{fig:te_top}
\end{figure}

\begin{figure}[H]
    \centering
    \includegraphics[width=1\textwidth]{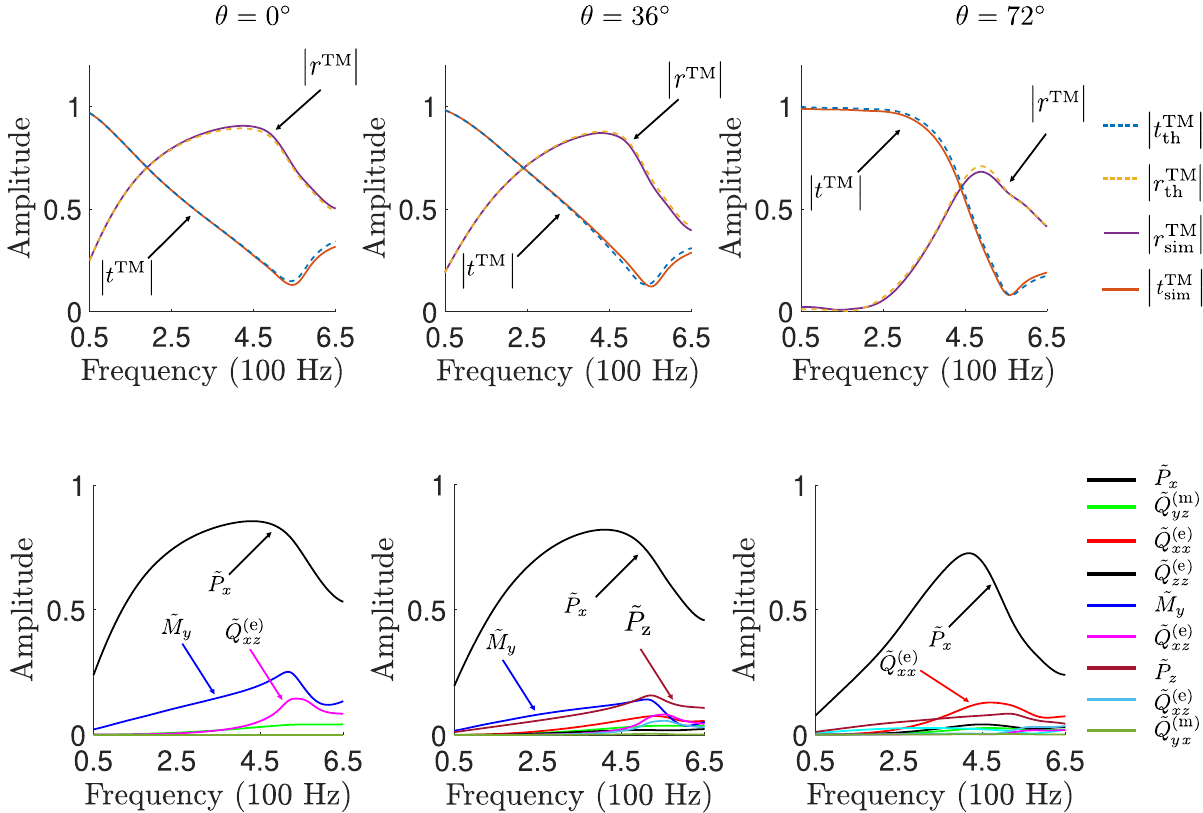} % Replace 'example-image' with your file name
    \caption{Simulation and theoretical results for gold nano-cylinders for $P=225$ nm and $H=85$ nm and $r=100$ nm under TM illumination for $\theta=0^{~\circ}, 27^{~\circ}, 63^{~\circ} $.}
    \label{tM_85} % Label for referencing the figure
\end{figure}

\begin{figure}[H]
    \centering
    \includegraphics[width=1\textwidth]{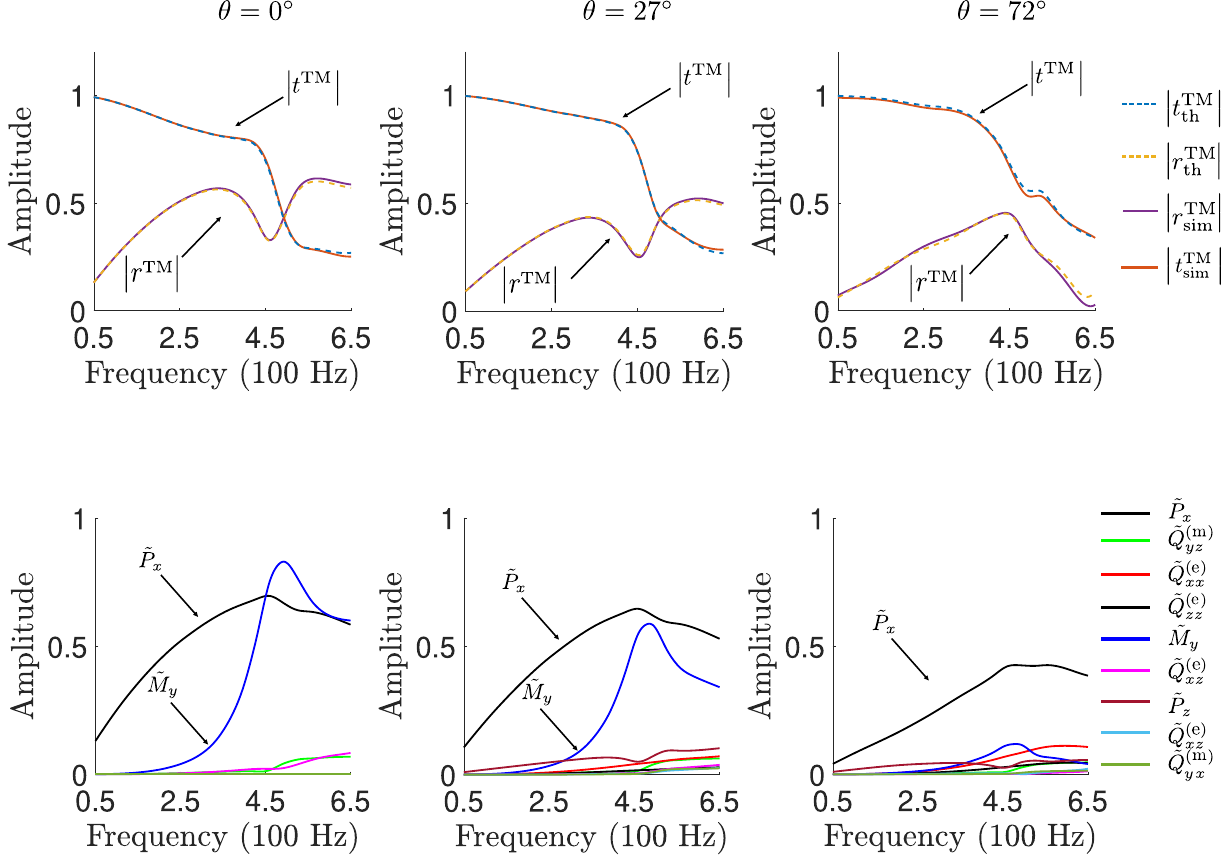} % Replace 'example-image' with your file name
    \caption{Simulation and theoretical results for amorphous silicon  nano-cylinders for $P=225$ nm and $H=85$~nm and $r=100$~nm under TM illumination for $\theta=0^{~\circ}, 27^{~\circ}, 63^{~\circ}$.}
    \label{tM_SIAM_85} % Label for referencing the figure
\end{figure}

\newpage
\renewcommand{\theequation}{G\arabic{equation}}
\section{\label{Bic}Appendix G: Bound state in continuum in free space}
\begin{figure}[h!]
    \centering
    \includegraphics[width=0.85\textwidth]{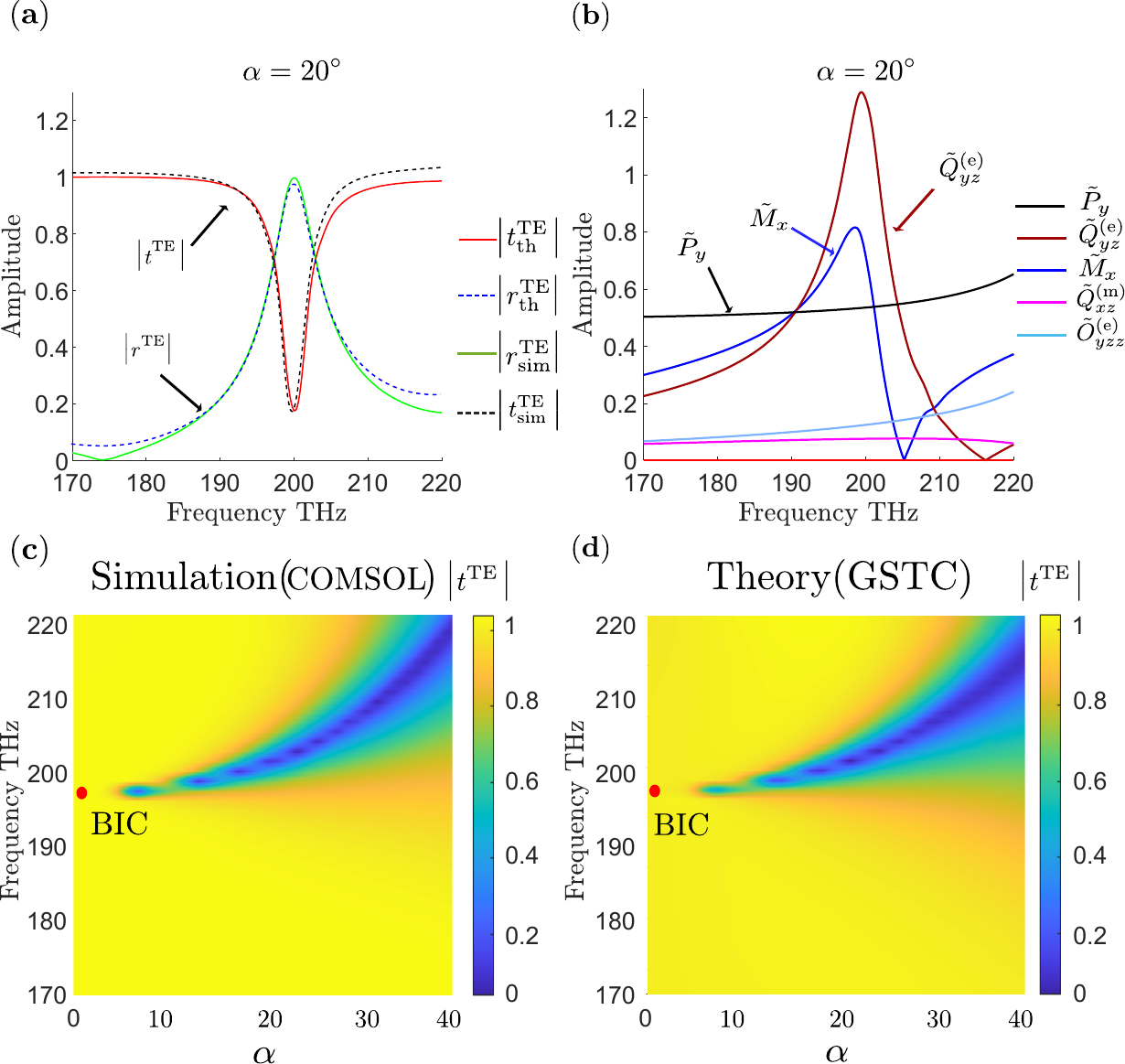} 
    \caption{Scattering analysis for the metasurface in ~Fig~.8a of main text. (a)~Reflection and transmission coefficients for $\alpha=20^\circ$ obtained via COMSOL and our GSTC method. (b)~Corresponding multipolar decomposition. (c)-(d)~Transmission amplitudes for different geometrical perturbations ($\alpha$) obtained via COMSOL and our GSTC method, respectively.}
    \label{simp_bic}
\end{figure}
 By comparing the results in Fig.~\ref{simp_bic}b and ~Fig~.9b of main text, it is clear that embedding the unit cell inside PMMA leads to the emergence of a second BIC. This phenomenon is referred to in~\cite{liu2018extreme} as E-QBIC and M-QBIC. The reasoning behind this naming is the common assumption that breaking inversion symmetry induces bianisotropy, which in turn leads to the generation of magnetic dipoles. While this explanation is valid, it does not fully capture the underlying physics.

A more general interpretation is given in~\cite{poleva2023multipolar}, which attributes this effect to the behavior of even and odd modes. These modes are distinguished based on whether they remain in phase or acquire a \( \pi \)-phase shift under inversion symmetry. The study in~\cite{poleva2023multipolar} shows that when inversion symmetry is preserved, only one set of parity modes exists. However, breaking this symmetry enables the coexistence of both even and odd parity modes, resulting in a richer multipolar response.

Further comparison between Fig.~\ref{simp_bic}b and ~Fig~.9b of main text shows that in Fig.~\ref{simp_bic}b, where inversion symmetry is maintained, the dominant contributions arise from the magnetic dipole (MD) and electric quadrupole (EQ) modes. In contrast, in ~Fig~.9b of main text, where inversion symmetry is broken by embedding the unit cell in PMMA, a combination of electric dipole (ED), magnetic dipole (MD), electric quadrupole (EQ), electric octupole (EO), and magnetic quadrupole (MQ) modes emerges, as expected. Additional multipoles are also present due to coupling effects in the metasurface and the phase asymmetry introduced between the top and bottom regions of the unit cell.

\begin{figure}[H]
    \centering
    \includegraphics[width=1\textwidth]{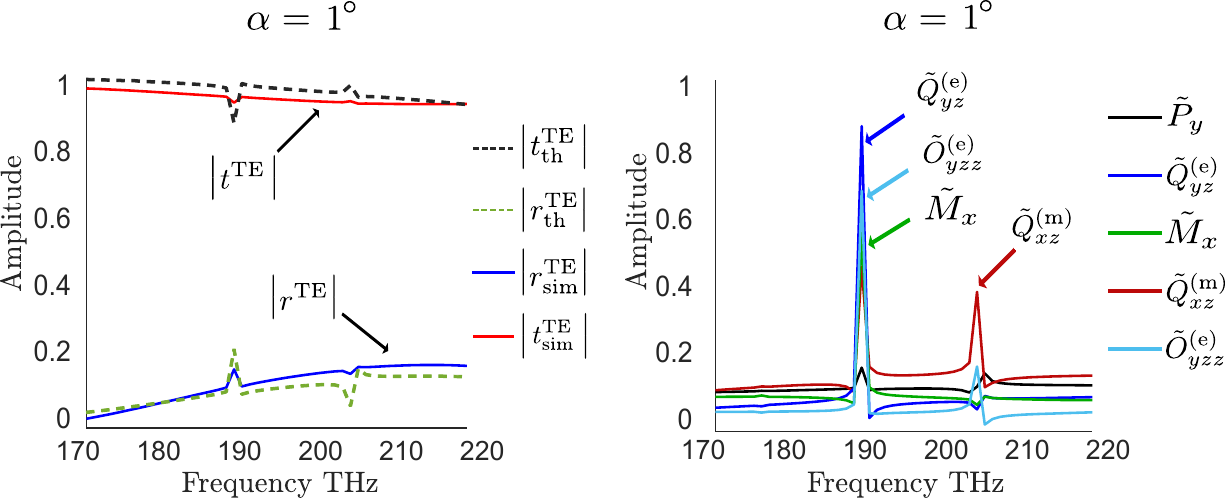} % Replace 'example-image' with your file name
    \caption{Bound state in continuum for the case when $\alpha=1^\circ$ (close to pure BIC) in ~Fig~.9 of main text.}
    \label{single_bic} % Label for referencing the figure
\end{figure}

In Fig.~\ref{single_bic}, the emergence of sharp multipolar features is clearly observed at \( \alpha = 1^\circ \) for the PMMA-embedded case. This highlights that conventional reflection and transmission spectra may not fully capture such resonance phenomena.

\begin{acknowledgement}
We acknowledge funding from the Swiss National Science Foundation (project TMSGI2\_218392).
\end{acknowledgement}

\bibliography{name}

\providecommand{\latin}[1]{#1}
\makeatletter
\providecommand{\doi}
  {\begingroup\let\do\@makeother\dospecials
  \catcode`\{=1 \catcode`\}=2 \doi@aux}
\providecommand{\doi@aux}[1]{\endgroup\texttt{#1}}
\makeatother
\providecommand*\mcitethebibliography{\thebibliography}
\csname @ifundefined\endcsname{endmcitethebibliography}  {\let\endmcitethebibliography\endthebibliography}{}
\begin{mcitethebibliography}{51}
\providecommand*\natexlab[1]{#1}
\providecommand*\mciteSetBstSublistMode[1]{}
\providecommand*\mciteSetBstMaxWidthForm[2]{}
\providecommand*\mciteBstWouldAddEndPuncttrue
  {\def\EndOfBibitem{\unskip.}}
\providecommand*\mciteBstWouldAddEndPunctfalse
  {\let\EndOfBibitem\relax}
\providecommand*\mciteSetBstMidEndSepPunct[3]{}
\providecommand*\mciteSetBstSublistLabelBeginEnd[3]{}
\providecommand*\EndOfBibitem{}
\mciteSetBstSublistMode{f}
\mciteSetBstMaxWidthForm{subitem}{(\alph{mcitesubitemcount})}
\mciteSetBstSublistLabelBeginEnd
  {\mcitemaxwidthsubitemform\space}
  {\relax}
  {\relax}

\bibitem[Li \latin{et~al.}(2023)Li, Zhang, Cheng, \latin{et~al.} others]{li2023universal}
Li,~R.; Zhang,~Y.; Cheng,~H.; others A universal metasurface antenna for electromagnetic wave control. \emph{Nature Communications} \textbf{2023}, \emph{14}, 5466\relax
\mciteBstWouldAddEndPuncttrue
\mciteSetBstMidEndSepPunct{\mcitedefaultmidpunct}
{\mcitedefaultendpunct}{\mcitedefaultseppunct}\relax
\EndOfBibitem
\bibitem[Lin \latin{et~al.}(2014)Lin, Fan, Hasman, and Brongersma]{lin2014dielectric}
Lin,~D.; Fan,~P.; Hasman,~E.; Brongersma,~M.~L. Dielectric gradient metasurface optical elements. \emph{science} \textbf{2014}, \emph{345}, 298--302\relax
\mciteBstWouldAddEndPuncttrue
\mciteSetBstMidEndSepPunct{\mcitedefaultmidpunct}
{\mcitedefaultendpunct}{\mcitedefaultseppunct}\relax
\EndOfBibitem
\bibitem[Achouri and Caloz(2021)Achouri, and Caloz]{achouri2021}
Achouri,~K.; Caloz,~C. \emph{Electromagnetic Metasurfaces: Theory and Applications}; Wiley-IEEE Press: Hoboken, NJ, 2021\relax
\mciteBstWouldAddEndPuncttrue
\mciteSetBstMidEndSepPunct{\mcitedefaultmidpunct}
{\mcitedefaultendpunct}{\mcitedefaultseppunct}\relax
\EndOfBibitem
\bibitem[Kim \latin{et~al.}(2024)Kim, Lu, \latin{et~al.} others]{kim2024multi}
Kim,~M.; Lu,~Z.; others A multi-resonant metasurface platform for broadband wavefront engineering. \emph{Nature Communications} \textbf{2024}, \emph{15}, 957\relax
\mciteBstWouldAddEndPuncttrue
\mciteSetBstMidEndSepPunct{\mcitedefaultmidpunct}
{\mcitedefaultendpunct}{\mcitedefaultseppunct}\relax
\EndOfBibitem
\bibitem[Simovskij(2018)]{simovski2018}
Simovskij,~K.~R. \emph{Composite Media with Weak Spatial Dispersion}; Pan Stanford Publishing: Singapur, 2018\relax
\mciteBstWouldAddEndPuncttrue
\mciteSetBstMidEndSepPunct{\mcitedefaultmidpunct}
{\mcitedefaultendpunct}{\mcitedefaultseppunct}\relax
\EndOfBibitem
\bibitem[Pfeiffer and Grbic(2013)Pfeiffer, and Grbic]{pfeiffer2013metamaterial}
Pfeiffer,~C.; Grbic,~A. Metamaterial Huygens’ surfaces: tailoring wave fronts with reflectionless sheets. \emph{Physical Review Letters} \textbf{2013}, \emph{110}, 197401\relax
\mciteBstWouldAddEndPuncttrue
\mciteSetBstMidEndSepPunct{\mcitedefaultmidpunct}
{\mcitedefaultendpunct}{\mcitedefaultseppunct}\relax
\EndOfBibitem
\bibitem[Silveirinha(2014)]{silveirinha2014nonlocal}
Silveirinha,~M.~G. Nonlocal homogenization model for a periodic array of epsilon-negative rods. \emph{Physical Review B} \textbf{2014}, \emph{89}, 075127\relax
\mciteBstWouldAddEndPuncttrue
\mciteSetBstMidEndSepPunct{\mcitedefaultmidpunct}
{\mcitedefaultendpunct}{\mcitedefaultseppunct}\relax
\EndOfBibitem
\bibitem[Zhou \latin{et~al.}(2020)Zhou, \latin{et~al.} others]{zhou2020nonlocal}
Zhou,~H.; others Angular-momentum selective metasurfaces via momentum-space engineering. \emph{Nature Communications} \textbf{2020}, \emph{11}, 3029\relax
\mciteBstWouldAddEndPuncttrue
\mciteSetBstMidEndSepPunct{\mcitedefaultmidpunct}
{\mcitedefaultendpunct}{\mcitedefaultseppunct}\relax
\EndOfBibitem
\bibitem[Achouri and Martin(2020)Achouri, and Martin]{achouri2020a}
Achouri,~K.; Martin,~O. J.~F. Angular {{Scattering Properties}} of {{Metasurfaces}}. \emph{IEEE Trans. Antennas Propagat.} \textbf{2020}, \emph{68}, 432--442\relax
\mciteBstWouldAddEndPuncttrue
\mciteSetBstMidEndSepPunct{\mcitedefaultmidpunct}
{\mcitedefaultendpunct}{\mcitedefaultseppunct}\relax
\EndOfBibitem
\bibitem[Achouri \latin{et~al.}(2015)Achouri, Salem, and Caloz]{achouri2015general}
Achouri,~K.; Salem,~M.~A.; Caloz,~C. General metasurface synthesis based on susceptibility tensors. \emph{IEEE Transactions on Antennas and Propagation} \textbf{2015}, \emph{63}, 2977--2991\relax
\mciteBstWouldAddEndPuncttrue
\mciteSetBstMidEndSepPunct{\mcitedefaultmidpunct}
{\mcitedefaultendpunct}{\mcitedefaultseppunct}\relax
\EndOfBibitem
\bibitem[Zhen \latin{et~al.}(2014)Zhen, Hsu, Lu, Stone, and Soljačić]{zhen2014topological}
Zhen,~B.; Hsu,~C.~W.; Lu,~L.; Stone,~A.~D.; Soljačić,~M. Topological nature of optical bound states in the continuum. \emph{Physical Review Letters} \textbf{2014}, \emph{113}, 257401\relax
\mciteBstWouldAddEndPuncttrue
\mciteSetBstMidEndSepPunct{\mcitedefaultmidpunct}
{\mcitedefaultendpunct}{\mcitedefaultseppunct}\relax
\EndOfBibitem
\bibitem[Doeleman(2018)]{doeleman2018experimental}
Doeleman,~H. M. e.~a. Experimental observation of a polarization vortex at an optical bound state in the continuum. \emph{Nature Photonics} \textbf{2018}, \emph{12}, 397--401\relax
\mciteBstWouldAddEndPuncttrue
\mciteSetBstMidEndSepPunct{\mcitedefaultmidpunct}
{\mcitedefaultendpunct}{\mcitedefaultseppunct}\relax
\EndOfBibitem
\bibitem[Shamkhi \latin{et~al.}(2019)Shamkhi, Baryshnikova, Sayanskiy, Kapitanova, Terekhov, Belov, Karabchevsky, Evlyukhin, Kivshar, and Shalin]{PhysRevLett.122.193905}
Shamkhi,~H.~K.; Baryshnikova,~K.~V.; Sayanskiy,~A.; Kapitanova,~P.; Terekhov,~P.~D.; Belov,~P.; Karabchevsky,~A.; Evlyukhin,~A.~B.; Kivshar,~Y.; Shalin,~A.~S. Transverse Scattering and Generalized Kerker Effects in All-Dielectric Mie-Resonant Metaoptics. \emph{Phys. Rev. Lett.} \textbf{2019}, \emph{122}, 193905\relax
\mciteBstWouldAddEndPuncttrue
\mciteSetBstMidEndSepPunct{\mcitedefaultmidpunct}
{\mcitedefaultendpunct}{\mcitedefaultseppunct}\relax
\EndOfBibitem
\bibitem[Y{\"u}cel \latin{et~al.}(2024)Y{\"u}cel, Cuesta, and Achouri]{yucel2024angle}
Y{\"u}cel,~M.; Cuesta,~F.~S.; Achouri,~K. Angle-Invariant Scattering in Metasurfaces. \emph{arXiv preprint arXiv:2412.13105} \textbf{2024}, \relax
\mciteBstWouldAddEndPunctfalse
\mciteSetBstMidEndSepPunct{\mcitedefaultmidpunct}
{}{\mcitedefaultseppunct}\relax
\EndOfBibitem
\bibitem[Gong \latin{et~al.}(2023)Gong, Liu, Ge, Xiang, and Han]{gong2023multipolar}
Gong,~M.; Liu,~J.; Ge,~L.; Xiang,~H.; Han,~D. Multipolar perspective on unidirectional guided resonances. \emph{Physical Review A} \textbf{2023}, \emph{108}, 013522\relax
\mciteBstWouldAddEndPuncttrue
\mciteSetBstMidEndSepPunct{\mcitedefaultmidpunct}
{\mcitedefaultendpunct}{\mcitedefaultseppunct}\relax
\EndOfBibitem
\bibitem[Liu and Miroshnichenko(2017)Liu, and Miroshnichenko]{liu2017beam}
Liu,~W.; Miroshnichenko,~A.~E. Beam steering with dielectric metalattices. \emph{Acs Photonics} \textbf{2017}, \emph{5}, 1733--1741\relax
\mciteBstWouldAddEndPuncttrue
\mciteSetBstMidEndSepPunct{\mcitedefaultmidpunct}
{\mcitedefaultendpunct}{\mcitedefaultseppunct}\relax
\EndOfBibitem
\bibitem[Momeni \latin{et~al.}(2019)Momeni, Rajabalipanah, Abdolali, and Achouri]{momeni2019generalized}
Momeni,~A.; Rajabalipanah,~H.; Abdolali,~A.; Achouri,~K. Generalized optical signal processing based on multioperator metasurfaces synthesized by susceptibility tensors. \emph{Physical Review Applied} \textbf{2019}, \emph{11}, 064042\relax
\mciteBstWouldAddEndPuncttrue
\mciteSetBstMidEndSepPunct{\mcitedefaultmidpunct}
{\mcitedefaultendpunct}{\mcitedefaultseppunct}\relax
\EndOfBibitem
\bibitem[Abdolali \latin{et~al.}(2019)Abdolali, Momeni, Rajabalipanah, and Achouri]{abdolali2019parallel}
Abdolali,~A.; Momeni,~A.; Rajabalipanah,~H.; Achouri,~K. Parallel integro-differential equation solving via multi-channel reciprocal bianisotropic metasurface augmented by normal susceptibilities. \emph{New Journal of Physics} \textbf{2019}, \emph{21}, 113048\relax
\mciteBstWouldAddEndPuncttrue
\mciteSetBstMidEndSepPunct{\mcitedefaultmidpunct}
{\mcitedefaultendpunct}{\mcitedefaultseppunct}\relax
\EndOfBibitem
\bibitem[Jang \latin{et~al.}(2021)Jang, Lee, Sung, and Lee]{jang2021}
Jang,~J.; Lee,~G.-Y.; Sung,~J.; Lee,~B. Independent Multichannel Wavefront Modulation for Angle Multiplexed Meta-Holograms. \emph{Advanced Optical Materials} \textbf{2021}, \emph{9}, 2100678\relax
\mciteBstWouldAddEndPuncttrue
\mciteSetBstMidEndSepPunct{\mcitedefaultmidpunct}
{\mcitedefaultendpunct}{\mcitedefaultseppunct}\relax
\EndOfBibitem
\bibitem[Alaee \latin{et~al.}(2018)Alaee, Rockstuhl, and Fernandez-Corbaton]{alaee2018electromagnetic}
Alaee,~R.; Rockstuhl,~C.; Fernandez-Corbaton,~I. An electromagnetic multipole expansion beyond the long-wavelength approximation. \emph{Optics Communications} \textbf{2018}, \emph{407}, 17--21\relax
\mciteBstWouldAddEndPuncttrue
\mciteSetBstMidEndSepPunct{\mcitedefaultmidpunct}
{\mcitedefaultendpunct}{\mcitedefaultseppunct}\relax
\EndOfBibitem
\bibitem[Allayarov \latin{et~al.}(2024)Allayarov, Evlyukhin, and Cal{\`a}~Lesina]{allayarov2024multiresonant}
Allayarov,~I.; Evlyukhin,~A.~B.; Cal{\`a}~Lesina,~A. Multiresonant all-dielectric metasurfaces based on high-order multipole coupling in the visible. \emph{Optics Express} \textbf{2024}, \emph{32}, 5641--5658\relax
\mciteBstWouldAddEndPuncttrue
\mciteSetBstMidEndSepPunct{\mcitedefaultmidpunct}
{\mcitedefaultendpunct}{\mcitedefaultseppunct}\relax
\EndOfBibitem
\bibitem[Rahimzadegan \latin{et~al.}(2022)Rahimzadegan, Karamanos, Alaee, Lamprianidis, Beutel, Boyd, and Rockstuhl]{rahimzadegan2022comprehensive}
Rahimzadegan,~A.; Karamanos,~T.~D.; Alaee,~R.; Lamprianidis,~A.~G.; Beutel,~D.; Boyd,~R.~W.; Rockstuhl,~C. A comprehensive multipolar theory for periodic metasurfaces. \emph{Advanced Optical Materials} \textbf{2022}, \emph{10}, 2102059\relax
\mciteBstWouldAddEndPuncttrue
\mciteSetBstMidEndSepPunct{\mcitedefaultmidpunct}
{\mcitedefaultendpunct}{\mcitedefaultseppunct}\relax
\EndOfBibitem
\bibitem[Babicheva and Evlyukhin(2017)Babicheva, and Evlyukhin]{babicheva2017resonant}
Babicheva,~V.~E.; Evlyukhin,~A.~B. Resonant lattice Kerker effect in metasurfaces with electric and magnetic optical responses. \emph{Laser \& Photonics Reviews} \textbf{2017}, \emph{11}, 1700132\relax
\mciteBstWouldAddEndPuncttrue
\mciteSetBstMidEndSepPunct{\mcitedefaultmidpunct}
{\mcitedefaultendpunct}{\mcitedefaultseppunct}\relax
\EndOfBibitem
\bibitem[Savinov \latin{et~al.}(2014)Savinov, Fedotov, and Zheludev]{savinov2014toroidal}
Savinov,~V.; Fedotov,~V.; Zheludev,~N.~I. Toroidal dipolar excitation and macroscopic electromagnetic properties of metamaterials. \emph{Physical Review B} \textbf{2014}, \emph{89}, 205112\relax
\mciteBstWouldAddEndPuncttrue
\mciteSetBstMidEndSepPunct{\mcitedefaultmidpunct}
{\mcitedefaultendpunct}{\mcitedefaultseppunct}\relax
\EndOfBibitem
\bibitem[Albooyeh \latin{et~al.}(2011)Albooyeh, Morits, and Simovski]{albooyeh2011}
Albooyeh,~M.; Morits,~D.; Simovski,~C. Electromagnetic characterization of substrated metasurfaces. \emph{Metamaterials} \textbf{2011}, \emph{5}, 178--205\relax
\mciteBstWouldAddEndPuncttrue
\mciteSetBstMidEndSepPunct{\mcitedefaultmidpunct}
{\mcitedefaultendpunct}{\mcitedefaultseppunct}\relax
\EndOfBibitem
\bibitem[Cuesta and Achouri(2025)Cuesta, and Achouri]{cuesta2025}
Cuesta,~F.~S.; Achouri,~K. A General Expression for Homogeneous Metasurface Scattering. \emph{arXiv preprint arXiv:2502.19013} \textbf{2025}, \relax
\mciteBstWouldAddEndPunctfalse
\mciteSetBstMidEndSepPunct{\mcitedefaultmidpunct}
{}{\mcitedefaultseppunct}\relax
\EndOfBibitem
\bibitem[Nanz(2016)]{nanz2016toroidal}
Nanz,~S. \emph{Toroidal multipole moments in classical electrodynamics: an analysis of their emergence and physical significance}; Springer, 2016\relax
\mciteBstWouldAddEndPuncttrue
\mciteSetBstMidEndSepPunct{\mcitedefaultmidpunct}
{\mcitedefaultendpunct}{\mcitedefaultseppunct}\relax
\EndOfBibitem
\bibitem[Prokhorov \latin{et~al.}(2022)Prokhorov, Terekhov, Gubin, Shesterikov, Ni, Tuz, and Evlyukhin]{prokhorov2022resonant}
Prokhorov,~A.~V.; Terekhov,~P.~D.; Gubin,~M.~Y.; Shesterikov,~A.~V.; Ni,~X.; Tuz,~V.~R.; Evlyukhin,~A.~B. Resonant light trapping via lattice-induced multipole coupling in symmetrical metasurfaces. \emph{ACS Photonics} \textbf{2022}, \emph{9}, 3869--3875\relax
\mciteBstWouldAddEndPuncttrue
\mciteSetBstMidEndSepPunct{\mcitedefaultmidpunct}
{\mcitedefaultendpunct}{\mcitedefaultseppunct}\relax
\EndOfBibitem
\bibitem[Achouri \latin{et~al.}(2023)Achouri, Tiukuvaara, and Martin]{achouri2023spatial}
Achouri,~K.; Tiukuvaara,~V.; Martin,~O.~J. Spatial symmetries in nonlocal multipolar metasurfaces. \emph{Advanced Photonics} \textbf{2023}, \emph{5}, 046001--046001\relax
\mciteBstWouldAddEndPuncttrue
\mciteSetBstMidEndSepPunct{\mcitedefaultmidpunct}
{\mcitedefaultendpunct}{\mcitedefaultseppunct}\relax
\EndOfBibitem
\bibitem[Hsu \latin{et~al.}(2016)Hsu, Zhen, Stone, Joannopoulos, and Soljačić]{hsu2016bound}
Hsu,~C.~W.; Zhen,~B.; Stone,~A.~D.; Joannopoulos,~J.~D.; Soljačić,~M. Bound states in the continuum. \emph{Nature Reviews Materials} \textbf{2016}, \emph{1}, 16048\relax
\mciteBstWouldAddEndPuncttrue
\mciteSetBstMidEndSepPunct{\mcitedefaultmidpunct}
{\mcitedefaultendpunct}{\mcitedefaultseppunct}\relax
\EndOfBibitem
\bibitem[Kildishev \latin{et~al.}(2023)Kildishev, Achouri, and Smirnova]{kildishev2023art}
Kildishev,~A.~V.; Achouri,~K.; Smirnova,~D. The art of finding the optimal scattering center (s). \emph{Advanced Optical Materials} \textbf{2023}, 2402787\relax
\mciteBstWouldAddEndPuncttrue
\mciteSetBstMidEndSepPunct{\mcitedefaultmidpunct}
{\mcitedefaultendpunct}{\mcitedefaultseppunct}\relax
\EndOfBibitem
\bibitem[Paknys(2016)]{paknys2016}
Paknys,~R. \emph{Applied {{Frequency}}-{{Domain Electromagnetics}}}, 1st ed.; Wiley, 2016\relax
\mciteBstWouldAddEndPuncttrue
\mciteSetBstMidEndSepPunct{\mcitedefaultmidpunct}
{\mcitedefaultendpunct}{\mcitedefaultseppunct}\relax
\EndOfBibitem
\bibitem[Idemen(2011)]{idemen2011discontinuities}
Idemen,~M.~M. \emph{Discontinuities in the electromagnetic field}; John Wiley \& Sons, 2011\relax
\mciteBstWouldAddEndPuncttrue
\mciteSetBstMidEndSepPunct{\mcitedefaultmidpunct}
{\mcitedefaultendpunct}{\mcitedefaultseppunct}\relax
\EndOfBibitem
\bibitem[Achouri \latin{et~al.}(2022)Achouri, Tiukuvaara, and Martin]{achouri2022multipolar}
Achouri,~K.; Tiukuvaara,~V.; Martin,~O.~J. Multipolar modeling of spatially dispersive metasurfaces. \emph{IEEE Transactions on Antennas and Propagation} \textbf{2022}, \emph{70}, 11946--11956\relax
\mciteBstWouldAddEndPuncttrue
\mciteSetBstMidEndSepPunct{\mcitedefaultmidpunct}
{\mcitedefaultendpunct}{\mcitedefaultseppunct}\relax
\EndOfBibitem
\bibitem[Tiukuvaara \latin{et~al.}(2023)Tiukuvaara, Martin, and Achouri]{tiukuvaara2023quadrupolar}
Tiukuvaara,~V.; Martin,~O.~J.; Achouri,~K. Quadrupolar susceptibility modeling of substrated metasurfaces with application to the generalized Brewster effect. \emph{Optics Express} \textbf{2023}, \emph{31}, 22982--22996\relax
\mciteBstWouldAddEndPuncttrue
\mciteSetBstMidEndSepPunct{\mcitedefaultmidpunct}
{\mcitedefaultendpunct}{\mcitedefaultseppunct}\relax
\EndOfBibitem
\bibitem[Evlyukhin(2024)]{10.1117/12.3016150}
Evlyukhin,~A.~B. {Multipole coupling mechanisms of light propagation control and trapping by dielectric nanoparticle structures}. Metamaterials XIV. 2024; p PC129900W\relax
\mciteBstWouldAddEndPuncttrue
\mciteSetBstMidEndSepPunct{\mcitedefaultmidpunct}
{\mcitedefaultendpunct}{\mcitedefaultseppunct}\relax
\EndOfBibitem
\bibitem[Papas(1988)]{papas1988}
Papas,~C.~H. \emph{Theory of Electromagnetic Wave Propagation}; Dover Books on Physics and Chemistry; Dover Publications: New York, 1988\relax
\mciteBstWouldAddEndPuncttrue
\mciteSetBstMidEndSepPunct{\mcitedefaultmidpunct}
{\mcitedefaultendpunct}{\mcitedefaultseppunct}\relax
\EndOfBibitem
\bibitem[Albooyeh \latin{et~al.}(2016)Albooyeh, Tretyakov, and Simovski]{albooyeh2016electromagnetic}
Albooyeh,~M.; Tretyakov,~S.; Simovski,~C. Electromagnetic characterization of bianisotropic metasurfaces on refractive substrates: General theoretical framework. \emph{Annalen der Physik} \textbf{2016}, \emph{528}, 721--737\relax
\mciteBstWouldAddEndPuncttrue
\mciteSetBstMidEndSepPunct{\mcitedefaultmidpunct}
{\mcitedefaultendpunct}{\mcitedefaultseppunct}\relax
\EndOfBibitem
\bibitem[Shalin(2010)]{shalin2010broadband}
Shalin,~A.~S. Broadband blooming of a medium modified by an incorporated layer of nanocavities. \emph{JETP letters} \textbf{2010}, \emph{91}, 636--642\relax
\mciteBstWouldAddEndPuncttrue
\mciteSetBstMidEndSepPunct{\mcitedefaultmidpunct}
{\mcitedefaultendpunct}{\mcitedefaultseppunct}\relax
\EndOfBibitem
\bibitem[Shalin and Moiseev(2009)Shalin, and Moiseev]{shalin2009optical}
Shalin,~A.; Moiseev,~S. Optical properties of nanostructured layers on the surface of an underlying medium. \emph{Optics and Spectroscopy} \textbf{2009}, \emph{106}, 916--925\relax
\mciteBstWouldAddEndPuncttrue
\mciteSetBstMidEndSepPunct{\mcitedefaultmidpunct}
{\mcitedefaultendpunct}{\mcitedefaultseppunct}\relax
\EndOfBibitem
\bibitem[Miroshnichenko \latin{et~al.}(2015)Miroshnichenko, Evlyukhin, Kivshar, and Chichkov]{evlyukhin2015substrate}
Miroshnichenko,~A.~E.; Evlyukhin,~A.~B.; Kivshar,~Y.~S.; Chichkov,~B.~N. Substrate-Induced Resonant Magnetoelectric Effects for Dielectric Nanoparticles. \emph{ACS Photonics} \textbf{2015}, \emph{2}, 1423--1428\relax
\mciteBstWouldAddEndPuncttrue
\mciteSetBstMidEndSepPunct{\mcitedefaultmidpunct}
{\mcitedefaultendpunct}{\mcitedefaultseppunct}\relax
\EndOfBibitem
\bibitem[Ustimenko \latin{et~al.}(2025)Ustimenko, Rockstuhl, and Kildishev]{ustimenko2025optimal}
Ustimenko,~N.; Rockstuhl,~C.; Kildishev,~A.~V. Optimal multipole center for subwavelength acoustic scatterers. \emph{arXiv preprint arXiv:2501.06168} \textbf{2025}, \relax
\mciteBstWouldAddEndPunctfalse
\mciteSetBstMidEndSepPunct{\mcitedefaultmidpunct}
{}{\mcitedefaultseppunct}\relax
\EndOfBibitem
\bibitem[Novotny and Hecht(2012)Novotny, and Hecht]{novotny2012principles}
Novotny,~L.; Hecht,~B. \emph{Principles of nano-optics}; Cambridge university press, 2012\relax
\mciteBstWouldAddEndPuncttrue
\mciteSetBstMidEndSepPunct{\mcitedefaultmidpunct}
{\mcitedefaultendpunct}{\mcitedefaultseppunct}\relax
\EndOfBibitem
\bibitem[Liu and Choi(2018)Liu, and Choi]{liu2018extreme}
Liu,~M.; Choi,~D.-Y. Extreme Huygens’ metasurfaces based on quasi-bound states in the continuum. \emph{Nano letters} \textbf{2018}, \emph{18}, 8062--8069\relax
\mciteBstWouldAddEndPuncttrue
\mciteSetBstMidEndSepPunct{\mcitedefaultmidpunct}
{\mcitedefaultendpunct}{\mcitedefaultseppunct}\relax
\EndOfBibitem
\bibitem[Poleva \latin{et~al.}(2023)Poleva, Frizyuk, Baryshnikova, Evlyukhin, Petrov, and Bogdanov]{poleva2023multipolar}
Poleva,~M.; Frizyuk,~K.; Baryshnikova,~K.; Evlyukhin,~A.; Petrov,~M.; Bogdanov,~A. Multipolar theory of bianisotropic response of meta-atoms. \emph{Physical Review B} \textbf{2023}, \emph{107}, L041304\relax
\mciteBstWouldAddEndPuncttrue
\mciteSetBstMidEndSepPunct{\mcitedefaultmidpunct}
{\mcitedefaultendpunct}{\mcitedefaultseppunct}\relax
\EndOfBibitem
\bibitem[Gladyshev \latin{et~al.}(2023)Gladyshev, Karamanos, Kuhn, Beutel, Weiss, Rockstuhl, and Bogdanov]{gladyshev2023inverse}
Gladyshev,~S.; Karamanos,~T.~D.; Kuhn,~L.; Beutel,~D.; Weiss,~T.; Rockstuhl,~C.; Bogdanov,~A. Inverse design of all-dielectric metasurfaces with accidental bound states in the continuum. \emph{Nanophotonics} \textbf{2023}, \emph{12}, 3767--3779\relax
\mciteBstWouldAddEndPuncttrue
\mciteSetBstMidEndSepPunct{\mcitedefaultmidpunct}
{\mcitedefaultendpunct}{\mcitedefaultseppunct}\relax
\EndOfBibitem
\bibitem[Gurvitz \latin{et~al.}(2019)Gurvitz, Ladutenko, Dergachev, Evlyukhin, Miroshnichenko, and Shalin]{gurvitz2019high}
Gurvitz,~E.~A.; Ladutenko,~K.~S.; Dergachev,~P.~A.; Evlyukhin,~A.~B.; Miroshnichenko,~A.~E.; Shalin,~A.~S. The high-order toroidal moments and anapole states in all-dielectric photonics. \emph{Laser \& Photonics Reviews} \textbf{2019}, \emph{13}, 1800266\relax
\mciteBstWouldAddEndPuncttrue
\mciteSetBstMidEndSepPunct{\mcitedefaultmidpunct}
{\mcitedefaultendpunct}{\mcitedefaultseppunct}\relax
\EndOfBibitem
\bibitem[Normand and Raynal(1982)Normand, and Raynal]{normand1982relations}
Normand,~J.-M.; Raynal,~J. Relations between Cartesian and spherical components of irreducible Cartesian tensors. \emph{Journal of Physics A: Mathematical and General} \textbf{1982}, \emph{15}, 1437\relax
\mciteBstWouldAddEndPuncttrue
\mciteSetBstMidEndSepPunct{\mcitedefaultmidpunct}
{\mcitedefaultendpunct}{\mcitedefaultseppunct}\relax
\EndOfBibitem
\bibitem[Riccardi \latin{et~al.}(2022)Riccardi, Kiselev, Achouri, and Martin]{riccardi2022}
Riccardi,~M.; Kiselev,~A.; Achouri,~K.; Martin,~O. J.~F. Multipolar Expansions for Scattering and Optical Force Calculations beyond the Long Wavelength Approximation. \emph{Phys. Rev. B} \textbf{2022}, \emph{106}, 115428\relax
\mciteBstWouldAddEndPuncttrue
\mciteSetBstMidEndSepPunct{\mcitedefaultmidpunct}
{\mcitedefaultendpunct}{\mcitedefaultseppunct}\relax
\EndOfBibitem
\bibitem[Tretyakov(2003)]{9100112}
Tretyakov,~S.~A. \emph{Analytical Modeling in Applied Electromagnetics}; Artech {{House}} Electromagnetic Analysis Series; Artech House: Boston, Mass., 2003\relax
\mciteBstWouldAddEndPuncttrue
\mciteSetBstMidEndSepPunct{\mcitedefaultmidpunct}
{\mcitedefaultendpunct}{\mcitedefaultseppunct}\relax
\EndOfBibitem
\end{mcitethebibliography}

\end{document}